\documentclass[pdflatex,sn-nature]{sn-jnl}
\usepackage{geometry}
\usepackage{graphicx}
\usepackage{multirow}
\usepackage{amsmath,amssymb,amsfonts}
\usepackage{amsthm}
\usepackage{mathrsfs}
\usepackage[title]{appendix}
\usepackage{xcolor}
\usepackage{textcomp}
\usepackage{manyfoot}
\usepackage{booktabs}
\usepackage{algorithm}
\usepackage{algorithmicx}
\usepackage{algpseudocode}
\usepackage{listings}
\usepackage{float}
\usepackage{url}
\usepackage{longtable}
\usepackage{array}
\usepackage{hyperref}
\usepackage{xurl}

\theoremstyle{thmstyleone}%
\theoremstyle{thmstyletwo}%

\theoremstyle{thmstylethree}%

\begin{document}
	\title[Article Title]{Retweeting patterns reveal partisan asymmetries in information sharing and signal state-level electoral outcomes in the 2020 US presidential election}

	\author[1]{\fnm{Dawei} \sur{Li}}
	\equalcont{These authors contributed equally to this work.}
	
	\author[1]{\fnm{Qinjun} \sur{Wan}}
	\equalcont{These authors contributed equally to this work.}
	
	\author*[2,4,5]{\fnm{Xin} \sur{Wang}}
	\email{wangxin\_1993@buaa.edu.cn}
	
	\author[2]{\fnm{Dahe} \sur{Jiang}}
	
	\author[2]{\fnm{Junyu} \sur{Lu}}
	
	\author[2]{\fnm{Shitong} \sur{Wang}}
	
	\author[1]{\fnm{Zhelun} \sur{Zhang}}
	
	\author[1]{\fnm{Zhenyu} \sur{Guan}}
	
	\author[2,3,6,7]{\fnm{Shaoting} \sur{Tang}}
	
	\author[8,9,10]{\fnm{Feng} \sur{Fu}}

	\affil[1]{\orgname{School of Cyber Science and Technology,
			Beihang University, Beijing, 100191, China}}
	
	\affil[2]{\orgname{School of Artificial Intelligence,
			Beihang University, Beijing, 100191, China}}
	
	\affil[3]{\orgname{Hangzhou International Innovation Institute,
			Beihang University, Hangzhou, 311115, China}}
	
	\affil[4]{\orgname{Key Laboratory of Mathematics, Informatics and Behavioral Semantics,
			Beihang University, Beijing, 100191, China}}
	
	\affil[5]{\orgname{Beijing Advanced Innovation Center for Future Blockchain and Privacy Computing,
			Beihang University, Beijing, 100191, China}}
	
	\affil[6]{\orgname{Institute of Trustworthy Artificial Intelligence,
			Zhejiang Normal University, Hangzhou, 310012, China}}
	
	\affil[7]{\orgname{Shandong Key Lab of Complex Medical Intelligence and Aging,
			Shandong Medical and Pharmaceutical University, Yantai, 264003, China}}
	
	\affil[8]{\orgname{Department of Mathematics,
			Dartmouth College, Hanover, NH, USA}}
	
	\affil[9]{\orgname{Department of Biomedical Data Science,
			Geisel School of Medicine at Dartmouth, Lebanon, NH 03756, USA}}
	
	\affil[10]{\orgname{Department of Applied Mathematics,
			School of Engineering \& Applied Science, Yale University,
			New Haven, CT 06520, USA}}

	\abstract{As social media has become a primary source of political information, concerns are growing that online information consumption may intensify political polarization and reshape democratic processes. Yet how large-scale online information consumption organizes polarized political ecosystems and relates to offline voting remains unclear. Here we reconstruct partisan information ecosystems from more than 500 million tweets generated during the 2020 US presidential election, integrating information diffusion across media outlets, influencers, and social peers. Contrary to the conventional expectation that broader information exposure promotes moderation, we find that retweeting a wider range of political information is associated with stronger political polarization. We further show systematic asymmetries in political information ecosystems: the Republican ecosystem is highly concentrated and ideologically biased, whereas the Democratic ecosystem is more dispersed and primarily organized around relatively moderate left-leaning sources. Finally, we demonstrate that several shared retweeting signatures robustly map onto state-level electoral outcomes despite cross-state heterogeneity. Our findings reveal asymmetric political information ecosystems as a key organizing principle of online political communication and identify collective retweeting patterns as a scalable signal of real-world voting behaviours.}

	\maketitle
	\section{Introduction}\label{sec1}
	Social media platforms have transformed political communication from a largely broadcast into an interactive environment where users actively select, redistribute, and amplify political information\cite{lazer2009computational}. Political information now circulates through interconnected layers of media organizations, political figures and influential accounts, and social peers, while users actively select which messages to endorse and amplify\cite{king2017news,bond2012experiment,onnela2010spontaneous}. On Twitter, retweeting is a central mechanism of this participatory circulation: by forwarding messages to their own followers, users publicly express agreement with selected content while increasing its visibility and reach\cite{guess2023reshares}. When repeated across millions of users, these individual choices collectively shape which sources and narratives attract attention, where influence becomes concentrated, and how political information diffuses across communities\cite{flamino2023political,brady2017emotion,vosoughi2018spread,appel2026deceptive}.

	Over the past decade, research has increasingly focused on patterns of political information exposure on social media and their implications for collective cognition \cite{bakshy2015exposure,flaxman2016filter,yang2020exposure,liu2020homogeneity}. Among the most prominent concerns are the emergence of echo chambers and the intensification of political polarization \cite{cinelli2020selective,goldenberg2023homophily,del2016echo,liu2021modeling}. Studies of political echo chambers on Twitter, Facebook, TikTok and other platforms have consistently identified ideological segregation and partisan clustering, showing that users with similar political orientations tend to interact within relatively closed information environments \cite{schmidt2017anatomy,cota2019quantifying,del2016spreading}. Algorithmic recommendation systems may further reinforce such echo-chamber dynamics by prioritizing content aligned with users' existing preferences and reducing exposure to opposing viewpoints \cite{lazer2015rise,gauthier2026political,santos2021link,huszar2022algorithmic,ibrahim2026systematic}. Furthermore, external campaign efforts can interact with endogenous opinion dynamics in online networks, intensifying the polarization of swing voters and potentially giving rise to more nuanced dynamics, such as backfire effects \cite{wang2020public,liu2026evolution}. These patterns have raised fundamental concerns about democratic elections in the digital age\cite{lazer2018science,bakcoleman2021stewardship,levin2021dynamics,watts2021measuring}. Democratic decision-making depends on citizens’ ability to form political judgments based on diverse and accessible information, yet polarized online environments may reduce opportunities for cross-cutting exposure, reinforcing group identities and political polarization\cite{rathje2021out,iyengar2008selective,hart2009feeling}.
	
	A natural solution to these concerns is to broaden users' exposure to dissimilar information. However, whether exposure to a broader range of political information moderates or reinforces political polarization remains contested \cite{bail2018exposure,nyhan2023like,allcott2024deactivation}. A conventional view holds that cross-cutting exposure broadens awareness of alternative viewpoints, reduces ideological rigidity, and thereby promotes moderation \cite{mutz2002crosscutting,balietti2021reducing,jost2022cognitive}. This convergence intuition is also reflected in classical models of opinion dynamics: repeated exchange and mutual adjustment of opinions can drive opinions toward a common or intermediate position when individuals remain open to one another \cite{degroot1974consensus,deffuant2000mixing,hegselmann2002opinion}. The same logic has motivated interventions that seek to reduce polarization by diversifying information exposure and increasing contact with politically dissimilar views\cite{balietti2021reducing}. A competing view suggests that broader exposure may not lead to moderation, but instead strengthen existing political positions and intensify polarization\cite{baumann2020modeling,dandekar2013biased,haghtalab2021belief}. Exposure to opposing information may trigger motivated reasoning and biased assimilation, leading people to question conflicting views while more readily accepting information that supports their existing beliefs, thereby strengthening their original political positions \cite{taber2006motivated,jost2022cognitive,weeks2015emotions,rathje2023accuracy}. Correspondingly, opinion dynamics models that incorporate biased assimilation show that greater interaction does not necessarily produce consensus and can instead intensify disagreement \cite{dandekar2013biased}. A Twitter field experiment provides direct evidence of such a backfire pattern, showing that Republican participants exposed to messages from opposing political elites became more conservative \cite{bail2018exposure}. We test these competing views using retweeting behavior during the 2020 US presidential election. We find that broader engagement with political information is associated with stronger political polarization, supporting the reinforcement account.
	
	Political information is neither produced nor circulated at a single level. Professional media organizations introduce and frame political narratives, influencers select and interpret them while concentrating public attention, and social peers further redistribute these narratives through interpersonal networks\cite{king2017news,wojcieszak2022elites,mosleh2021partisanship}. Previous studies have shown that left and right users are embedded in systematically different political information environments, with partisan differences in the concentration of attention, the ideological composition of information sources, patterns of elite influence, and the circulation of low-credibility content \cite{flamino2023political,grinberg2019fake,guess2020exposure,bovet2019influence,mosleh2024differences}. Using more than 500 million tweets during the 2020 US presidential election, we reconstruct information diffusion across three levels, media outlets, political influencers and social peers. We find a systematic asymmetry between the two partisan ecosystems that is evident in both top-down and bottom-up information diffusion. From the top down, the Republican ecosystem is concentrated around a small set of strongly right-leaning and ideologically extreme influencers, whereas the Democratic ecosystem is more dispersed and organized around relatively moderate left-leaning sources. From the bottom up, right-leaning users engage more with cross-partisan information, while left users are more concentrated within their own partisan networks. These patterns show that partisan asymmetry operates across both elite-driven and peer-driven political information diffusion.
	
	Finally, we raise an important question: can collective online information behaviour reveal collective offline voting behaviour? Election research has traditionally relied on surveys and telephone polls\cite{jennings2018polling,galesic2018socialcircles}, but these approaches are costly, difficult to scale, and can be affected by uneven or non-representative sampling \cite{bovet2018validation,bradley2021unrepresentative}. Large-scale digital traces offer a different source of behavioural information\cite{allcott2017social,grover2019polarization,bail2020assessing,lazer2020computational}, yet online activity cannot be interpreted directly as individual vote choice\cite{bovet2018validation,lorenzspreen2023systematic}. Previous studies have shown that social-media data can reveal political attitudes, ideological structure, political participation and election-related dynamics \cite{waller2021quantifying,eady2023exposure,green2022online}. However, it remains unclear whether collective patterns of online information sharing contain a stable signal of collective electoral outcomes across regions with different populations and political environments. We address this question by examining retweeting behaviour across US states during the 2020 presidential election. We identify several shared retweeting signatures that recur across states and robustly correspond to state-level election outcomes despite cross-state heterogeneity. Our results indicate that large-scale social-media behaviour may provide a scalable, complementary signal for understanding collective electoral outcomes.

	%
	%
	\section{Results}\label{sec2}
	We tracked election-related Twitter discourse during the 2020 U.S. presidential election from December 2019 to February 2021. We used a publicly available Tweet ID dataset collected for the 2020 U.S. presidential election using the names of presidential candidates as query terms \cite{echen2020dataset}. We retrieved tweet information with these Tweet IDs through the Twitter Developer API and obtained 542,994,806 available, non-reply tweets from an initial collection of approximately 1.6 billion Tweet IDs. These valid records consist of retweets and quote tweets, capturing how users consumed election-related information on Twitter. Quote tweets only accounted for 6.83\% of these interactions. Stance analysis showed that explicitly oppositional quote tweets accounted for only a very small share of the overall interaction records. We pooled retweets and quote tweets in the subsequent analyses, with the detailed proportions and stance-analysis procedure reported in Methods. Because the dataset was reconstructed from publicly available Tweet IDs, it should be interpreted as a large-scale sample of election-related Twitter activity rather than a complete archive of all Twitter activity.
	
	To characterize users' political orientations, we constructed a media-based ideology measure from their retweeting behavior. Retweeting on Twitter is commonly treated as a strong behavioral signal of endorsement and ideological alignment, and has been widely used to infer users' political ideology from the information they redistribute \cite{flamino2023political}. Following previous approaches that infer user leaning from the political orientation of shared news sources \cite{cinelli2021echo}, we assigned identified media outlets to seven ideological categories based on Media Bias/Fact Check (MBFC), ranging from Extreme Left to Extreme Right. These categories were mapped onto a numerical scale from $-3$ to $3$. For each user within a given analysis period, we averaged the ideological scores of the media sources associated with that user's retweets to obtain a continuous ideology score. Detailed procedures for media identification, source matching, and score aggregation are described in Methods. All data and code used in this paper are available in Ref.~\cite{retweeting_code}.

	\begin{figure}[htbp]
		\centering
		\includegraphics[width=1\textwidth]{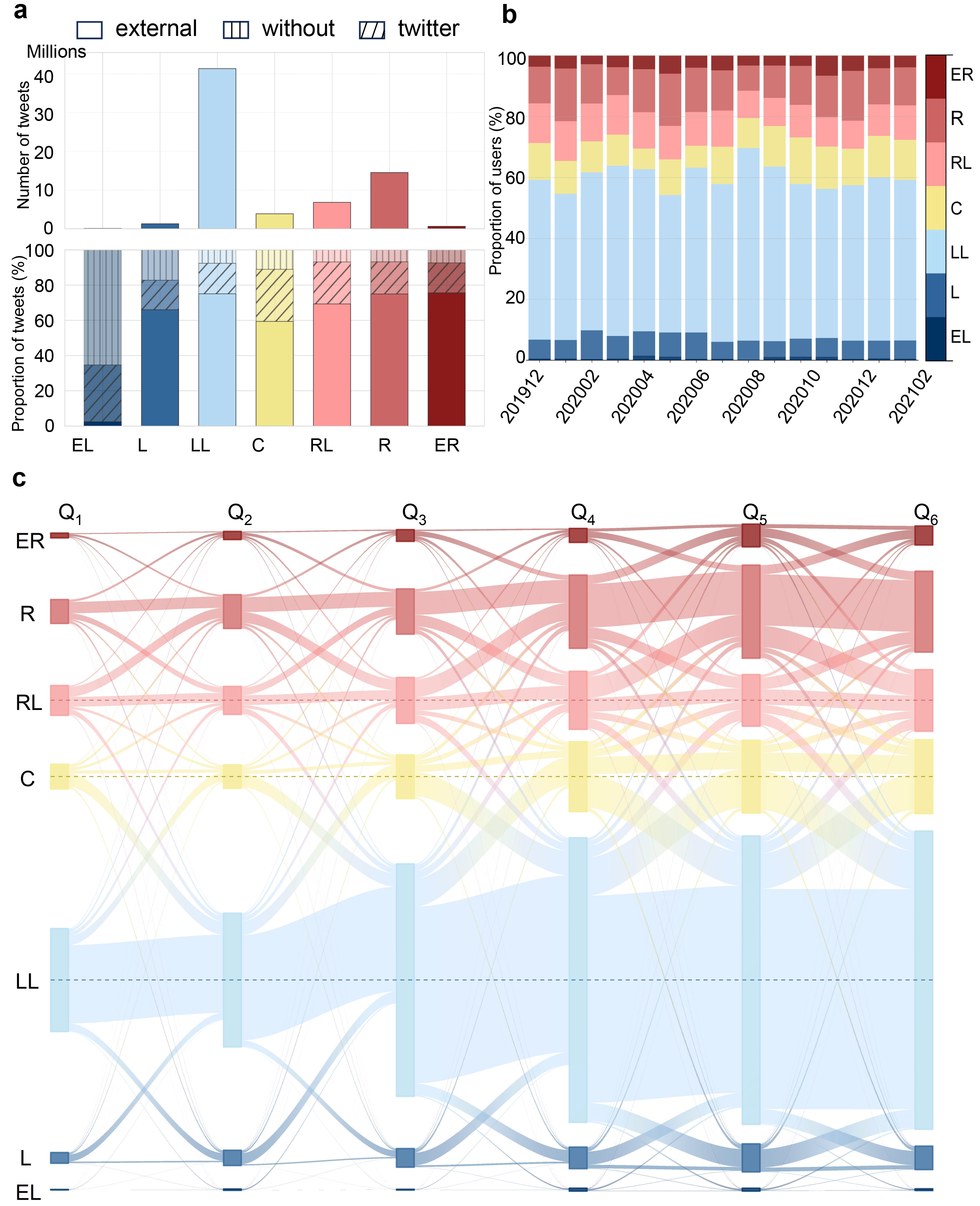}
		\caption{Temporal dynamics and ideological shifts of Twitter users during the 2020 U.S. election. 
			\textbf{a,} Distribution of media-based retweeting records across ideological groups. The upper panel shows the total number of media-based retweeting records for each ideological group, aggregated across External URLs, Twitter URLs, and Without URLs. The lower panel shows the proportional distribution of these records across the three source channels within each ideological group. Colors indicate seven ideological categories: Extreme Left (EL), Left (L), Left-Leaning (LL), Center (C), Right-Leaning (RL), Right (R), and Extreme Right (ER).
			\textbf{b,} Monthly distribution of users across ideological categories from December 2019 to February 2021, based on the media-based ideology scoring system. Values represent the monthly proportion of users in each ideological category. Users active in multiple months are counted once in each month.
			\textbf{c,} Sankey diagram depicting user-level ideological transitions across six consecutive analysis periods. Nodes represent ideological categories in each analysis period, and flow width represents the number of users transitioning between adjacent periods. Users observed in only one of two adjacent periods contribute to node size but not to transition flows.
			\label{fig1}}
	\end{figure}
	
	\subsection{Ideological evolution on Twitter}\label{subsec1}
	We divided the dataset into three mutually exclusive subsets: External URLs, Twitter URLs, and Without URLs. External URLs include retweeting records containing links to domains outside Twitter, capturing users’ spread of information from external news outlets, campaign websites, blogs, and other platform sources. Twitter URLs include records containing links pointing to Twitter itself, capturing internal circulation in which users redirected attention to Twitter-native content. Without URLs include retweeting records that contain no URLs, capturing information redistribution without link sharing. The media-based ideology measure introduced above captured ideological signals from media sources across all three channels with different coverage. Under the media-based scoring system, 5,306,383 tweets in Without URLs, 13,008,910 in Twitter URLs, and 50,313,771 in External URLs received ideological scores, representing 1.47\%, 13.58\%, and 61.18\% of the records in the respective subsets. Media-based scoring leverages the ideological leaning of linked sources, which is most effective in the external subset. For comparison, we constructed a parallel politician-based scoring system, because political information on Twitter is conveyed not only through media sources but also through political figures. Political figures were classified into three ideological categories: Left, Center, and Right. The politician-based system showed the opposite coverage pattern: it assigned ideological scores to 214,495,937 tweets in Without URLs and 56,595,626 tweets in Twitter URLs, accounting for 59.30\% and 59.09\% of the two subsets, respectively, but covered only 3,745,948 records in External URLs (4.55\%). Thus, media-based scoring captured ideological signals primarily through externally linked media sources, whereas politician-based scoring was more effective for Twitter-native and non-linked content. Monthly tweet volumes across the three subsets are summarized in Table S2. Detailed coverage statistics of the media- and politician-based scoring systems are respectively provided in Tables S3 and S7. The media and politician labels used for ideological scoring are provided in Tables S13–S16.

	We first examined who contributed to media-based information consumption and through what types of sources. Media-based information consumption was unevenly distributed across ideological groups. In Fig.~\ref{fig1}a, the upper panel reports the total number of media-based consumption records for each ideological group, summed across External URLs, Twitter URLs, and Without URLs, while the lower panel shows the percentage composition of these three source types. The upper panel shows that media-based consumption was concentrated among Left-Leaning users, who accounted for 41,339,675 consumption records, 60.26\% of all media-based consumption. Right users formed the second-largest group, with 14,522,749 records, accounting for 21.17\%. Right-Leaning and Center users accounted for 6,842,785 records (9.97\%) and 3,865,634 records (5.64\%), respectively, while Left, Extreme Right, and Extreme Left users contributed much smaller shares. Despite this uneven ideological composition, most ideological groups shared the same dominant source type. The lower panel of Fig.~\ref{fig1}a shows that External URLs dominated most ideological groups. Except for the small Extreme Left group, External URLs were the largest information source across all ideological categories, ranging from 59.34\% among Center users to 75.52\% among Extreme Right users. The two largest consumption groups, Left-Leaning and Right users, relied heavily on External URLs, which respectively accounted for 75.02\% and 74.95\% of their consumption. Center users also consumed mainly through External URLs, but showed a higher Twitter URL share of 29.66\%. By contrast, Extreme Left users accounted for only 0.11\% of all media-based consumption and showed a different source composition, with Without URLs as the dominant source. Source-specific distributions of scored records for the media- and politician-based scoring systems are respectively reported in Tables S4–S6 and S8–S10. Overall, media-based information consumption differed in who contributed to it, while the dominant types of sources through which information was consumed were comparatively consistent across ideological groups.

	The monthly ideological composition of users captured by the media-based scoring system was also highly uneven, with Left-Leaning users forming the dominant category throughout the observation period. Figure~\ref{fig1}b reports the monthly distribution of users associated with media-based information consumption, with each stacked bar representing one month and colored segments indicating the proportion of users in each ideological category. Left-Leaning users accounted for the largest share in every month from December 2019 to February 2021, ranging from 45.20\% in May 2020 to 63.29\% in August 2020, with a monthly average of 52.72\%. Across the full observation period, each user contributed one record per active month, yielding a total of 12,074,513 user-month records. Left-Leaning category accounted for 6,395,395 records, representing 52.97\% of the total. The other left-side categories were much smaller, with Extreme Left accounting for 90,315 records (0.75\%) and Left accounting for 759,090 records (6.29\%). Center category accounted for 1,497,387 records (12.40\%). On the right side, Right-Leaning category accounted for 1,275,519 records (10.56\%), Right category accounted for 1,539,060 records (12.75\%), and Extreme Right category accounted for 517,747 records (4.29\%). The three left-side categories, including EL, L, LL, together accounted for 7,244,800 of 12,074,513 records, representing 60.00\% of the total, while the three right-side categories, including ER, R, RL, accounted for 3,332,326 records, representing 27.60\% of the total. Monthly right-side presence fluctuated. The combined share of Right-Leaning, Right, and Extreme Right category  reached 34.51\% in January 2020 and 34.05\% in May 2020, declined to its lowest level of 20.50\% in August 2020, and then increased to 29.81\% in November 2020 and 30.55\% in December 2020. The monthly share of Center category ranged from 6.79\% in April 2020 to 15.30\% in October 2020. These patterns indicate that the user base involved in media-based retweeting behavior was persistently left-leaning in composition, while the right-side share became more visible again during the election and post-election months.

	Having characterized the temporal composition of ideological groups, we next shifted from group-level distributions to user-level ideological trajectories. Figure~\ref{fig1}c traces user transitions across six consecutive analysis periods using the media-based ideology categories. To understand the temporal shifts in political leanings, we divided the timeline into six consecutive analysis periods: pre-pandemic (December 2019 - February 2020), during the pandemic outbreak (March 2020 - May 2020), pre-election (June 2020 - August 2020), election(September 2020 - October 2020), election results announcement period (November 2020 - December 2020), and post-election(January 2021 - February 2021).  Left-Leaning (LL) users formed the most stable group. Their retention rate averaged 75.94\% across the five transitions, ranging from 72.78\% to 78.59\%. Movement from LL into the right-side categories (RL, R, and ER) was comparatively limited, averaging only 6.55\%. Center (C) users, by contrast, most frequently shifted toward LL, with an average C-to-LL transition rate of 53.66\%. Although C-to-C retention was lower, it increased from 13.11\% in Q1-Q2 to 24.64\% and 25.86\% in the final two transitions. LL remained the dominant destination for Center users. Right-Leaning (RL) users showed a broader spread across destination categories than LL users. Their average retention rate was 29.98\%, while an average of 32.07\% shifted to R. This rightward movement was strongest from Q3 to Q4, when the RL-to-R transition reached 38.60\%, and remained high at 37.12\% from Q4 to Q5. At the same time, RL users also showed substantial movement toward the left-side categories (EL, L, and LL), averaging 21.02\% across the five transitions. This cross-side movement was greater than the 6.55\% of LL users who moved into the right-side categories. R users were more stable than RL users, with an average retention rate of 51.44\%. Movement from R into the left-side categories was more limited, averaging 8.95\% across the five transitions. Extreme Right (ER) users were less likely to remain in ER, with an average retention rate of 17.06\%. ER users most frequently shifted to R, with an average ER-to-R transition rate of 41.19\%. Detailed transition patterns and the ideological composition of long-term-active users are reported in Fig.~S1 and Table~S11. Overall, these transition patterns show that ideological shift was asymmetric across the political spectrum. LL users formed a stable core, while Center users were repeatedly absorbed into LL. RL users showed greater dispersion and election-period movement toward R, and the right-side categories were organized around R.
	Together, Fig.~\ref{fig1}a-c reveal an asymmetric pattern from group-level composition to individual-level ideological change. At the group level, media-based retweeting activity was unevenly distributed across ideological categories, with Left-Leaning users constituting the largest group throughout the observation period. Moving from group-level composition to individual-level trajectories, we found that ideological change was characterized by asymmetric transition patterns across categories. Retention within the same ideological categories represented the dominant transition pattern for most groups, while users who changed categories exhibited distinct directional tendencies: LL users showed strong persistence with limited movement toward the right side, Center users predominantly shifted toward LL, and RL users displayed greater mobility with movements both toward R and across the left side.

	\subsection{Broader retweeting is associated with stronger political polarization}\label{subsec2}
	\begin{figure}[htbp]
		\centering
		\includegraphics[width=1\textwidth]{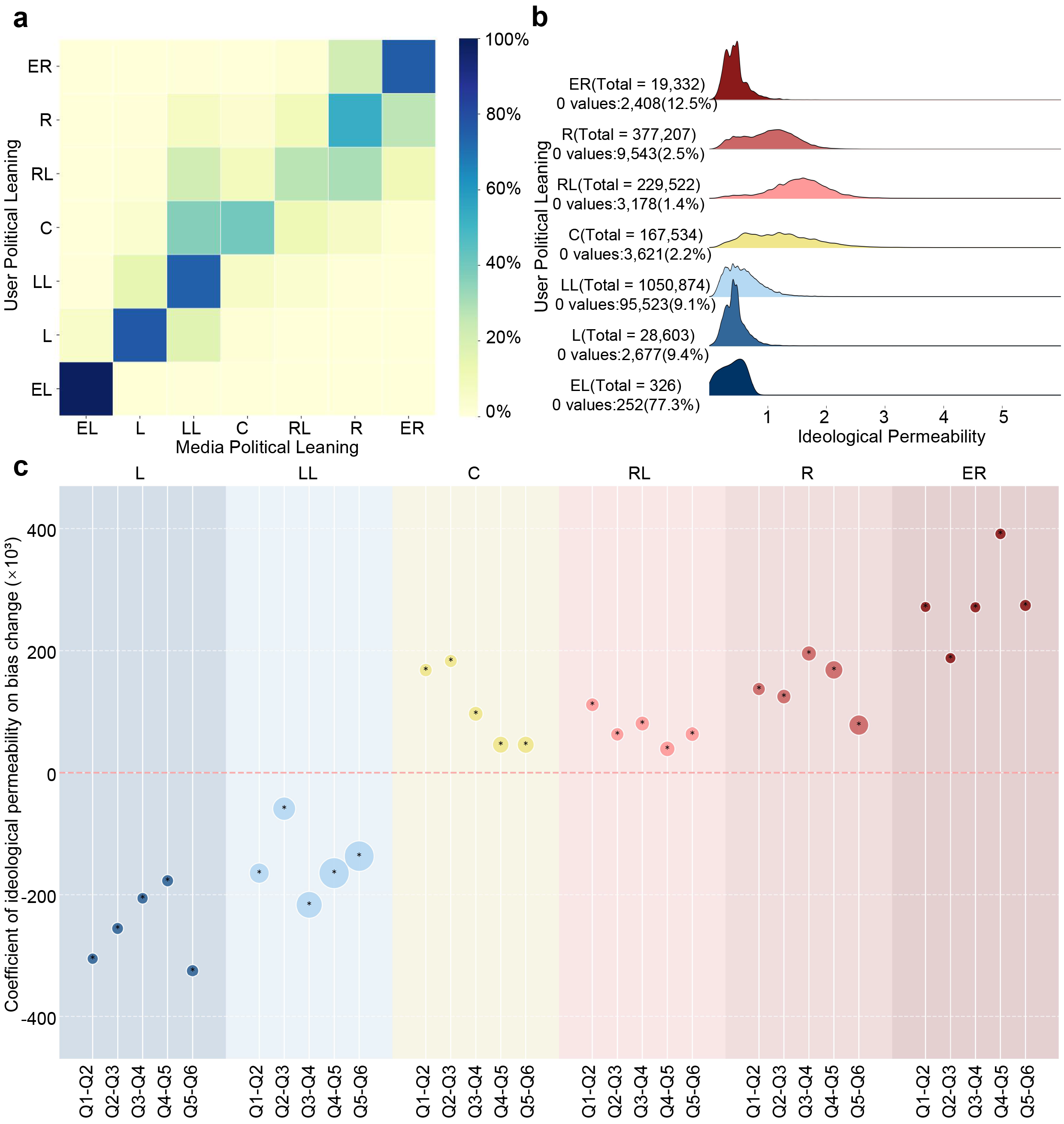}
		\caption{Broader retweeting is associated with stronger political polarization across ideological groups.  
			\textbf{a,} Cross-ideological information consumption heatmap during the 2020 U.S. presidential election. Rows represent user groups categorized by media-based political ideology, and columns represent the ideological leaning of consumed media content. The matrix was constructed by aggregating monthly consumption counts across all media-based information sources. Colour intensity represents the relative volume of cross-ideological media consumption. Center and Right-Leaning users consume information across a broader ideological spectrum.
			\textbf{b,} Ridge plots illustrating the distribution of non-zero ideological permeability across user categories, restricted to active users with an interaction frequency $\ge$ 5 in each quarter. Ideological permeability was calculated as the average scaled ideological distance between users' ideological score and the ideological leaning of consumed media content. Users with an absolute zero permeability score are excluded from the continuous distributions, with their exact counts and proportions annotated alongside. Center and Right-Leaning users exhibit higher ideological permeability.    
			\textbf{c,} Bubble plot illustrating the relationship between ideological permeability and bias change across six user categories over six quarters, excluding the Extreme Left category because of its limited sample size. For each political category and adjacent quarter transition, we fitted a robust linear model of bias change on ideological permeability. The x-axis indicates user political alignment and quarter, and the y-axis represents regression coefficients between ideological permeability and bias change, with positive values indicating a shift toward the right and negative values indicating a shift toward the left. For visual clarity, the plotted regression coefficients are multiplied by 1,000. Bubble size represents the number of observations included in each regression, and the symbols inside the bubbles indicate whether the corresponding coefficient is statistically significant at the 0.05 level. Results show Left-Leaning users tend to exhibit negative correlations, Right-Leaning users positive, with more ideologically extreme users showing stronger correlations. This analysis is restricted to active users with at least two recorded engagements in the starting quarter of each adjacent-quarter transition.
			\label{fig4}}
	\end{figure}

	Whether exposure to a broader range of political information reduces or reinforces political polarization remains contested. One perspective suggests that cross-cutting exposure may promote moderation by increasing awareness of alternative viewpoints and reducing ideological rigidity, whereas another perspective argues that exposure to opposing information may reinforce existing beliefs. To examine these competing possibilities using real-world Twitter behavior, we analyzed users' retweeting patterns and quantified the extent to which their retweeted media content extended beyond ideologically aligned sources.
	
	The heatmap in Figure \ref{fig4}a visualizes micro-level user retweeting patterns. The matrix reveals a diagonal concentration. Extreme Left, Left, Left-Leaning, Right, and Extreme Right users retweeted ideologically congruent media at shares of 98.28\%, 76.31\%, 74.46\%, 52.76\%, and 75.76\%, respectively. This indicates that ideologically aligned  information dominates users' media consumption. At the same time, exposure breadth was distinct across groups. Right-Leaning users exhibited the broadest retweeting pattern, with exposure distributed across Left-Leaning, Center, Right-Leaning, Right, and Extreme Right media categories, accounting for 21.85\%, 8.68\%, 27.30\%, 29.85\%, and 10.80\% of their retweeting activity, respectively. For Right-Leaning users, Right-Leaning and Right media together accounted for 57.15\%, while left-side media categories (Extreme Left, Left, and Left-Leaning) together accounted for 23.36\% of their retweeting activity. For Center users, Center and Left-Leaning media together accounted for 76.67\% of their retweeting activity, while right-side media categories (Right-Leaning, Right, and Extreme Right) accounted for 19.29\%. By contrast, Left-Leaning users showed narrower exposure than Right-Leaning users, with 89.60\% of their retweeting activity concentrated in left-side media categories and only 4.33\% in right-side media categories. The corresponding raw information-consumption matrix is provided in Fig.~S2. Overall, Fig.~\ref{fig4}a suggests that media retweeting was dominated by ideological alignment, but the extent of cross-ideological exposure was uneven, with Right-Leaning users showing the broadest exposure range.

To quantify the extent to which users consumed ideologically distant media, Fig.~\ref{fig4}b shows the distribution of ideological permeability scores at the user-quarter level. Each observation represents one user in one quarter and is included only if the user had at least five matched information-consumption engagements during that quarter, summed across External URLs, Twitter URLs, and Without URLs. Thus, all statistics reported for Fig.~\ref{fig4}b are calculated at the user-quarter level rather than the unique-user level. For each ideological group, the percentage of user-quarter observations with a score of zero is displayed alongside the group label, whereas the ridge density is estimated using observations with scores greater than zero. The proportion of zero-permeability observations varies across ideological groups. Extreme Left observations have the highest zero-permeability share, at 77.3\%, followed by Extreme Right observations at 12.5\%, Left observations at 9.4\%, and Left-Leaning observations at 9.1\%. Zero-permeability is much less common in the Right, Center, and Right-Leaning groups, accounting for only 2.5\%, 2.2\%, and 1.4\% of the corresponding user-quarter observations, respectively. The Right-Leaning and Center groups show the broadest distributions toward higher levels of ideological permeability. Right-Leaning observations have the highest median non-zero permeability score of 1.550, with an interquartile interval of [1.225, 1.833], and 15.10\% of all Right-Leaning user-quarter observations have scores greater than or equal to 2. Center observations also show a relatively broad distribution, with a median non-zero score of 1.167, an interquartile interval of [0.733, 1.600], and 10.11\% of all Center user-quarter observations reaching scores greater than or equal to 2. Right observations occupy an intermediate position, with a median non-zero permeability score of 1.037. The share of Right user-quarter observations with permeability scores greater than or equal to 2 is 1.17\%, which is lower than the corresponding shares in the Right-Leaning and Center groups but higher than those in the Extreme Left, Left, Left-Leaning, and Extreme Right groups. In contrast, the left-side groups show more limited ideological permeability. The median non-zero scores for Extreme Left, Left, and Left-Leaning observations are 0.404, 0.418, and 0.533, respectively, and the shares of all user-quarter observations with scores greater than or equal to 2 in these groups are 0.00\%, 0.03\%, and 0.12\%, respectively. Extreme Right observations also display a compact non-zero distribution, with a median of 0.417 and an interquartile interval of [0.300, 0.500]. Overall, Fig.~\ref{fig4}b shows an asymmetry in cross-ideological exposure at the user-quarter level. Observations in the Center and Right-Leaning groups are more likely to exhibit high ideological permeability scores, whereas observations in the left-side and extreme groups are more concentrated near lower scores, indicating consumption of ideologically more proximate information.
Regression analysis in Fig.~\ref{fig4}c examines the association between ideological permeability and subsequent changes in political leaning. Positive coefficients indicate that higher ideological permeability is associated with relatively more rightward subsequent changes, while negative coefficients indicate relatively more leftward subsequent changes. For visual clarity, the regression coefficients plotted in Fig.~\ref{fig4}c are multiplied by 1,000, whereas the coefficients reported in the text retain their original scale. Bubble size represents the number of observations included in each regression, and the symbols inside the bubbles indicate whether the corresponding coefficient is statistically significant at the 0.05 level. Extreme Left users are not shown because of the small sample size. For Left and Left-Leaning users, all 10 coefficients were negative, and all 10 reached statistical significance. The largest negative coefficients appear for Left users in Q5-Q6 ($\beta=-0.325$) and Q1-Q2 ($\beta=-0.305$), and for Left-Leaning users in Q3-Q4 ($\beta=-0.217$). This indicates that among left-side users, higher ideological permeability is associated with relatively more leftward subsequent changes. The pattern is reversed on the right side of the spectrum. All coefficients for Right-Leaning, Right, and Extreme Right users were positive, and all 15 reached statistical significance. Right-Leaning users show positive coefficients across all five transitions, ranging from $\beta=0.039$ to $\beta=0.111$. Right users show positive coefficients in all transitions, with larger coefficients in Q3-Q4 ($\beta=0.195$) and Q4-Q5 ($\beta=0.168$). Extreme Right users show the largest positive coefficients, with $\beta=0.271$, $\beta=0.391$, and $\beta=0.274$ from Q3-Q4 to Q5-Q6. Center users also show positive and significant coefficients across all five transitions, indicating that higher permeability among Center users is associated with relatively more rightward subsequent changes. The robustness of these regressions under alternative weighting multipliers and minimum information-consumption thresholds is examined in Figs. S3 and S4. To further link the user-level ideological transitions observed in Fig.~\ref{fig1}c with the differences in ideological permeability shown in Fig.~\ref{fig4}b, we compared subsequent bias changes between users in the top and bottom quartiles of ideological permeability within the Left-Leaning, Center, and Right-Leaning groups. Consistent with the regression results in Fig.~\ref{fig4}c, higher ideological permeability was associated with relatively more leftward subsequent changes among Left-Leaning users, but relatively more rightward subsequent changes among Center and Right-Leaning users. Detailed distributions of subsequent bias changes and corresponding statistical tests for the top- and bottom-quartile comparison are reported in Fig.~S5 and Table~S12. Overall, Fig.~\ref{fig4}c shows that higher ideological permeability is associated with relatively more leftward subsequent changes among Left and Left-Leaning users, and relatively more rightward subsequent changes among Center, Right-Leaning, Right, and Extreme Right users. This pattern is consistent with Bail et al., who show that exposure to opposing political accounts on Twitter made Republicans more conservative \cite{bail2018exposure}. It also aligns with Thurner et al., whose model shows that increased social connectivity can move societies past a critical threshold into strong polarization \cite{thurner2025more}. These results in our data suggest that broader retweeting across ideological boundaries is associated with a more polarized pattern of subsequent bias changes, with higher ideological permeability linked to relatively more leftward changes among left-side users and relatively more rightward changes among Center and right-side users.
\subsection{Asymmetric political information ecosystems}\label{subsec3}
\begin{figure}[htbp]
	\centering
	\includegraphics[width=1\textwidth]{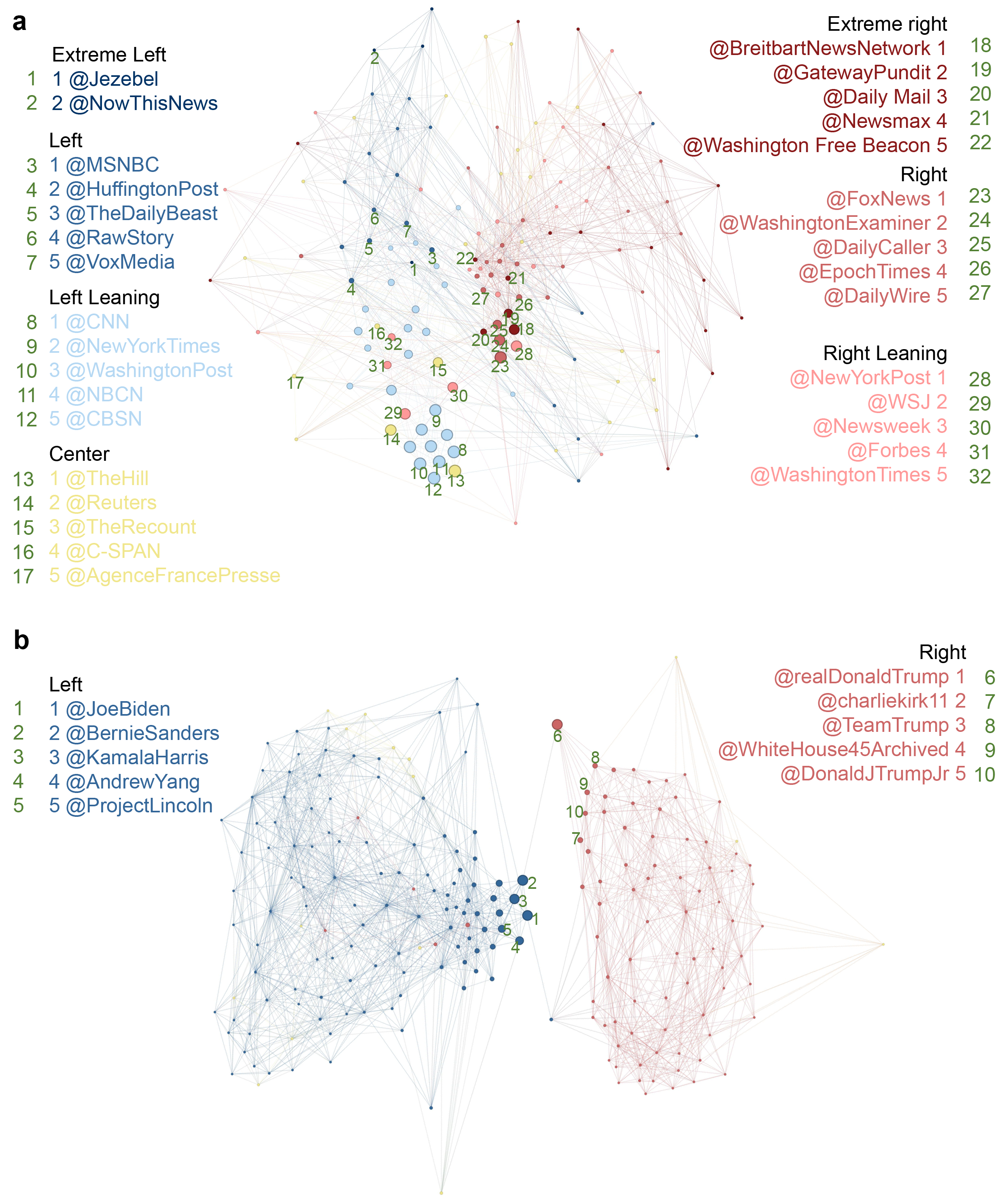}
	\caption{Co-exposure networks of media and political accounts during the 2020 U.S. presidential election. \textbf{a,} Media co-exposure network and \textbf{b,} politician co-exposure network. Node size indicates retweet volume, whereas edge weight captures audience overlap based on the Jaccard similarity between the user sets that retweeted the two accounts. For the media network, we retained the top 25 media accounts with the largest audience size in each ideological category. For the politician network, accounts were ranked by audience size, and the top 80\% formed the initial node set. Edges capture the extent to which two accounts reached overlapping audiences and are weighted by the Jaccard similarity between their influenced user sets. In both networks, we retained the union of each node's 10 edges with the highest Jaccard similarities, removed isolated nodes, and extracted the largest connected component. The resulting media co-exposure network contains 152 nodes and 1,257 edges, and the politician co-exposure network contains 207 nodes and 1,782 edges. Community structure was detected using the Louvain algorithm in Gephi. Node colour indicates the ideological category, with deep blue denoting Extreme Left, blue for Left, light blue for Left-Leaning, yellow for Center, light red for Right-Leaning, red for Right, and deep red for Extreme Right. Text beside each network lists the top five influencers in each ideological category, ranked by retweet count from highest to lowest, with green numbers indicating the corresponding labelled nodes in the network.\label{fig2}}
\end{figure}

The previous analyses revealed that users' retweeting patterns were shaped by ideological alignment, while ideological trajectories showed directional shifts across groups, consistent with previous studies showing that selective exposure and homophilous interactions on social media contribute to the formation of echo chambers \cite{cinelli2021echo}. These retweeting patterns and ideological trajectories raise a further question: how are such echo-chamber structures reflected in political information ecosystems? To address this question, we examine three major components of political information ecosystems: media, political actors, and peers. Media and politicians represent top-down information sources that shape users' exposure environments through co-exposure structures, whereas peers represent bottom-up social interactions formed through user-to-user retweeting networks.

%
%
%
%
%
The media co-exposure network revealed a polarized and asymmetric structure in which media accounts clustered by political leaning but differed in audience overlap and retweet volume. As shown in Figure~\ref{fig2}a, we constructed a media co-exposure network in which nodes represent media accounts, weighted edges represent audience overlap based on Jaccard similarity, and node size is proportional to the total number of retweets received by each account. Communities were identified using the Louvain algorithm, which detected seven communities in the media co-exposure network, with a modularity score of 0.561. The network layout was generated using the ForceAtlas 2 algorithm. Nodes are colored in varying shades of blue and red to represent the intensity of Democratic and Republican leanings. Media accounts exhibited a polarized structure. Representative accounts across the seven ideological categories included @Jezebel (Extreme Left), @MSNBC (Left), @CNN (Left-Leaning), @Reuters (Center), @NewYorkPost (Right-Leaning), @FoxNews (Right), and @BreitbartNewsNetwork (Extreme Right). Left and Left-Leaning outlets, represented by @MSNBC and @CNN, formed a dense area of shared audiences, while Right and Extreme Right outlets, represented by @FoxNews and @BreitbartNewsNetwork, were positioned on the opposite side of the network. Based on node-size comparisons, Left-Leaning nodes were generally larger than Right-Leaning and Right nodes, while nodes in the Right category were larger  than those in the Left category. In the ForceAtlas 2 layout, Center outlets appeared closer to the Left-Leaning region, whereas Right-Leaning outlets appeared more dispersed in the layout.
%

The politician co-exposure network also showed a left–right partition. As shown in Fig.~\ref{fig2}b, the co-exposure network of 207 political accounts contained three communities identified by the Louvain algorithm, with a modularity score of 0.542. In the ForceAtlas 2 layout, the network appears as two spatially separated regions dominated by left and right accounts, with relatively sparse connections between them. This indicates that audiences for Democratic and Republican political accounts had limited overlap. On the Republican side, @realDonaldTrump and nearby accounts appeared visually prominent, whereas Democratic accounts appeared more dispersed around several political figures, including @JoeBiden, @BernieSanders and @KamalaHarris. Taken together, these findings show that top-down political information sources, including media and politicians, form an asymmetric information ecosystem.
To relate this structure to the user-level ideological permeability observed in Fig.~\ref{fig4}b, we further examined shared audiences between left- and right-side media categories. After normalizing for differences in user-group size, Right-Leaning and Right users showed the highest shared-audience rates in combinations involving Left-Leaning media (42.03–53.38\%), whereas rates among Left and Left-Leaning users remained lower across all four combinations (7.46–29.65\%).  Detailed compositions of shared audiences across cross-camp media-category combinations are reported in Fig. S6. The ideological composition of these shared audiences provides complementary evidence consistent with the higher ideological permeability observed among Right and Right-Leaning users in Fig.~\ref{fig4}b.

\begin{figure}[htbp]
\centering
\includegraphics[width=1 \textwidth]{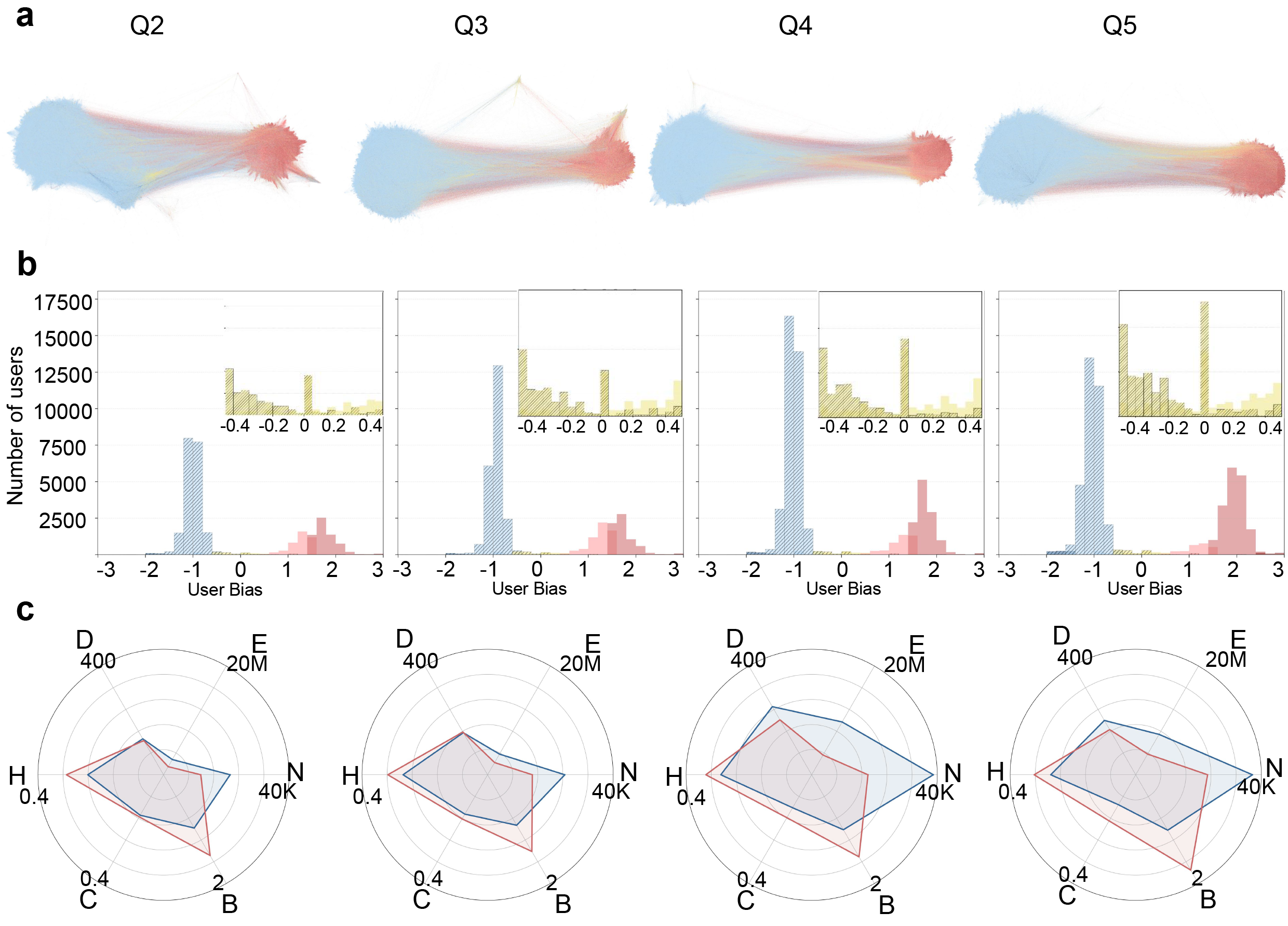}
\caption{Temporal evolution of core retweet networks, ideological distributions, and community-level structural statistics across four quarters.
\textbf{a,} For each quarter, we built a weighted undirected user-level retweet graph, in which nodes represent users, edges represent retweet relations between user pairs, and edge weights indicate the total frequency of retweet interactions after aggregating duplicate user pairs. After extracting the largest connected component, we retained users in the top 1\% of the unweighted degree distribution and visualized the resulting core network. Louvain community detection was applied to the weighted undirected graph with resolution set to 1.0 and random seed set to 42. Node colours indicate the ideological category. 
\textbf{b,} The ideological distributions of users within the two largest Louvain communities are shown in the main histograms, and the distribution for the third-largest community is shown in the inset at the upper right of each panel. The horizontal axis represents users' media-based bias scores on the same scale from $-3$ to $3$ as the main histograms, and the vertical axis represents the number of users.
\textbf{c,} Network-level statistics characterize the top two largest communities each quarter, which include the number of nodes $N$, the total count of edges $E$, the average clustering coefficients $C$, the edge-to-node ratio ($D=E/N$), degree heterophily $H$, and absolute average media bias scores $B$. Regarding degree heterophily $H$, values were calculated as the negative of the degree assortativity coefficient, such that larger values indicate a more heterogeneous network structure. The signed value of $B$ indicates ideological direction, while $|B|$ is shown in the figure to represent ideological intensity. These metrics quantitatively reflect the growth, cohesion, and ideological segregation within these communities over time.
\label{fig3}
}
\end{figure}

Following the top-down information structures revealed by media and political-actor co-exposure networks, we next examine the bottom-up organization of users through peer retweet interactions. We focus on four analysis periods, Q2–Q5, spanning March–December 2020 to capture the main run-up to the 2020 U.S. presidential election, the election itself, and its immediate aftermath. Q1 represents an earlier phase, whereas Q6 falls after the election. Constrained by main-text layout, we present these four core quarters in Fig.~\ref{fig3}. Across the four analysis quarters from Q2 to Q5, the core user retweet network consistently exhibits a bipolar structure in Fig.~\ref{fig3}a. Consistent with the co-exposure networks, the user-level retweet network shows that echo-chamber structures also emerge through peer interactions among users. We quantified the number and proportion of users with valid numerical media-based bias scores in each of the three largest communities. Across Q2–Q5, the number of users with valid scores in the largest community ranged from 20,092 to 37,059, corresponding to coverage rates of 93.75–95.84\%. In the second-largest community, 10,881–21,498 users had valid scores, with coverage rates of 91.59–93.90\%. In the third-largest community, coverage showed greater variation, with 1,443 to 6,997 users having valid scores, representing 68.08\% to 92.72\% of the total community members. The ideological distributions in Fig.~\ref{fig3}b characterize this structural separation. Users in the largest community are concentrated around left-wing  media scores, whereas users in the second largest community are concentrated around right-wing scores. Across quarters, the left-wing community remains numerically dominant, while the right-wing community shows a more ideologically concentrated distribution. This suggests that the two largest communities are not only topologically separated, but also ideologically distinct.
The community-level statistics for the two largest communities in each quarter, shown in Fig.~\ref{fig3}c, provide a quantitative characterization of these structural asymmetries. Across Q2-Q5, the left-wing cluster consistently contains more nodes and edges than the right-wing cluster, forming the larger and more connected component of the core retweet structure. The edge-to-node ratio $D$ further indicates stronger connectivity in the left-wing community, with values ranging from 132.0794 to 250.5765 across quarters, compared with 125.6768 to 202.0732 for the right-wing community. The only exception occurs in Q3, where the right-wing community has a slightly higher edge-to-node ratio. In contrast, the second-largest community remains smaller but more ideologically concentrated. The signed average media bias score B shows that the largest community was consistently left-wing, with values ranging from -1.019713 to -0.928507, whereas the second largest community was consistently right-wing with stronger ideological intensity, with values ranging from 1.4105639 to 1.7536387. The right-wing community also exhibited larger heterophily $H$ across all quarters, corresponding to heterophily coefficients ranging from 0.3094 to 0.3372, compared with 0.2413 to 0.2894 for the left-wing community. This indicates stronger degree disassortativity and a greater tendency for highly connected users to interact with users of different degree levels. The average clustering coefficient C was consistently higher in the right-wing community (0.1548–0.1697) than in the left-wing community (0.1102–0.1488), indicating greater local clustering among users in the right-wing community. Overall, the left-wing community constitutes the larger and more connected part of the core retweet structure, while the right-wing community is smaller, more ideologically polarized, and more strongly organized around hub users.

Together, Figs.~\ref{fig1}-\ref{fig3} reveal that political polarization emerges through individual, informational, and network-level processes. At the individual level, ideological groups showed different transition patterns: while most users remained within their original categories, those who changed categories exhibited asymmetric shifts, with some moving toward less extreme categories and others shifting in different ideological directions. These transitions did not indicate a uniform convergence toward the center. At the information level, users engaged with ideologically dissimilar content, particularly among right-leaning groups with broader cross-ideological exposure. However, higher ideological permeability was associated with relatively more leftward subsequent bias changes among Left and Left-Leaning users, but relatively more rightward subsequent bias changes among Center, Right-Leaning, Right, and Extreme Right users. At the ecosystem level, both top-down information sources, including media and politicians, and bottom-up peer interactions formed ideologically clustered structures, indicating the emergence of echo-chamber structures across multiple layers of political information ecosystems.

\subsection{Retweeting patterns signal state-level electoral outcomes}\label{subsec4}

\begin{figure}[htbp]
\centering
\includegraphics[width=1 \textwidth]{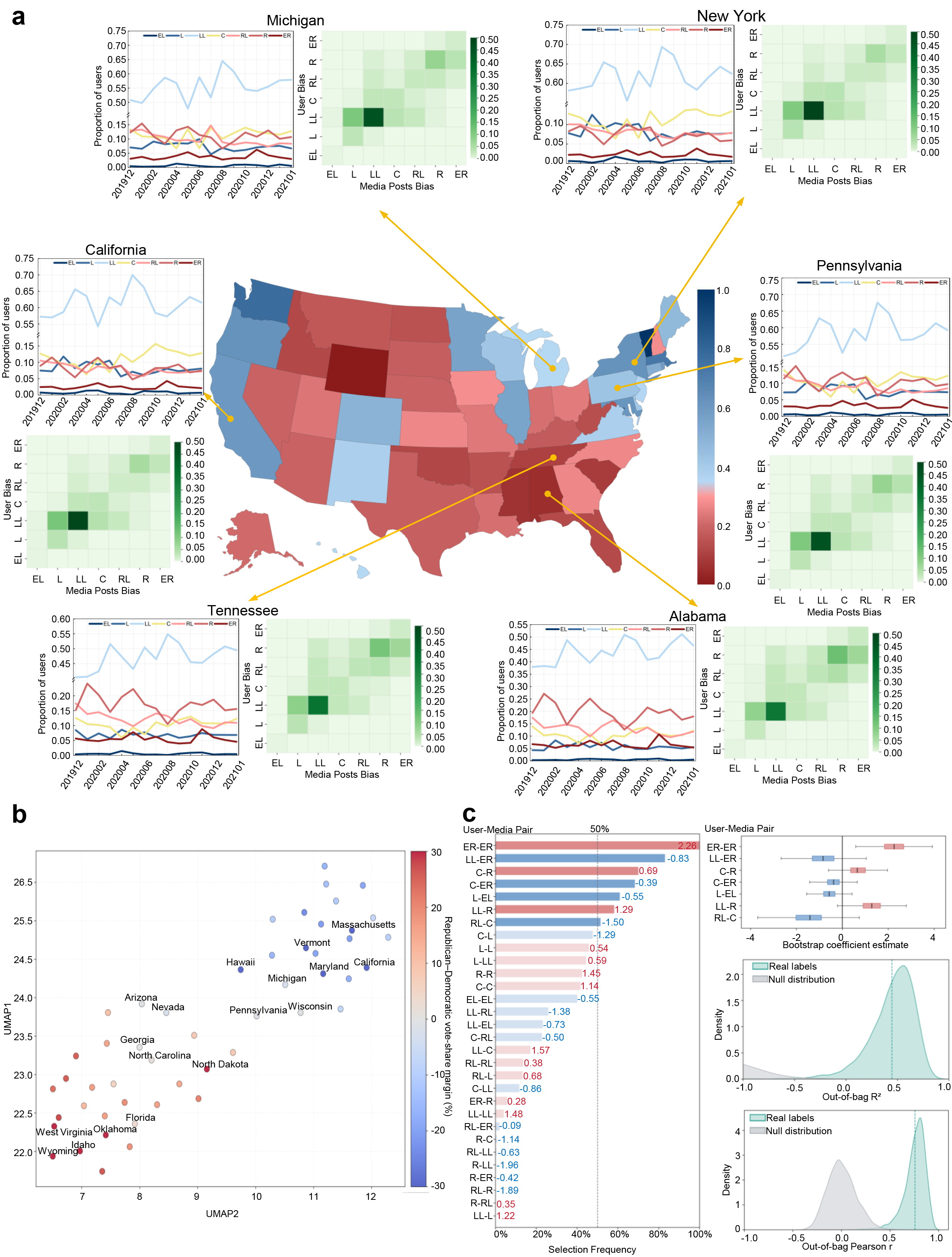}
\caption{State-Level Retweeting Patterns and Electoral Predictions.
\textbf{a}, Geographic distribution of state-level standardized left-to-right user ratio and its correspondence with U.S. presidential election outcomes. For representative states, line graphs illustrate the temporal evolution of users across ideological categories, and heatmaps present the specific retweeting matrices between user bias and media bias.
\textbf{b}, Two-dimensional UMAP projection of state-level retweeting matrices. Each point represents a state, colored by the party vote-share margin. The supervised UMAP projection shows separation between Republican and Democratic states.
\textbf{c}, Identification and validation of stable user-media interaction predictors of state-level electoral margins. The left panel shows the selection stability of each interaction feature across repeated resampling, where the vertical dashed line indicates the 50\% stability threshold and numbers denote the mean post-Lasso OLS coefficients across iterations in which each feature was selected. The upper right panel displays the post-Lasso OLS coefficient distributions for stable predictors retained above this threshold. Positive coefficients indicate larger Republican margins and negative coefficients larger Democratic margins. The lower right panels compare the out-of-bag $R^2$ and Pearson correlation coefficient under the true labels against the null distributions obtained by label permutation, with dashed lines marking the observed means.
\label{fig5}}
\end{figure}
Can collective online information behaviour reveal collective offline voting behaviour at the state level? Firstly, our spatial analysis shows that the relative sizes of different online camps at state level align with the macro-political landscape. As shown in Fig.~\ref{fig5}a, we calculated a standardized left-to-right user ratio for each state by dividing the number of left-side users, including EL, L, and LL users, by the number of right-side users, including RL, R, and ER users, and then normalizing this ratio across states. As shown in  Fig. S7, the normalized state-level user ratio illustrates the relationship between the ideological composition of Twitter users and state-level electoral outcomes, with states ordered according to Democratic vote share in the 2020 U.S. presidential election from highest to lowest. The ratio generally decreases along this electoral gradient, where Democratic-leaning states exhibit higher proportions of left-to-right users and Republican-leaning states exhibit lower ratios. This metric aligns closely with political outcomes across the 50 states. The four states marked with asterisks, New Hampshire, Nevada, Georgia, and Arizona, represent cases where the online user ratio falls below the empirical threshold despite being won by Biden, highlighting specific swing states where online user composition diverges from offline voting results. Across states, this ratio is correlated with the Biden minus Trump vote margin ($r=0.841$, $P=1.99\times10^{-14}$). The ratio also shows high classification consistency with the election outcome, with an ROC-AUC of 0.962. Using the ROC-based Youden's J criterion, the optimal classification threshold was 0.314, corresponding to an original left-to-right user ratio of approximately 2.88. At this threshold, the ratio classified Republican- and Democratic-winning states with 92.0\% accuracy. All 25 Trump-winning states fell below the threshold, while 21 of the 25 Biden-winning states fell above it. The four Biden-winning states below this threshold are Nevada, Arizona, Georgia, and New Hampshire, among which Nevada, Arizona, and Georgia were close swing states in the 2020 election. Such unexpected alignment revealed by this simple statistic also offers independent evidence for the high quality of our dataset and the validity of our data-processing procedures, including the approximate classification of individual political leanings.

%
Representative state-level retweeting matrices  in Fig.~\ref{fig5}a further show that the electoral signal is not only reflected in the relative size of left- and right-wing user groups, but also in the structure of user-media interactions. In Democratic states such as California and New York, interactions between Left-Leaning users and Left-Leaning media account for 49.0\% and 49.6\% of all interactions in the state-level matrix, respectively, while the right-side block, defined as RL, R, and ER users consuming RL, R, and ER media, accounts for only 15.5\% and 15.2\%. Thus, the proportion of Left-Leaning information consumption is approximately three times larger than right-wing consumption in these states. In Republican states, this balance changes. The LL-to-LL consumption block decreases to 35.9\% in Tennessee and 34.2\% in Alabama, while the right-side block rises to 31.6\% and 34.5\%, respectively. The relative strengths of these two aligned consumption structures become nearly balanced in Republican states. Swing states fall between these two patterns. Michigan and Pennsylvania retain relatively strong LL-to-LL consumption patterns, at 45.3\% and 45.5\%, but their right-side block shares increase to 20.4\% in both states. The accompanying monthly line plots show the monthly proportions of users across ideological categories in each representative state. The temporal evolution of ideological groups across states follows similar patterns, consistent with the user-level dynamics shown in Fig.~\ref{fig1}. 

We further examined whether state-level retweeting patterns could account for the variation in 2020 presidential election outcomes across states. Interaction patterns accounting for less than 0.1\% of all user-media interactions were removed, resulting in a 30-dimensional user-media interaction feature space. After applying the centered log-ratio transformation and feature standardization to these remaining features, we then projected the 30-dimensional state-level user-media interaction patterns into a two-dimensional space using supervised UMAP. As shown in Fig.~\ref{fig5}b, Democratic and Republican states occupy two distinguishable regions. Democratic states, such as Vermont, Massachusetts, Maryland, California, and Hawaii, cluster on one side of the projection, while Republican states, such as Wyoming, West Virginia, North Dakota, Oklahoma, and Idaho, cluster on the opposite side. Electoral swing states, including Michigan, Nevada, Pennsylvania, Wisconsin, Arizona, Georgia, North Carolina, and Florida, fall between these two regions. This supervised projection shows that state-level retweeting patterns can be mapped into a low-dimensional structure consistent with the 2020 electoral divide.

Having shown that the state-level retweeting feature space separates along the electoral divide, we next examined which specific user–media interaction patterns are most strongly associated with vote-share differences across states. Identifying these patterns poses two statistical challenges. First, the number of candidate features is 30, which is close to the number of states available for estimation, with 38 states retained after screening as described in Methods 4.5. The features are mutually correlated because they are components of the same composition. An ordinary least squares, or OLS, regression using all features would fit the training states almost perfectly but generalize poorly, resulting in overfitting. Second, interaction volumes are highly imbalanced. Aligned patterns such as LL-LL account for nearly half of all interactions in some states, whereas most other patterns account for only a small fraction. A selection rule based on absolute interaction volume would favor dominant patterns regardless of whether they vary with electoral outcomes.

To address these challenges, we combined a compositional transformation, regularized regression, bootstrap resampling, and out-of-bag validation, following previous studies that used regularized regression to identify stable predictors from high-dimensional, correlated feature sets \cite{yang2023enhanced}. Each feature was first CLR-transformed and then standardized across states. After this step, each feature represents the relative enrichment or depletion of an interaction pattern within a state's overall consumption structure, and all features enter the model on a common scale. A pattern with a low absolute frequency can therefore still carry predictive information if its relative variation across states consistently tracks electoral margins. We then fitted a Lasso regression of the transformed Republican-minus-Democratic vote-share margin on the 30 features. The L1 penalty shrinks the coefficients of uninformative features exactly to zero, yielding a sparse model and limiting overfitting. The penalty strength was determined by 10-fold cross-validation and set to $\alpha = 0.1$ after rounding. Lasso selection alone can be unstable when the sample is small and the features are correlated. We therefore repeated the fit on 10,000 bootstrap resamples of the states, recorded how often each feature was selected, and refitted the selected features by OLS to obtain coefficients free of Lasso shrinkage. This selection frequency measures how robustly a feature is selected when the sample is perturbed, rather than how prevalent the feature is. To evaluate generalization and further guard against overfitting, we used out-of-bag, or OOB, validation, a resampling-based form of cross-validation in which each model is evaluated on the states left out of its resample. Finally, we compared the OOB performance with null distributions obtained by rerunning the entire pipeline on shuffled outcome labels.

The left panel in Fig.~\ref{fig5}c shows the bootstrap selection frequency of each user–media interaction pattern. The dashed line marks the 50\% threshold, and the numbers indicate the mean post-Lasso OLS coefficient, denoted by $\bar{\beta}$, across the resamples in which each feature was selected. Because the predictors were CLR-transformed and standardized, coefficient magnitudes should be compared within the model and should not be interpreted as changes in raw retweeting shares or vote-share percentage points. Positive coefficients indicate associations with larger Republican margins, whereas negative coefficients indicate associations with larger Democratic margins. Seven patterns exceeded the 50\% threshold, namely ER-ER, LL-ER, C-R, C-ER, L-EL, LL-R, and RL-C. The upper-right panel of Fig.~\ref{fig5}c shows the distributions of the bootstrap coefficient estimates for these seven features. The most stable feature was ER-ER, representing Extreme Right users retweeting Extreme Right media. It was selected in 99.91\% of resamples and was associated with Republican advantage, with a mean coefficient of $\bar{\beta} = 2.26$. The second most stable feature was LL-ER, representing Left-Leaning users retweeting Extreme Right media. It was selected in 83.05\% of resamples, with a mean coefficient of $\bar{\beta} = -0.83$, and was associated with Democratic advantage. In contrast, LL-R represents the same user group retweeting less extreme right-wing media and was associated with Republican advantage, with a selection frequency of 58.18\% and a mean coefficient of $\bar{\beta} = 1.29$. This contrast is consistent with the bounded confidence model of opinion dynamics [36, 37]. In this model, opposing information within an acceptable ideological distance can move opinions toward it, whereas opposing information beyond that distance tends to reinforce existing orientations. Patterns involving Center users, whose preferences are less firmly anchored to either party, showed a similar contrast. C-R was associated with Republican advantage, with a selection frequency of 70.02\% and a mean coefficient of $\bar{\beta} = 0.69$, whereas C-ER was associated with Democratic advantage, with a selection frequency of 68.31\% and a mean coefficient of $\bar{\beta} = -0.39$. The remaining stable features, L-EL and RL-C, were both associated with Democratic advantage. L-EL had a selection frequency of 60.93\% and a mean coefficient of $\bar{\beta} = -0.55$, while RL-C had a selection frequency of 51.48\% and a mean coefficient of $\bar{\beta} = -1.50$.

The lower-right panels of Fig.~\ref{fig5}c report the OOB predictive performance. Under the true outcome labels, the pipeline achieved a mean OOB $R^2$ of 0.447 and a mean OOB Pearson correlation of 0.757 between predicted and observed margins. None of the 10,000 label permutations produced a higher value for either metric, with $P = 1.0 \times 10^{-4}$ for both. Because each model was evaluated on states excluded from fitting, these results show that the predictive signal carries over to held-out states rather than reflecting overfitting to the training sample.

We then entered the seven features with selection frequencies above 50\% into a single post-selection OLS model fitted to all 38 states. The model showed good in-sample fit, with $R^2 = 0.829$ and adjusted $R^2 = 0.789$, and was statistically significant overall, with an $F$ statistic of 20.75 on 7 and 30 degrees of freedom and $P = 7.08 \times 10^{-10}$. The predictors showed no severe multicollinearity, with variance inflation factors ranging from 1.79 to 5.87. As expected, the in-sample $R^2$ was higher than the OOB $R^2$ of 0.447, and the OOB value provides a more conservative estimate of predictive performance. Fig. S8 reports the post-selection coefficients. ER-ER had the largest positive coefficient at +2.27, while LL-ER had the largest negative coefficient at -1.78. The remaining coefficients were +1.25 for LL-R, -0.67 for L-EL, -0.55 for RL-C, -0.23 for C-ER, and +0.06 for C-R. The signs of all seven coefficients in the final model agreed with those of the corresponding mean bootstrap coefficients shown in Fig.~\ref{fig5}c, indicating that the direction of each association was robust. Moreover, for each of the seven features, more than 84\% of the resamples in which it was selected yielded a coefficient with the same sign as in the final model.

Fig. S9 compares the observed vote-share margin of each state with the margin estimated by the post-selection model. The estimated margins preserve the ordering of states along the electoral spectrum, with a Spearman correlation coefficient of $\rho = 0.92$, and the model correctly predicts the winning party in 47 of the 50 states. The three misclassified states, Michigan, Nevada, and Arizona, were all decided by margins of less than 3 percentage points. Together, these results indicate that aggregated retweeting patterns contain fine-grained signals of state-level electoral differences, with ER-ER providing the strongest signal associated with Republican advantage.

\section{Discussion}\label{sec12}	
Using retweeting behaviour during the 2020 US presidential election, we examined three questions: whether broader political information exposure promotes moderation, how political information is organized within a multilayer information ecosystem, and whether collective online information behaviour can reveal collective offline voting behaviour. We find that broader information breadth did not necessarily promote ideological moderation, as users with greater exposure to diverse information could still shift toward more extreme positions depending on their initial ideology. Republican and Democratic information ecosystems exhibited asymmetric structures, with differences in how influential accounts and peer interactions were organized across layers, including more concentrated attention patterns in right-side networks and more distributed structures in left-side networks. Extending these online patterns to the offline scale, common retweeting patterns across states distinguished Democratic and Republican states, and specific user-media interaction patterns were associated with state-level voting outcomes. These findings show how individual retweeting choices aggregate into partisan information ecosystems and how the resulting collective patterns are associated with political polarization and carry information about offline voting behaviour.

Our findings contribute to understanding the challenges of bridging political polarization by revealing why broader information exposure does not necessarily promote ideological moderation. The association between ideological permeability and subsequent ideological change varied with users’ prior political positions and showed no general convergence towards the center. These findings are relevant for platform environments, where recommendation systems shape which political information users encounter and amplify. Many depolarization interventions assume that increasing cross-partisan exposure will weaken existing biases. Experimental interventions that altered cross-partisan exposure or information flows have not consistently produced moderation. Exposure to opposing political views increased conservatism among Republicans in a Twitter field experiment, whereas large-scale platform experiments found little detectable change in political attitudes after reducing like-minded content and changing feed ranking \cite{nyhan2023like,guess2023social,guess2023reshares}. One possible explanation is that users interpret politically discordant information through their prior commitments, consistent with motivated reasoning and biased assimilation \cite{jost2022cognitive,dandekar2013biased}. Our results also show that exposure breadth alone is insufficient to produce moderation. The complicated effects of cross-partisan exposure may need to be understood in relation to the recipient, the source and the context in which information circulates.

Our results systematically reveal how individuals across political groups are embedded within distinct multilayer information ecosystems spanning media outlets, political influencers, and social peers. From the top down, Republican information flows were more concentrated around a small set of strongly right-wing media and politicians, while Democratic attention was more broadly distributed. From the bottom up, greater permeability on the right coexisted with a more concentrated peer network, while the left-side network was larger and more diffuse. These patterns show that partisan asymmetry extends across both elite-driven and peer-driven information diffusion. Previous research has documented ideological segregation, asymmetric exposure and algorithmic amplification, often at individual layers of the information environment \cite{gonzalez2023asymmetric,flamino2023political}. Viewing these processes together highlights the importance of ecosystem structure in political information diffusion. In contemporary platform environments, where recommendation and resharing continually reorganize information flows, assessments may benefit from considering source concentration, cross-partisan permeability and network structure. Together, our findings suggest that echo-chamber structures may provide an important foundation for political polarization, but broader cross-partisan exposure does not necessarily reduce polarization. In particular, greater cross-partisan information exposure among some right-leaning users was associated with subsequent shifts toward more extreme ideological positions.

Finally, we unveil that at the state level, user composition and user–media retweeting patterns contained offline electoral signals. This aggregate relationship is interesting because individual retweeting behaviour cannot be directly mapped onto individual voting behaviour. However, when retweeting behaviour is aggregated by state, user choices form a collective signature of regional political attention. Digital traces can contain information about collective electoral outcomes. Previous work has shown that collective social information can improve election estimates and that aggregate Twitter opinion trends can track conventional polling results \cite{galesic2018socialcircles,bovet2018validation}. Our results identify a complementary form of electoral signal at the state level, embedded in how different ideological groups interact with different media sources. These shared state-level patterns provide new insights into the relationship between information exposure structures and electoral outcomes, and may offer a useful direction for future efforts to understand and monitor the evolution of public opinion during electoral processes.

This study has several limitations. Twitter users may not be representative of the entire electorate. Media- and politician-based ideological scores provide approximate measures of political orientation and do not capture all dimensions of individual political beliefs. Future advances in large language models may enable more fine-grained natural language analysis of user-generated content, providing more accurate assessments of individual political orientations. The permeability regressions identify associations rather than causal effects. State geolocation relies on user-provided location metadata, which covers only a small subset of users and may not accurately reflect their actual locations, introducing uncertainty into state-level analyses. Similarly, the mapping between online retweeting patterns and offline electoral outcomes remains correlational. Future work could investigate whether these collective patterns can be further developed into predictive models for electoral outcomes. Finally, our analysis focuses on one platform during one U.S. presidential election. Comparative research across platforms, countries and elections is needed to determine how broadly these patterns generalize.

\section{Methods}\label{sec11}
\subsection{Group Ideology Scoring}\label{methods_subsec1}
Our dataset was derived from a publicly available Tweet ID collection for the 2020 U.S. presidential election, which was constructed using presidential candidate names as query terms. The dataset is released at the project repository \cite{echen2020dataset}. The data span from December 2019 to February 2021. To comply with Twitter’s developer terms of service, the public dataset contains Tweet IDs only. We hydrated these Tweet IDs through the Twitter Developer API and obtained 542,994,806 non-reply records from an initial collection of approximately 1.6 billion Tweet IDs. These valid records consist of retweets and quote tweets, capturing how users consumed election-related information through retweeting or quoting. Many previous studies have relied on the external URL subset of social media datasets to infer political leaning, because links to news outlets can be matched to external media-bias databases. However, this subset usually accounts for only a small fraction of the full dataset, often around 10\%, and therefore excludes a large volume of Twitter-internal links and non-linked native content. To address this, we partitioned the dataset into four groups according to the number and destination of URLs contained in the metadata. External URLs refer to links pointing to third-party websites and are used to track information flowing from sources outside the platform. External URLs account for 15.15\% of the data, comprising 82,242,667 tweets. Twitter URLs refer to internal hyperlinks that reference existing content within the Twitter platform, accounting for 17.64\% of the data and 95,775,521 tweets. Without URLs refer to retweets or quote tweets for which the referenced original post contained no expanded URL, which capture direct user interaction. This group represents the largest share, accounting for 66.61\% of the data and 361,690,197 tweets. Multiple URLs account for 0.61\% of the data, including 3,286,421 tweets containing multiple links. Detailed statistics are provided in Table S2.

In the subsequent ideological analysis, retweets and quote tweets were treated as information-consumption events because both capture users' engagement with previously posted political content. To assess whether oppositional quoting could substantially bias this treatment, we conducted a stratified stance analysis of 100,000 quote tweets. Explicit opposition accounted for 12.98\% of quote tweets and was estimated to represent approximately 0.89\% of all interaction records. These results indicate that oppositional quote tweeting constituted only a small fraction of the overall interaction dataset and was unlikely to materially affect the aggregate ideological analysis. We therefore pooled retweets and quote tweets in the subsequent analyses. Detailed sampling and stance-classification procedures are provided in Supplementary Methods S1.2.

To capture ideological scores from a broader range of retweet and quote records, we developed an entity-level labeling framework. Relying exclusively on media domains would restrict the analysis to records containing identifiable links to news websites and would fail to capture ideological signals associated with political figures. We expanded the annotation framework to cover both media and political influencers, each identified through either standardized website domains or Twitter usernames. Media entities were assigned seven-category ideological labels based on Media Bias/Fact Check. Political entities were classified as Left, Center, or Right according to party affiliation, which was identified with GPT assistance. Four label dictionaries were constructed: media-domain labels, media-username labels, politician-related domain labels, and politician-username labels. Candidate entities were ranked by their frequency of occurrence, allowing more frequently occurring entities to be prioritized during annotation. The final dictionaries contained 1,061 media domains, 663 media usernames, 35 politician-related domains, and 4,023 politician usernames. Detailed entity identification and annotation procedures are provided in Supplementary Methods S1.3.


After the four label dictionaries were constructed, ideological labels were assigned to retweet and quote records by matching the referenced website domain or Twitter username against the corresponding media or politician label dictionary. Records containing multiple URLs were excluded from ideological scoring because they accounted for only 0.61\% of the dataset and could not be assigned to a unique information source without introducing additional assumptions. After this exclusion, each remaining record was assigned to one of three mutually exclusive information-source types: External URLs, Twitter URLs, and Without URLs. The remaining records were evaluated separately in the media and politician dimensions.

For records containing an external URL, the hostname was extracted and standardized using the same procedure applied during label construction. The standardized hostname was matched separately against the media-domain dictionary and the politician-related domain dictionary. For records containing a Twitter URL, the first path segment following "twitter.com/" was extracted as the username of the linked account and matched separately against the media-username dictionary and the politician-username dictionary. For records without URLs, the user ID of the author of the original retweeted or quoted post was extracted and matched against the account identifiers retained in the media-username and politician-username dictionaries. For each successful match, the political-orientation label associated with the matched entity was retrieved from the corresponding dictionary and converted into a numerical score, which was then assigned to that retweet or quote-tweet. Media labels were mapped to a seven-point scale. Extreme Bias Left was assigned as -3, Left as -2, Left-Leaning as -1, Center as 0, Right-Leaning as 1, Right as 2, and Extreme Right as 3. Politician labels were mapped to a three-point scale, with Left assigned -1, Center 0, and Right 1. Media-based and politician-based matching was performed independently. A record was included in the media-based calculation whenever a media match was found and in the politician-based calculation whenever a politician match was found. Failure to match in one dimension did not affect eligibility in the other. Records that could not be matched to a labeled entity in a given dimension were excluded from both the score total and the valid count for that dimension. For each user, media-based and politician-based ideology scores were calculated separately at the monthly and quarterly levels by averaging the scores of all successfully matched interaction records within the corresponding period. If a user had no successfully matched record in a given dimension and period, the corresponding ideology score was treated as undefined rather than zero. Detailed record-matching and score-aggregation procedures are provided in Supplementary Methods S1.4.

The resulting user-level average scores were categorized independently for each dimension and period. Media-based ideology scores were classified into seven categories: Extreme Left for scores in $[-3,-2.5)$, Left for $[-2.5,-1.5)$, Left-Leaning for $[-1.5,-0.5)$, Center for $[-0.5,0.5)$, Right-Leaning for $[0.5,1.5)$, Right for $[1.5,2.5)$, and Extreme Right for $[2.5,3]$. Politician-based ideology scores were classified into three categories: Left for scores in \([-1,-\frac{1}{3})\), Center for scores in \([-\frac{1}{3},\frac{1}{3}]\), and Right for scores in \((\frac{1}{3},1]\). These categories constituted the final ideological labels assigned to users in each monthly or quarterly period. To construct Fig.~\ref{fig1}b, we counted the number of users assigned to each media-based ideological category in each month. The monthly proportion for each ideological category was calculated as the number of users classified into that category divided by the total number of users with a defined media-based ideology score in the corresponding month. Each user was counted at most once within a given month. The resulting monthly proportions were displayed as a stacked bar chart to show temporal changes in the ideological distribution of users.

To characterize information-consumption patterns in Fig.~\ref{fig1}a, we first grouped retweet and quote-tweet records in each quarter according to information-source type and the user's media-based ideological category in the corresponding quarter. Each record was assigned a media-based ideological score through the source-matching and scoring procedure, and records without a valid media-based score were excluded. For each ideological category, the numbers of records from External URLs, Twitter URLs, and Without URLs were then accumulated across all quarters. The upper panel shows the total record volume for each ideological category, calculated by summing the accumulated record counts from the three information-source types. The lower panel shows the source composition within each ideological category, calculated as the proportion of its total record volume contributed by each of the three information-source types.

To examine temporal changes in users' ideological classifications, we constructed the multi-stage Sankey diagram shown in Fig.~\ref{fig1}c using media-based ideology scores. The study period was divided into six consecutive analysis periods: December 2019 to February 2020, March to May 2020, June to August 2020, September to October 2020, November to December 2020, and January to February 2021. Within each period, users with a defined media-based ideology score were assigned to one of the seven ideological categories defined above. The total number of users in each period was defined as the number of unique users with a valid media-based ideology score in that period. For each pair of adjacent periods, users with valid scores in both periods were matched by user ID, and the number of users transitioning between each pair of ideological categories determined the width of the corresponding Sankey flow. For each pair of adjacent periods, unmatched users were defined as those with a valid score in only one of the two periods. These users were excluded from the transition flows and transition-rate denominators but retained in the corresponding period's category counts. For each adjacent-period transition, the transition rate from an origin category was calculated as the number of matched users moving to a given destination category divided by the total number of matched users in that origin category across all destination categories. The reported average transition rates were calculated as the arithmetic means of the corresponding percentages across the five adjacent-period transitions.

\subsection{Micro-level Retweeting Patterns and Ideological Permeability}\label{methods_subsec2}
For each month and each of the three information-source types, we constructed a $7 \times 7$ information-consumption matrix in which rows represented users' media-based ideological categories and columns represented the ideological categories of the media entities they consumed. Each successfully matched retweet or quote-tweet record contributed one count to the corresponding user-media ideological combination. Across the 15-month study period and three information-source types, this procedure yielded 45 matrices. The corresponding cells were then summed across months and source types to obtain a single aggregate count matrix for the full study period. Because the seven user groups differed in their total numbers of media-consumption events, the aggregate matrix was normalized within each user ideological category. Each cell in the normalized matrix therefore represents the proportion of all media-consumption events generated by a given user group that were directed toward a particular media ideological category, and the values within each row sum to one. The resulting row-normalized matrix was visualized in Fig.~\ref{fig4}a. The corresponding unnormalized count matrix is provided in Fig.~S2.


To measure the extent to which users consume political information that differs from their own ideological positions, we constructed an analysis-period measure of ideological permeability. We treated each tweet that was successfully matched to a labeled media entity as one information-consumption event. The ideological distance associated with each information-consumption event was measured as the absolute difference between the user's continuous ideological score and the discrete ideological score of the corresponding media entity. Let $\operatorname{MediaScore}_{u,q,s} \in [-3,3]$ denote the continuous media-based ideological score of user $u$ in analysis period $q$, calculated for information-source type $s$. Negative values indicate left-wing positions, positive values indicate right-wing positions, and values closer to zero indicate more center positions. Let $\operatorname{MediaEntityScore}_{r} \in \{-3,-2,-1,0,1,2,3\}$ denote the discrete ideological score of the media entity associated with information-consumption record $r$ generated by user $u$ in the same analysis period and information-source type. The ideological distance for record $r$ was therefore defined as $\delta_{u,q,s,r}=\left|\operatorname{MediaScore}_{u,q,s}-\operatorname{MediaEntityScore}_{r}\right|$. The seven media categories were mapped onto this scale from extreme left $-3$ to extreme right $3$. To assign greater weight to ideologically distant exposure, the distance was multiplied by 2 under either of two conditions. For center users, defined as those with $\operatorname{MediaScore}_{u,q,s} \in [-0.5,0.5)$, the distance was doubled when the associated media entity had a clearly partisan ideological score, $\left|\operatorname{MediaEntityScore}_{r}\right|>1$. For users outside the center interval, the distance was doubled when the user and the media entity were located on opposite sides of the ideological spectrum. In all other cases, the weight was 1.

The ideological permeability score of user $u$ in quarter $q$ was calculated as follows:

\begin{equation}
\operatorname{IdeoP}_{u,q}
=
\frac{
\displaystyle
\sum_{s \in \mathcal{S}}
\sum_{r \in \mathcal{R}^{\mathrm{M}}_{u,q,s}}
w_{u,q,s,r}
\delta_{u,q,s,r}
}{
\displaystyle
\sum_{s \in \mathcal{S}}
n_{u,q,s}
},
\end{equation}

where $n_{u,q,s}$ is the number of valid information-consumption events generated by user $u$ in analysis period $q$ for information-source type $s$, defined as $n_{u,q,s}=\left|\mathcal{R}^{\mathrm{M}}_{u,q,s}\right|$. The weight $w_{u,q,s,r}$ is defined as:

\begin{equation}
w_{u,q,s,r}
=
\begin{cases}
2,
&
\text{if }
\operatorname{MediaScore}_{u,q,s}\in[-0.5,0.5)
\text{ and }
\left|\operatorname{MediaEntityScore}_{r}\right|>1,
\\[6pt]
2,
&
\text{if }
\operatorname{MediaScore}_{u,q,s}\notin[-0.5,0.5)
\text{ and }
\operatorname{MediaScore}_{u,q,s}
\operatorname{MediaEntityScore}_{r}<0,
\\[6pt]
1,
&
\text{otherwise}.
\end{cases}
\end{equation}

The calculation was initially performed separately for the three information-source types defined above: External URLs, Twitter URLs, and Without URLs. For each user and quarter, the weighted ideological distances and the corresponding numbers of information-consumption events were then summed across the three source types. The aggregate ideological permeability score was obtained by dividing the total weighted distance by the total number of valid information-consumption events. Therefore, the three information-source types contributed to the final measure in proportion to their observed numbers of consumption records, rather than being assigned equal weights. If a user had no valid information-consumption event in any of the three source types in a quarter, no ideological permeability score was assigned. By contrast, a score of zero was retained when valid consumption events existed but all corresponding weighted ideological distances were zero. A higher value of $\operatorname{IdeoP}_{u,q}$ indicates that the media information consumed by a user was, on average, more ideologically distant from the user's own position. A lower value indicates that the user's retweeting activity was concentrated primarily on media entities whose ideological positions were similar to the user's own position.

Building on this metric, we examined the association between ideological permeability and subsequent shifts in political leaning using robust linear regression models. We estimated the models using the RLM estimator in Statsmodels, which fits robust linear models through iteratively reweighted least squares using the default Huber's T norm. All other fitting options were retained at their Statsmodels defaults. Let $\operatorname{MediaScore}_{u,q} \in [-3,3]$ denote the overall media-based ideological score of user $u$ in analysis period $q$, calculated using information from all three source types. This aggregate score was used to calculate subsequent ideological change and to determine the user's political category for the subgroup regression. The regression analysis was conducted only for adjacent analysis-period transitions. For each transition from analysis period $q$ to analysis period $q+1$, users were included if they had valid overall media-based ideological scores in both periods. We further retained users whose ideological permeability score in the initial analysis period $q$ was based on at least two valid information-consumption events. For each user $u$, the ideological shift was defined as:

\begin{equation}
\Delta\operatorname{MediaScore}_{u,q}
=
\operatorname{MediaScore}_{u,q+1}
-
\operatorname{MediaScore}_{u,q},
\end{equation}

where a positive value denotes movement toward the right end of the ideological spectrum, and a negative value denotes movement toward the left.

Before model estimation, observations with missing or non-finite values in ideological permeability or ideological change were removed. The outlier-screening procedure was then applied to the full set of eligible user-transition observations, pooling observations across all five adjacent analysis-period transitions and all seven aggregate political categories in the initial analysis period. First, we calculated the first quartile, third quartile, and interquartile range of $\operatorname{IdeoP}_{u,q}$ using the full eligible sample. Observations with ideological permeability scores outside the interval $\left[Q_1\left(\operatorname{IdeoP}\right)-1.5\,\operatorname{IQR}\left(\operatorname{IdeoP}\right),\;Q_3\left(\operatorname{IdeoP}\right)+1.5\,\operatorname{IQR}\left(\operatorname{IdeoP}\right)\right]$ were removed. Second, using only the observations retained after the first filtering step, we separately calculated the first quartile, third quartile, and interquartile range of $\Delta\operatorname{MediaScore}_{u,q}$. Observations with ideological changes outside the interval $\left[Q_1\left(\Delta\operatorname{MediaScore}\right)-1.5\,\operatorname{IQR}\left(\Delta\operatorname{MediaScore}\right),\;Q_3\left(\Delta\operatorname{MediaScore}\right)+1.5\,\operatorname{IQR}\left(\Delta\operatorname{MediaScore}\right)\right]$ were then removed.

After these two sequential filtering steps, the retained observations were divided into 35 regression subsets defined by the combination of the five adjacent analysis-period transitions and the seven aggregate political categories in the initial analysis period $q$. A separate regression model was estimated for each subset. Models were estimated only for groups containing at least 10 observations, and all 35 groups met this criterion. For users belonging to aggregate political category $c$ in the initial analysis period $q$, the model was specified as:
\begin{equation}
\Delta\operatorname{MediaScore}_{u,q}
=
\beta_{0,c,q}
+
\beta_{1,c,q}
\operatorname{IdeoP}_{u,q}
+
\epsilon_{u,q},
\end{equation}

where $\Delta\operatorname{MediaScore}_{u,q}$ represents the change in the overall media-based ideological score from analysis period $q$ to analysis period $q+1$, and $\operatorname{IdeoP}_{u,q}$ is the aggregate ideological permeability score constructed from the user's media consumption during analysis period $q$. The subscript $c$ indexes the user's aggregate political category, represented by $\operatorname{MediaCat}_{u,q}$ in the initial analysis period $q$. In this specification, $\beta_{1,c,q}$ is the primary coefficient of interest, indicating the association between ideological permeability in analysis period $q$ and subsequent ideological movement from analysis period $q$ to analysis period $q+1$. A positive coefficient indicates a more rightward subsequent movement, whereas a negative coefficient indicates a more leftward subsequent movement. The intercept $\beta_{0,c,q}$ captures the expected ideological change when ideological permeability equals zero, and $\epsilon_{u,q}$ denotes the error term.

To examine whether subsequent ideological changes differed between users with relatively high and low ideological permeability, we linked each user's ideological permeability in quarter \(t\) to the change in their media-based ideological score between quarter \(t\) and the subsequent quarter \(t+1\). Only user-quarter observations for which the user had valid ideological scores in both consecutive quarters were retained. Subsequent bias change was defined as

$$
\Delta Bias_{i,t}=Bias_{i,t+1}-Bias_{i,t},
$$

where positive values indicate a rightward change in the subsequent quarter and negative values indicate a leftward change. Within each ideological group, user-quarter observations were ranked according to their ideological permeability scores, and the top 25\% and bottom 25\% were defined as the high- and low-permeability groups, respectively. We then compared the distributions and mean values of subsequent bias change between these two groups. A higher mean bias change in the high-permeability group indicates a relatively more rightward subsequent change associated with higher ideological permeability, whereas a lower mean indicates a relatively more leftward subsequent change. The resulting distributions are presented in Fig.~S5.

To assess the robustness of the relationship between ideological permeability and subsequent ideological change, we conducted two non-parametric robustness analyses based on the comparison between users in the top and bottom quartiles of ideological permeability. First, we performed Wilcoxon rank-sum tests to examine whether the distributions of subsequent bias change differed significantly between high- and low-permeability users within each ideological group. Second, we conducted permutation tests by randomly shuffling permeability-group labels while preserving the original group sizes and recalculating the difference in mean bias change between the top and bottom quartiles. The permutation procedure was repeated 5,000 times to generate a null distribution of group differences. The empirical difference was compared with this null distribution to evaluate whether the observed directional differences could arise by chance.

\subsection{Co-exposure Network}\label{methods_subsec3}
We constructed co-exposure networks to examine the extent to which media sources and political accounts reached overlapping audiences during the 2020 U.S. presidential election as shown in Fig.~\ref{fig2}. Each node represents a media account or political account, and its audience was defined as the set of users who retweeted the corresponding account. Nodes were selected based on audience size. For the media co-exposure network, we retained the top 25 accounts with the largest audience size within each ideological category. For the politician co-exposure network, we first ranked political accounts by audience size, defined as the number of distinct users who retweeted each account, and selected the top 80\% of accounts as the initial node set. Within this node set, we retained the union of each account's top 10 strongest incident edges ranked by Jaccard similarity. We then removed isolated nodes and extracted the largest connected component to obtain the final network. Edge weights $\operatorname{JaccardWeight}_{a,b}$ between nodes $a$ and $b$ were calculated using the Jaccard similarity of their respective audience sets $\mathcal{A}_{a}$ and $\mathcal{A}_{b}$:

\begin{equation}
\operatorname{JaccardWeight}_{a,b}
=
\frac{
\left|
\mathcal{A}_{a}
\cap
\mathcal{A}_{b}
\right|
}{
\left|
\mathcal{A}_{a}
\cup
\mathcal{A}_{b}
\right|
}.
\end{equation}

This measure captures the proportion of shared audiences between two accounts relative to their combined audience. To reduce network density and improve visual interpretability, we applied a K-nearest-neighbour sparsification strategy. For each node, we identified its top 10 strongest incident edges ranked by Jaccard weight and retained the union of these selected edges across all nodes. Isolated nodes were then removed, and the largest connected component was extracted for visualization and analysis. Nodes in both networks were coloured according to their ideological labels. After exporting the networks, we used the Louvain algorithm in Gephi to detect community structure. The ForceAtlas 2 algorithm in Gephi was used to generate the network layout for visualization.

For each cross-camp media combination, we first identified the unique users belonging to its shared audience over the observation period. This shared-audience user set was then matched to users' quarter-specific ideological classifications across the six analysis periods. For each ideological group, we counted the number of shared-audience users assigned to that group in each quarter and the total number of users assigned to the same ideological group in that quarter. The normalization was performed at the user-quarter level across the full observation period. The pooled normalized co-exposure rate was calculated by summing the shared-audience counts across all six quarters and dividing this by the corresponding sum of ideological-group user counts across the six quarters.

\subsection{Top-degree Retweet Network}\label{methods_subsec4}
To examine the temporal evolution of user-level retweet structures, we constructed quarterly top-degree retweet networks using the full set of retweet records. For each quarterly period, the corresponding monthly files were merged, and each retweet record was represented as an interaction between the retweeting user and the retweeted original user within a quarter. For each period, we constructed a weighted undirected user-level retweet graph, in which nodes represent users, undirected edges connect user pairs involved in retweet interactions, and edge weights represent the frequency of retweet interactions. Repeated interactions between the same user pair were aggregated into edge weights, and self-loops were removed. We extracted the largest connected component of this undirected graph and calculated the unweighted degree of each node. Users whose unweighted degrees were at or above the 99th-percentile threshold were retained to form the top 1\% of users by degree. The retained users and the edges among them formed the quarterly top-degree retweet network. We then applied the Louvain method to the weighted undirected network, using edge weights corresponding to retweet frequencies, a resolution parameter of 1.0, and a random seed of 42. The resulting Louvain community labels were assigned to users and used to identify the largest communities in each quarter. In Figure~\ref{fig3}a, the quarterly top-degree retweet networks are visualized with node colours indicating users' ideological categories.

To characterize the ideological composition of major communities, we matched users in the top-degree retweet networks with quarterly media-based bias scores. Media-based user ideology was measured on a continuous scale of $[-3, 3]$, and users were further classified into seven ideological categories: Extreme Left, Left, Left-Leaning, Center, Right-Leaning, Right, and Extreme Right. For Figure~\ref{fig3}b, we focused on the three largest Louvain communities in each quarter and plotted their media-based ideological distributions. Score coverage was calculated as the proportion of users in each community with a valid numerical media-based bias score, relative to the total number of users in that community. Users without a valid score were excluded from the corresponding ideological distributions. 

For the structural analysis in Figure~\ref{fig3}c, we calculated community-level network statistics for the two largest Louvain communities in each quarter. These statistics include the number of nodes ($N$), the number of undirected edges ($E$), the edge-to-node ratio ($D=E/N$), the average clustering coefficient ($C$), the heterophily coefficient ($H$), and the average media bias score ($B$). All network structural metrics were calculated on the undirected community subgraphs. Each undirected edge was counted once in calculating $E$, and the edge-to-node ratio was defined as the number of undirected edges divided by the number of nodes. The heterophily coefficient (H) was defined as the negative of the degree assortativity coefficient, with larger values indicating stronger degree disassortative mixing. These metrics were used to quantify changes in community size, cohesion, structural mixing, and ideological composition over time.

\subsection{Retweeting Patterns in Each State}\label{methods_subsec5}
To assign users to U.S. states, we performed a geolocation procedure based on user-provided location metadata. The dataset contained a total of 11,031,814 location records, comprising 2,310,502 unique location strings. Because the same user could appear multiple times in the raw records, these records did not correspond to 11,031,814 unique users. For geolocation, each user ID was associated with a single profile-location value. These profile-location values were primarily in free-text format, with a minority represented as geographic coordinates, and were mapped to U.S. states. We first applied spaCy, a named entity recognition model, to extract location entities from the text-based location field. A dictionary was then constructed, including U.S. state full names, abbreviations, and a curated list of major counties and municipalities. Textual location entities were matched to this dictionary to infer the corresponding state. A subset of ambiguous or unmatched entities was manually annotated to enhance coverage and accuracy. This procedure resulted in a mapping between user-provided location strings and U.S. states. Users with empty or ambiguous location entries were excluded from the geolocation mapping. The resulting subset enabled the construction of state-level user populations and corresponding user ID sets for downstream analyses.

We quantified each user's political orientation based on their content interaction history. These scores were then associated with users’ geographic locations. For each state, we aggregated information consumption across all months into a $7 \times 7$ matrix, representing interactions between user bias and media bias across the seven ideological categories.

To examine whether the geographic distribution of political information consumption on Twitter corresponded to the offline political landscape, we aggregated geolocated users by U.S. state and constructed a state-level left-to-right user ratio. The same seven media-based ideological categories defined above were used in this analysis. For each state $g$, users classified as Extreme Left, Left, or Left-Leaning were grouped as left-side users, whereas users classified as Right-Leaning, Right, or Extreme Right were grouped as right-side users. Center users were not included in either side of the ratio. The numbers of left- and right-side users were denoted by $N^{\mathrm{L}}_{g}$ and $N^{\mathrm{R}}_{g}$, respectively. The state-level left-to-right user ratio was calculated as \begin{equation} \operatorname{LRRatio}_{g} = \frac{N^{\mathrm{L}}_{g}} {N^{\mathrm{R}}_{g}}. \end{equation} Because the magnitude of this ratio varied across geographic units, it was rescaled to the interval $[0,1]$ using min-max normalization: \begin{equation} \operatorname{StateScore}_{g} = \frac{ \operatorname{LRRatio}_{g} - \min(\operatorname{LRRatio}) }{ \max(\operatorname{LRRatio}) - \min(\operatorname{LRRatio}) }. \end{equation} A larger $\operatorname{StateScore}_{g}$ therefore indicates a greater relative prevalence of left-side users, whereas a smaller value indicates a greater relative prevalence of right-side users. We compared the Twitter-derived state-level score with the 2020 U.S. presidential election results. The election-based evaluation was restricted to the 50 U.S. states, denoted by $\mathcal{G}$; the District of Columbia, Puerto Rico, and the United States Virgin Islands were excluded from this evaluation. For each state $g \in \mathcal{G}$, the observed electoral margin was calculated as the difference between the Biden and Trump vote percentages: \begin{equation} \operatorname{VoteMargin}_{g} = P_{g,\mathrm{Biden}} - P_{g,\mathrm{Trump}}. \end{equation} Pearson's correlation coefficient was used to quantify the association between $\operatorname{StateScore}_{g}$ and $\operatorname{VoteMargin}_{g}$ across the 50 states. To evaluate whether the state-level Twitter score could distinguish Democratic- and Republican-winning states without imposing a manually selected red-blue cutoff, we performed receiver operating characteristic (ROC) analysis. The observed election outcome was coded as \begin{equation} Y_g = \begin{cases} 1, & P_{g,\mathrm{Biden}} > P_{g,\mathrm{Trump}},\\ 0, & P_{g,\mathrm{Biden}} < P_{g,\mathrm{Trump}}, \end{cases} \end{equation} where $Y_g=1$ represents a Biden-winning state and $Y_g=0$ represents a Trump-winning state. The discriminatory ability of $\operatorname{StateScore}_{g}$ was quantified using the area under the ROC curve (ROC-AUC). Rather than manually specifying the transition point between red and blue states, the classification threshold was selected using Youden's $J$ statistic. For each candidate threshold $\tau$, we calculated \begin{equation} J(\tau) = \mathrm{Sensitivity}(\tau) + \mathrm{Specificity}(\tau) -1 = \mathrm{TPR}(\tau) - \mathrm{FPR}(\tau). \end{equation} The optimal threshold was defined as \begin{equation} \tau^{*} = \arg\max_{\tau} J(\tau). \end{equation} Each state was subsequently classified according to \begin{equation} \widehat{Y}_{g} = \begin{cases} 1, & \operatorname{StateScore}_{g} \geq \tau^{*},\\ 0, & \operatorname{StateScore}_{g} < \tau^{*}. \end{cases} \end{equation} The inferred state classifications were compared with the observed election outcomes across the 50 states. Classification accuracy was calculated as the proportion of states for which $\widehat{Y}_{g}=Y_g$. Sensitivity, specificity, and balanced accuracy were additionally calculated from the corresponding confusion matrix. Because $\tau^{*}$ was estimated using the same set of states included in the primary classification analysis, we additionally conducted leave-one-state-out cross-validation as a robustness check. In each iteration, one state was held out, the Youden-optimal threshold was re-estimated using the remaining 49 states, and the held-out state was classified using the resulting threshold. This procedure was repeated for all 50 states, and the proportion of correctly classified held-out states was used to assess out-of-sample classification consistency. Finally, $\operatorname{StateScore}_{g}$ was visualized using a U.S. choropleth map with an Albers USA projection. A continuous color scale ranging from dark red to dark blue was used to represent the magnitude of the state-level score. The transition between the red and blue portions of the color scale was positioned at the data-driven threshold $\tau^{*}$ rather than at a manually specified value. Thus, the map preserves the continuous variation in state-level political composition while providing a visually interpretable reference to the classification boundary derived from the observed election outcomes.

This analysis supports Figure 5b and was designed to visualize the relationship between state-level information-consumption patterns and electoral outcomes in the 2020 U.S. presidential election. We first calculated the two-party vote-share margin for each state as

\begin{equation}
M_s = V_{\mathrm{Rep},s} - V_{\mathrm{Dem},s},
\end{equation}

where $V_{\mathrm{Rep},s}$ and $V_{\mathrm{Dem},s}$ denote the Republican and Democratic vote percentages in state $s$, respectively. Positive values of $M_s$ indicate a Republican-leading vote margin, whereas negative values indicate a Democratic-leading vote margin. The untransformed vote-share margin $M_s$ was used as the continuous supervision target in the supervised UMAP analysis.

The state-level information-consumption features were derived from the $7 \times 7$ user-media ideological interaction matrix, which initially yielded 49 interaction patterns. Interaction patterns accounting for less than 0.1\% of all user-media interactions across states were removed, resulting in a final $30$-dimensional feature vector for each of the $50$ states. Each feature corresponds to the retweeting count between one specific user ideological category and one media ideological category. Zero counts occurring in only a subset of states within retained features were preserved because they represented valid state-specific observations. These retained zero values were handled by adding a small pseudocount during the subsequent centered log-ratio transformation. Because the analysis focused on the relative composition of different user-media interaction types within each state, rather than solely on the absolute volume of interactions, the retained interaction-count vectors were treated as compositional data. Direct use of compositional vectors may introduce closure effects because each component is interpreted relative to the complete composition. We therefore applied the Aitchison centered log-ratio transformation, hereafter referred to as the CLR transformation.

Let $\mathbf{x}_s=(x_{s1},x_{s2},\ldots,x_{sp})$ denote the retained retweeting-count vector for state $s$, where $p=30$. To avoid taking the logarithm of zero, a small pseudocount of $\epsilon=10^{-6}$ was added to every retained count entry, such that the adjusted value was $x_{sj}^{*}=x_{sj}+\epsilon$. The geometric mean of the adjusted feature vector for state $s$ was calculated as $g(\mathbf{x}_s^{*})=\left[\prod_{j=1}^{p}(x_{sj}+\epsilon)\right]^{1/p}$. The CLR-transformed value of feature $j$ for state $s$ was then calculated as

\begin{equation}
x_{sj}^{\mathrm{CLR}}
=
\ln\left[
\frac{x_{sj}+\epsilon}
{g(\mathbf{x}_s^{*})}
\right].
\end{equation}

Thus, each adjusted feature value was expressed as its log-ratio relative to the geometric mean of all retained interaction features within the same state. This transformation reduced closure effects and mapped the compositional feature vectors into a Euclidean feature space suitable for distance-based dimensionality reduction. After the CLR transformation, each feature was standardized across the $50$ states to have mean $0$ and unit variance. Specifically, the standardized value was calculated as $z_{sj}=(x_{sj}^{\mathrm{CLR}}-\mu_j)/\sigma_j$, where $\mu_j$ and $\sigma_j$ denote the mean and standard deviation of the CLR-transformed values of feature $j$ across all states. The resulting input matrix therefore contained $50$ state observations and $30$ standardized state-level information-consumption features.

We then fitted a supervised UMAP model using the $30$-dimensional CLR-transformed and standardized feature vectors as the input feature space. All $50$ U.S. states were included in the analysis, and the untransformed Republican-minus-Democratic vote-share margin $M_s$ was supplied as a continuous supervision target. The projection therefore reflected both the neighborhood structure of the state-level information-consumption features and the similarity structure of the state-level vote margins. The UMAP model was fitted with $\texttt{n\_neighbors}=25$, $\texttt{min\_dist}=0.15$, and two output dimensions. Euclidean distance was used in the transformed feature space, while L2 distance was used to measure similarity in the continuous target space. The target weight was set to $\texttt{target\_weight}=0.7$, assigning substantial weight to the vote-margin structure while retaining information from the neighborhood structure of the information-consumption features. The random seed was fixed at $42$ to improve reproducibility.

The fitted UMAP model produced two coordinates for each state. In Figure 5b, each point represents one state and is colored according to the original, untransformed vote-share margin $M_s$, with Democratic-leading states shown in blue, Republican-leading states in red, and states with margins near zero in neutral colors. Because $M_s$ was used both as the continuous supervision target during UMAP fitting and as the point-color variable, the projection should be interpreted as a supervised visualization of the relationship between state-level information-consumption features and electoral outcomes, rather than as an unsupervised discovery of partisan clusters. Moreover, because UMAP embeddings are arbitrary with respect to rotation, reflection, translation, and scaling, UMAP1 and UMAP2 were not assigned substantive political meanings. Interpretation therefore focused on the relative positions and neighborhood relationships of states within the two-dimensional embedding.

For the subsequent regression analysis reported in Figure 5c, the vote-share margin was transformed as $Y_s=\operatorname{arcsinh}(M_s)$ to preserve its direction while reducing the influence of large absolute margins. The transformed outcome $Y_s$ was used only in the regression analysis.

Building on the $30$-dimensional interaction-feature set and the transformed vote-share outcome defined above, we next used a Bootstrap-Lasso pipeline followed by post-selection ordinary least squares (OLS) estimation to identify state-level user-media interaction features associated with vote-share differences.

The $50$ states were first ordered according to the transformed vote-share margin $Y_s$, and the middle $40$ states were retained to reduce the influence of electorally extreme observations. Because the inverse hyperbolic sine transformation is monotonic, this procedure produced the same ordering as the original vote-share margin $M_s$. The CLR transformation described above was then applied to the $30$-dimensional interaction-count vector of each retained state using the same pseudocount, $\epsilon=10^{-6}$. Features with a mean absolute CLR value not exceeding $0.0001$ across the $40$ states were designated for removal. All $30$ features exceeded this threshold and were retained. Potentially influential observations were assessed using Cook's distance calculated from an OLS diagnostic model fitted to the CLR-transformed features. Connecticut and Delaware had Cook's distance values greater than the prespecified threshold of $1$ and were therefore excluded. The remaining $38$ states constituted the sample used in the subsequent Bootstrap-Lasso analysis. After this observation-level screening, each CLR-transformed feature was standardized across the $38$ states to have mean $0$ and unit variance.

Lasso regression with an L1 penalty was used to obtain sparse predictor sets from the standardized CLR-transformed features. The regularization parameter was fixed at $\alpha=0.1$, and variable-selection stability was evaluated across $10{,}000$ bootstrap iterations. In each iteration, the $38$ states were sampled with replacement to form an in-bag training sample, while states not selected in that iteration formed the out-of-bag (OOB) sample. Iterations containing fewer than three OOB states were excluded from subsequent aggregation. Within each valid iteration, Lasso was fitted to the in-bag sample, and features with non-zero Lasso coefficients were recorded as selected. The coefficients of the selected features were subsequently re-estimated using OLS on the same in-bag sample, yielding post-Lasso coefficient estimates without the shrinkage imposed by the L1 penalty. The resulting post-Lasso OLS model was then used to predict $Y_s$ for the OOB states. Predictive performance in each iteration was evaluated using the OOB $R^2$ and the Pearson correlation between the predicted and observed values of $Y_s$. If Lasso selected no features in a given iteration, the mean value of $Y_s$ in the in-bag sample was used to predict the OOB observations.

For each feature, the selection frequency was calculated as the proportion of valid bootstrap iterations in which its Lasso coefficient was non-zero. Features selected in more than $50\%$ of the valid iterations were retained as stable predictors. These stable predictors were then entered into an OLS model fitted to all $38$ states. The final reported coefficients were obtained from the resulting post-selection OLS model. Variance inflation factors were additionally calculated to assess multicollinearity among the final predictors.

To assess whether the observed mean out-of-bag (OOB) predictive performance exceeded a shuffled-label baseline, we performed 10,000 permutations of the transformed outcome, \(Y_s=\operatorname{arcsinh}(M_s)\), across the retained states, holding the screened sample and preprocessed predictor matrix fixed. For each permutation, the outcome labels were shuffled once and then held fixed across 10,000 bootstrap resamples. The same bootstrap training samples and corresponding OOB sets were used for the original labels and all permutations; resamples containing fewer than three OOB states were excluded. Within each valid resample, Lasso feature selection (\(\alpha=0.1\)) was followed by OLS refitting on the selected predictors using the in-bag observations, and predictions were evaluated against the permuted outcomes in the OOB set. The OOB \(R^2\) and Pearson correlation were averaged separately across valid resamples, yielding one mean value for each metric per permutation. These permutation-specific means formed the null distributions against which the corresponding mean scores obtained using the original outcome labels were compared. One-sided empirical \(P\) values were calculated as \(P=(b+1)/(B+1)\), where \(B=10{,}000\) and \(b\) denotes the number of permutation-specific mean scores strictly greater than the corresponding observed mean score.

\section{Acknowledgements}

This paper is supported by the New Generation Artificial Intelligence-National Science and Technology Major Project (2025ZD0123704), the National Natural Science Foundation of China (U25A20425, 62372022, 12671643), and the Beijing Natural Science Foundation (L251040).


\begin{thebibliography}{10}
	\expandafter\ifx\csname url\endcsname\relax
	\def\url#1{\burl{#1}}\fi
	\expandafter\ifx\csname urlprefix\endcsname\relax\def\urlprefix{URL }\fi
	\providecommand{\bibinfo}[2]{#2}
	\providecommand{\eprint}[2][]{\url{#2}}
	\providecommand{\doi}[1]{\url{https://doi.org/#1}}
	\bibcommenthead
	
	\bibitem{lazer2009computational}
	\bibinfo{author}{Lazer, D.} \emph{et~al.}
	\newblock \bibinfo{title}{Computational social science}.
	\newblock \emph{\bibinfo{journal}{Science}} \textbf{\bibinfo{volume}{323}},
	\bibinfo{pages}{721--723} (\bibinfo{year}{2009}).
	
	\bibitem{king2017news}
	\bibinfo{author}{King, G.}, \bibinfo{author}{Schneer, B.} \&
	\bibinfo{author}{White, A.}
	\newblock \bibinfo{title}{How the news media activate public expression and
		influence national agendas}.
	\newblock \emph{\bibinfo{journal}{Science}} \textbf{\bibinfo{volume}{358}},
	\bibinfo{pages}{776--780} (\bibinfo{year}{2017}).
	
	\bibitem{bond2012experiment}
	\bibinfo{author}{Bond, R.~M.} \emph{et~al.}
	\newblock \bibinfo{title}{A 61-million-person experiment in social influence
		and political mobilization}.
	\newblock \emph{\bibinfo{journal}{Nature}} \textbf{\bibinfo{volume}{489}},
	\bibinfo{pages}{295--298} (\bibinfo{year}{2012}).
	
	\bibitem{onnela2010spontaneous}
	\bibinfo{author}{Onnela, J.-P.} \& \bibinfo{author}{Reed-Tsochas, F.}
	\newblock \bibinfo{title}{Spontaneous emergence of social influence in online
		systems}.
	\newblock \emph{\bibinfo{journal}{Proceedings of the National Academy of
			Sciences}} \textbf{\bibinfo{volume}{107}}, \bibinfo{pages}{18375--18380}
	(\bibinfo{year}{2010}).
	
	\bibitem{guess2023reshares}
	\bibinfo{author}{Guess, A.~M.} \emph{et~al.}
	\newblock \bibinfo{title}{Reshares on social media amplify political news but
		do not detectably affect beliefs or opinions}.
	\newblock \emph{\bibinfo{journal}{Science}} \textbf{\bibinfo{volume}{381}},
	\bibinfo{pages}{404--408} (\bibinfo{year}{2023}).
	
	\bibitem{flamino2023political}
	\bibinfo{author}{Flamino, J.} \emph{et~al.}
	\newblock \bibinfo{title}{Political polarization of news media and influencers
		on {Twitter} in the 2016 and 2020 {US} presidential elections}.
	\newblock \emph{\bibinfo{journal}{Nature Human Behaviour}}
	\textbf{\bibinfo{volume}{7}}, \bibinfo{pages}{904--916}
	(\bibinfo{year}{2023}).
	
	\bibitem{brady2017emotion}
	\bibinfo{author}{Brady, W.~J.}, \bibinfo{author}{Wills, J.~A.},
	\bibinfo{author}{Jost, J.~T.}, \bibinfo{author}{Tucker, J.~A.} \&
	\bibinfo{author}{Van~Bavel, J.~J.}
	\newblock \bibinfo{title}{Emotion shapes the diffusion of moralized content in
		social networks}.
	\newblock \emph{\bibinfo{journal}{Proceedings of the National Academy of
			Sciences}} \textbf{\bibinfo{volume}{114}}, \bibinfo{pages}{7313--7318}
	(\bibinfo{year}{2017}).
	
	\bibitem{vosoughi2018spread}
	\bibinfo{author}{Vosoughi, S.}, \bibinfo{author}{Roy, D.} \&
	\bibinfo{author}{Aral, S.}
	\newblock \bibinfo{title}{The spread of true and false news online}.
	\newblock \emph{\bibinfo{journal}{Science}} \textbf{\bibinfo{volume}{359}},
	\bibinfo{pages}{1146--1151} (\bibinfo{year}{2018}).
	
	\bibitem{appel2026deceptive}
	\bibinfo{author}{Appel, R.~E.} \emph{et~al.}
	\newblock \bibinfo{title}{How deceptive online networks reached millions in the
		{US} 2020 elections}.
	\newblock \emph{\bibinfo{journal}{Nature Human Behaviour}}
	\bibinfo{pages}{1--15} (\bibinfo{year}{2026}).
	
	\bibitem{bakshy2015exposure}
	\bibinfo{author}{Bakshy, E.}, \bibinfo{author}{Messing, S.} \&
	\bibinfo{author}{Adamic, L.~A.}
	\newblock \bibinfo{title}{Exposure to ideologically diverse news and opinion on
		{Facebook}}.
	\newblock \emph{\bibinfo{journal}{Science}} \textbf{\bibinfo{volume}{348}},
	\bibinfo{pages}{1130--1132} (\bibinfo{year}{2015}).
	
	\bibitem{flaxman2016filter}
	\bibinfo{author}{Flaxman, S.}, \bibinfo{author}{Goel, S.} \&
	\bibinfo{author}{Rao, J.~M.}
	\newblock \bibinfo{title}{Filter bubbles, echo chambers, and online news
		consumption}.
	\newblock \emph{\bibinfo{journal}{Public Opinion Quarterly}}
	\textbf{\bibinfo{volume}{80}}, \bibinfo{pages}{298--320}
	(\bibinfo{year}{2016}).
	
	\bibitem{yang2020exposure}
	\bibinfo{author}{Yang, T.}, \bibinfo{author}{Maj{\'o}-V{\'a}zquez, S.},
	\bibinfo{author}{Nielsen, R.~K.} \& \bibinfo{author}{Gonz{\'a}lez-Bail{\'o}n,
		S.}
	\newblock \bibinfo{title}{Exposure to news grows less fragmented with an
		increase in mobile access}.
	\newblock \emph{\bibinfo{journal}{Proceedings of the National Academy of
			Sciences}} \textbf{\bibinfo{volume}{117}}, \bibinfo{pages}{28678--28683}
	(\bibinfo{year}{2020}).
	
	\bibitem{liu2020homogeneity}
	\bibinfo{author}{Liu, L.} \emph{et~al.}
	\newblock \bibinfo{title}{Homogeneity trend on social networks changes
		evolutionary advantage in competitive information diffusion}.
	\newblock \emph{\bibinfo{journal}{New Journal of Physics}}
	\textbf{\bibinfo{volume}{22}}, \bibinfo{pages}{013019}
	(\bibinfo{year}{2020}).
	
	\bibitem{cinelli2020selective}
	\bibinfo{author}{Cinelli, M.} \emph{et~al.}
	\newblock \bibinfo{title}{Selective exposure shapes the {Facebook} news diet}.
	\newblock \emph{\bibinfo{journal}{{PloS One}}} \textbf{\bibinfo{volume}{15}},
	\bibinfo{pages}{e0229129} (\bibinfo{year}{2020}).
	
	\bibitem{goldenberg2023homophily}
	\bibinfo{author}{Goldenberg, A.} \emph{et~al.}
	\newblock \bibinfo{title}{Homophily and acrophily as drivers of political
		segregation}.
	\newblock \emph{\bibinfo{journal}{Nature Human Behaviour}}
	\textbf{\bibinfo{volume}{7}}, \bibinfo{pages}{219--230}
	(\bibinfo{year}{2023}).
	
	\bibitem{del2016echo}
	\bibinfo{author}{Del~Vicario, M.} \emph{et~al.}
	\newblock \bibinfo{title}{Echo chambers: Emotional contagion and group
		polarization on {Facebook}}.
	\newblock \emph{\bibinfo{journal}{Scientific Reports}}
	\textbf{\bibinfo{volume}{6}}, \bibinfo{pages}{37825} (\bibinfo{year}{2016}).
	
	\bibitem{liu2021modeling}
	\bibinfo{author}{Liu, L.}, \bibinfo{author}{Wang, X.}, \bibinfo{author}{Chen,
		X.}, \bibinfo{author}{Tang, S.} \& \bibinfo{author}{Zheng, Z.}
	\newblock \bibinfo{title}{Modeling confirmation bias and peer pressure in
		opinion dynamics}.
	\newblock \emph{\bibinfo{journal}{Frontiers in Physics}}
	\textbf{\bibinfo{volume}{9}}, \bibinfo{pages}{649852} (\bibinfo{year}{2021}).
	
	\bibitem{schmidt2017anatomy}
	\bibinfo{author}{Schmidt, A.~L.} \emph{et~al.}
	\newblock \bibinfo{title}{Anatomy of news consumption on {Facebook}}.
	\newblock \emph{\bibinfo{journal}{Proceedings of the National Academy of
			Sciences}} \textbf{\bibinfo{volume}{114}}, \bibinfo{pages}{3035--3039}
	(\bibinfo{year}{2017}).
	
	\bibitem{cota2019quantifying}
	\bibinfo{author}{Cota, W.}, \bibinfo{author}{Ferreira, S.~C.},
	\bibinfo{author}{Pastor-Satorras, R.} \& \bibinfo{author}{Starnini, M.}
	\newblock \bibinfo{title}{Quantifying echo chamber effects in information
		spreading over political communication networks}.
	\newblock \emph{\bibinfo{journal}{EPJ Data Science}}
	\textbf{\bibinfo{volume}{8}}, \bibinfo{pages}{35} (\bibinfo{year}{2019}).
	
	\bibitem{del2016spreading}
	\bibinfo{author}{Del~Vicario, M.} \emph{et~al.}
	\newblock \bibinfo{title}{The spreading of misinformation online}.
	\newblock \emph{\bibinfo{journal}{Proceedings of the National Academy of
			Sciences}} \textbf{\bibinfo{volume}{113}}, \bibinfo{pages}{554--559}
	(\bibinfo{year}{2016}).
	
	\bibitem{lazer2015rise}
	\bibinfo{author}{Lazer, D.}
	\newblock \bibinfo{title}{The rise of the social algorithm}.
	\newblock \emph{\bibinfo{journal}{Science}} \textbf{\bibinfo{volume}{348}},
	\bibinfo{pages}{1090--1091} (\bibinfo{year}{2015}).
	
	\bibitem{gauthier2026political}
	\bibinfo{author}{Gauthier, G.}, \bibinfo{author}{Hodler, R.},
	\bibinfo{author}{Widmer, P.} \& \bibinfo{author}{Zhuravskaya, E.}
	\newblock \bibinfo{title}{The political effects of {X}'s feed algorithm}.
	\newblock \emph{\bibinfo{journal}{Nature}} \textbf{\bibinfo{volume}{652}},
	\bibinfo{pages}{416--423} (\bibinfo{year}{2026}).
	
	\bibitem{santos2021link}
	\bibinfo{author}{Santos, F.~P.}, \bibinfo{author}{Lelkes, Y.} \&
	\bibinfo{author}{Levin, S.~A.}
	\newblock \bibinfo{title}{Link recommendation algorithms and dynamics of
		polarization in online social networks}.
	\newblock \emph{\bibinfo{journal}{Proceedings of the National Academy of
			Sciences}} \textbf{\bibinfo{volume}{118}}, \bibinfo{pages}{e2102141118}
	(\bibinfo{year}{2021}).
	
	\bibitem{huszar2022algorithmic}
	\bibinfo{author}{Husz{\'a}r, F.} \emph{et~al.}
	\newblock \bibinfo{title}{Algorithmic amplification of politics on {Twitter}}.
	\newblock \emph{\bibinfo{journal}{Proceedings of the National Academy of
			Sciences}} \textbf{\bibinfo{volume}{119}}, \bibinfo{pages}{e2025334119}
	(\bibinfo{year}{2022}).
	
	\bibitem{ibrahim2026systematic}
	\bibinfo{author}{Ibrahim, H.} \emph{et~al.}
	\newblock \bibinfo{title}{Systematic partisan content skews in {TikTok} during
		the 2024 {US} elections}.
	\newblock \emph{\bibinfo{journal}{Nature}} \bibinfo{pages}{1--8}
	(\bibinfo{year}{2026}).
	
	\bibitem{wang2020public}
	\bibinfo{author}{Wang, X.}, \bibinfo{author}{Sirianni, A.~D.},
	\bibinfo{author}{Tang, S.}, \bibinfo{author}{Zheng, Z.} \&
	\bibinfo{author}{Fu, F.}
	\newblock \bibinfo{title}{Public discourse and social network echo chambers
		driven by socio-cognitive biases}.
	\newblock \emph{\bibinfo{journal}{Physical Review X}}
	\textbf{\bibinfo{volume}{10}}, \bibinfo{pages}{041042}
	(\bibinfo{year}{2020}).
	
	\bibitem{liu2026evolution}
	\bibinfo{author}{Liu, Z.} \emph{et~al.}
	\newblock \bibinfo{title}{Evolution of swing voters under strategic campaigns
		in presidential elections}.
	\newblock \emph{\bibinfo{journal}{Physical Review Research}}
	\textbf{\bibinfo{volume}{8}}, \bibinfo{pages}{013306} (\bibinfo{year}{2026}).
	
	\bibitem{lazer2018science}
	\bibinfo{author}{Lazer, D.~M.} \emph{et~al.}
	\newblock \bibinfo{title}{The science of fake news}.
	\newblock \emph{\bibinfo{journal}{Science}} \textbf{\bibinfo{volume}{359}},
	\bibinfo{pages}{1094--1096} (\bibinfo{year}{2018}).
	
	\bibitem{bakcoleman2021stewardship}
	\bibinfo{author}{Bak-Coleman, J.~B.} \emph{et~al.}
	\newblock \bibinfo{title}{Stewardship of global collective behavior}.
	\newblock \emph{\bibinfo{journal}{Proceedings of the National Academy of
			Sciences}} \textbf{\bibinfo{volume}{118}}, \bibinfo{pages}{e2025764118}
	(\bibinfo{year}{2021}).
	
	\bibitem{levin2021dynamics}
	\bibinfo{author}{Levin, S.~A.}, \bibinfo{author}{Milner, H.~V.} \&
	\bibinfo{author}{Perrings, C.}
	\newblock \bibinfo{title}{The dynamics of political polarization}.
	\newblock \emph{\bibinfo{journal}{Proceedings of the National Academy of
			Sciences}} \textbf{\bibinfo{volume}{118}}, \bibinfo{pages}{e2116950118}
	(\bibinfo{year}{2021}).
	
	\bibitem{watts2021measuring}
	\bibinfo{author}{Watts, D.~J.}, \bibinfo{author}{Rothschild, D.~M.} \&
	\bibinfo{author}{Mobius, M.}
	\newblock \bibinfo{title}{Measuring the news and its impact on democracy}.
	\newblock \emph{\bibinfo{journal}{Proceedings of the National Academy of
			Sciences}} \textbf{\bibinfo{volume}{118}}, \bibinfo{pages}{e1912443118}
	(\bibinfo{year}{2021}).
	
	\bibitem{rathje2021out}
	\bibinfo{author}{Rathje, S.}, \bibinfo{author}{Van~Bavel, J.~J.} \&
	\bibinfo{author}{Van Der~Linden, S.}
	\newblock \bibinfo{title}{Out-group animosity drives engagement on social
		media}.
	\newblock \emph{\bibinfo{journal}{Proceedings of the National Academy of
			Sciences}} \textbf{\bibinfo{volume}{118}}, \bibinfo{pages}{e2024292118}
	(\bibinfo{year}{2021}).
	
	\bibitem{iyengar2008selective}
	\bibinfo{author}{Iyengar, S.}, \bibinfo{author}{Hahn, K.~S.},
	\bibinfo{author}{Krosnick, J.~A.} \& \bibinfo{author}{Walker, J.}
	\newblock \bibinfo{title}{Selective exposure to campaign communication: The
		role of anticipated agreement and issue public membership}.
	\newblock \emph{\bibinfo{journal}{The Journal of Politics}}
	\textbf{\bibinfo{volume}{70}}, \bibinfo{pages}{186--200}
	(\bibinfo{year}{2008}).
	
	\bibitem{hart2009feeling}
	\bibinfo{author}{Hart, W.} \emph{et~al.}
	\newblock \bibinfo{title}{Feeling validated versus being correct: A
		meta-analysis of selective exposure to information.}
	\newblock \emph{\bibinfo{journal}{Psychological Bulletin}}
	\textbf{\bibinfo{volume}{135}}, \bibinfo{pages}{555} (\bibinfo{year}{2009}).
	
	\bibitem{bail2018exposure}
	\bibinfo{author}{Bail, C.~A.} \emph{et~al.}
	\newblock \bibinfo{title}{Exposure to opposing views on social media can
		increase political polarization}.
	\newblock \emph{\bibinfo{journal}{Proceedings of the National Academy of
			Sciences}} \textbf{\bibinfo{volume}{115}}, \bibinfo{pages}{9216--9221}
	(\bibinfo{year}{2018}).
	
	\bibitem{nyhan2023like}
	\bibinfo{author}{Nyhan, B.} \emph{et~al.}
	\newblock \bibinfo{title}{Like-minded sources on {Facebook} are prevalent but
		not polarizing}.
	\newblock \emph{\bibinfo{journal}{Nature}} \textbf{\bibinfo{volume}{620}},
	\bibinfo{pages}{137--144} (\bibinfo{year}{2023}).
	
	\bibitem{allcott2024deactivation}
	\bibinfo{author}{Allcott, H.} \emph{et~al.}
	\newblock \bibinfo{title}{The effects of {Facebook} and {Instagram} on the 2020
		election: A deactivation experiment}.
	\newblock \emph{\bibinfo{journal}{Proceedings of the National Academy of
			Sciences}} \textbf{\bibinfo{volume}{121}}, \bibinfo{pages}{e2321584121}
	(\bibinfo{year}{2024}).
	
	\bibitem{mutz2002crosscutting}
	\bibinfo{author}{Mutz, D.~C.}
	\newblock \bibinfo{title}{Cross-cutting social networks: Testing democratic
		theory in practice}.
	\newblock \emph{\bibinfo{journal}{American Political Science Review}}
	\textbf{\bibinfo{volume}{96}}, \bibinfo{pages}{111--126}
	(\bibinfo{year}{2002}).
	
	\bibitem{balietti2021reducing}
	\bibinfo{author}{Balietti, S.}, \bibinfo{author}{Getoor, L.},
	\bibinfo{author}{Goldstein, D.~G.} \& \bibinfo{author}{Watts, D.~J.}
	\newblock \bibinfo{title}{Reducing opinion polarization: Effects of exposure to
		similar people with differing political views}.
	\newblock \emph{\bibinfo{journal}{Proceedings of the National Academy of
			Sciences}} \textbf{\bibinfo{volume}{118}}, \bibinfo{pages}{e2112552118}
	(\bibinfo{year}{2021}).
	
	\bibitem{jost2022cognitive}
	\bibinfo{author}{Jost, J.~T.}, \bibinfo{author}{Baldassarri, D.~S.} \&
	\bibinfo{author}{Druckman, J.~N.}
	\newblock \bibinfo{title}{Cognitive--motivational mechanisms of political
		polarization in social-communicative contexts}.
	\newblock \emph{\bibinfo{journal}{Nature Reviews Psychology}}
	\textbf{\bibinfo{volume}{1}}, \bibinfo{pages}{560--576}
	(\bibinfo{year}{2022}).
	
	\bibitem{degroot1974consensus}
	\bibinfo{author}{DeGroot, M.~H.}
	\newblock \bibinfo{title}{Reaching a consensus}.
	\newblock \emph{\bibinfo{journal}{Journal of the American Statistical
			Association}} \textbf{\bibinfo{volume}{69}}, \bibinfo{pages}{118--121}
	(\bibinfo{year}{1974}).
	
	\bibitem{deffuant2000mixing}
	\bibinfo{author}{Deffuant, G.}, \bibinfo{author}{Neau, D.},
	\bibinfo{author}{Amblard, F.} \& \bibinfo{author}{Weisbuch, G.}
	\newblock \bibinfo{title}{Mixing beliefs among interacting agents}.
	\newblock \emph{\bibinfo{journal}{Advances in Complex Systems}}
	\textbf{\bibinfo{volume}{3}}, \bibinfo{pages}{87--98} (\bibinfo{year}{2000}).
	
	\bibitem{hegselmann2002opinion}
	\bibinfo{author}{Hegselmann, R.} \& \bibinfo{author}{Krause, U.}
	\newblock \bibinfo{title}{Opinion dynamics and bounded confidence: models,
		analysis and simulation}.
	\newblock \emph{\bibinfo{journal}{Journal of Artificial Societies and Social
			Simulation}} \textbf{\bibinfo{volume}{5}} (\bibinfo{year}{2002}).
	\newblock \bibinfo{note}{Article 2}.
	
	\bibitem{baumann2020modeling}
	\bibinfo{author}{Baumann, F.}, \bibinfo{author}{Lorenz-Spreen, P.},
	\bibinfo{author}{Sokolov, I.~M.} \& \bibinfo{author}{Starnini, M.}
	\newblock \bibinfo{title}{Modeling echo chambers and polarization dynamics in
		social networks}.
	\newblock \emph{\bibinfo{journal}{Physical Review Letters}}
	\textbf{\bibinfo{volume}{124}}, \bibinfo{pages}{048301}
	(\bibinfo{year}{2020}).
	
	\bibitem{dandekar2013biased}
	\bibinfo{author}{Dandekar, P.}, \bibinfo{author}{Goel, A.} \&
	\bibinfo{author}{Lee, D.~T.}
	\newblock \bibinfo{title}{Biased assimilation, homophily, and the dynamics of
		polarization}.
	\newblock \emph{\bibinfo{journal}{Proceedings of the National Academy of
			Sciences}} \textbf{\bibinfo{volume}{110}}, \bibinfo{pages}{5791--5796}
	(\bibinfo{year}{2013}).
	
	\bibitem{haghtalab2021belief}
	\bibinfo{author}{Haghtalab, N.}, \bibinfo{author}{Jackson, M.~O.} \&
	\bibinfo{author}{Procaccia, A.~D.}
	\newblock \bibinfo{title}{Belief polarization in a complex world: A learning
		theory perspective}.
	\newblock \emph{\bibinfo{journal}{Proceedings of the National Academy of
			Sciences}} \textbf{\bibinfo{volume}{118}}, \bibinfo{pages}{e2010144118}
	(\bibinfo{year}{2021}).
	
	\bibitem{taber2006motivated}
	\bibinfo{author}{Taber, C.~S.} \& \bibinfo{author}{Lodge, M.}
	\newblock \bibinfo{title}{Motivated skepticism in the evaluation of political
		beliefs}.
	\newblock \emph{\bibinfo{journal}{American Journal of Political Science}}
	\textbf{\bibinfo{volume}{50}}, \bibinfo{pages}{755--769}
	(\bibinfo{year}{2006}).
	
	\bibitem{weeks2015emotions}
	\bibinfo{author}{Weeks, B.~E.}
	\newblock \bibinfo{title}{Emotions, partisanship, and misperceptions: How anger
		and anxiety moderate the effect of partisan bias on susceptibility to
		political misinformation}.
	\newblock \emph{\bibinfo{journal}{Journal of Communication}}
	\textbf{\bibinfo{volume}{65}}, \bibinfo{pages}{699--719}
	(\bibinfo{year}{2015}).
	
	\bibitem{rathje2023accuracy}
	\bibinfo{author}{Rathje, S.}, \bibinfo{author}{Roozenbeek, J.},
	\bibinfo{author}{Van~Bavel, J.~J.} \& \bibinfo{author}{Van Der~Linden, S.}
	\newblock \bibinfo{title}{Accuracy and social motivations shape judgements of
		(mis) information}.
	\newblock \emph{\bibinfo{journal}{Nature Human Behaviour}}
	\textbf{\bibinfo{volume}{7}}, \bibinfo{pages}{892--903}
	(\bibinfo{year}{2023}).
	
	\bibitem{wojcieszak2022elites}
	\bibinfo{author}{Wojcieszak, M.}, \bibinfo{author}{Casas, A.},
	\bibinfo{author}{Yu, X.}, \bibinfo{author}{Nagler, J.} \&
	\bibinfo{author}{Tucker, J.~A.}
	\newblock \bibinfo{title}{Most users do not follow political elites on
		{Twitter}; those who do show overwhelming preferences for ideological
		congruity}.
	\newblock \emph{\bibinfo{journal}{Science Advances}}
	\textbf{\bibinfo{volume}{8}}, \bibinfo{pages}{eabn9418}
	(\bibinfo{year}{2022}).
	
	\bibitem{mosleh2021partisanship}
	\bibinfo{author}{Mosleh, M.}, \bibinfo{author}{Martel, C.},
	\bibinfo{author}{Eckles, D.} \& \bibinfo{author}{Rand, D.~G.}
	\newblock \bibinfo{title}{Shared partisanship dramatically increases social tie
		formation in a {Twitter} field experiment}.
	\newblock \emph{\bibinfo{journal}{Proceedings of the National Academy of
			Sciences}} \textbf{\bibinfo{volume}{118}}, \bibinfo{pages}{e2022761118}
	(\bibinfo{year}{2021}).
	
	\bibitem{grinberg2019fake}
	\bibinfo{author}{Grinberg, N.}, \bibinfo{author}{Joseph, K.},
	\bibinfo{author}{Friedland, L.}, \bibinfo{author}{Swire-Thompson, B.} \&
	\bibinfo{author}{Lazer, D.}
	\newblock \bibinfo{title}{Fake news on {Twitter} during the 2016 {US}
		presidential election}.
	\newblock \emph{\bibinfo{journal}{Science}} \textbf{\bibinfo{volume}{363}},
	\bibinfo{pages}{374--378} (\bibinfo{year}{2019}).
	
	\bibitem{guess2020exposure}
	\bibinfo{author}{Guess, A.~M.}, \bibinfo{author}{Nyhan, B.} \&
	\bibinfo{author}{Reifler, J.}
	\newblock \bibinfo{title}{Exposure to untrustworthy websites in the 2016 {US}
		election}.
	\newblock \emph{\bibinfo{journal}{Nature Human Behaviour}}
	\textbf{\bibinfo{volume}{4}}, \bibinfo{pages}{472--480}
	(\bibinfo{year}{2020}).
	
	\bibitem{bovet2019influence}
	\bibinfo{author}{Bovet, A.} \& \bibinfo{author}{Makse, H.~A.}
	\newblock \bibinfo{title}{Influence of fake news in {Twitter} during the 2016
		{US} presidential election}.
	\newblock \emph{\bibinfo{journal}{Nature Communications}}
	\textbf{\bibinfo{volume}{10}}, \bibinfo{pages}{7} (\bibinfo{year}{2019}).
	
	\bibitem{mosleh2024differences}
	\bibinfo{author}{Mosleh, M.}, \bibinfo{author}{Yang, Q.},
	\bibinfo{author}{Zaman, T.}, \bibinfo{author}{Pennycook, G.} \&
	\bibinfo{author}{Rand, D.~G.}
	\newblock \bibinfo{title}{Differences in misinformation sharing can lead to
		politically asymmetric sanctions}.
	\newblock \emph{\bibinfo{journal}{Nature}} \textbf{\bibinfo{volume}{634}},
	\bibinfo{pages}{609--616} (\bibinfo{year}{2024}).
	
	\bibitem{jennings2018polling}
	\bibinfo{author}{Jennings, W.} \& \bibinfo{author}{Wlezien, C.}
	\newblock \bibinfo{title}{Election polling errors across time and space}.
	\newblock \emph{\bibinfo{journal}{Nature Human Behaviour}}
	\textbf{\bibinfo{volume}{2}}, \bibinfo{pages}{276--283}
	(\bibinfo{year}{2018}).
	
	\bibitem{galesic2018socialcircles}
	\bibinfo{author}{Galesic, M.} \emph{et~al.}
	\newblock \bibinfo{title}{Asking about social circles improves election
		predictions}.
	\newblock \emph{\bibinfo{journal}{Nature Human Behaviour}}
	\textbf{\bibinfo{volume}{2}}, \bibinfo{pages}{187--193}
	(\bibinfo{year}{2018}).
	
	\bibitem{bovet2018validation}
	\bibinfo{author}{Bovet, A.}, \bibinfo{author}{Morone, F.} \&
	\bibinfo{author}{Makse, H.~A.}
	\newblock \bibinfo{title}{Validation of {Twitter} opinion trends with national
		polling aggregates: {Hillary Clinton} vs {Donald Trump}}.
	\newblock \emph{\bibinfo{journal}{Scientific Reports}}
	\textbf{\bibinfo{volume}{8}}, \bibinfo{pages}{8673} (\bibinfo{year}{2018}).
	
	\bibitem{bradley2021unrepresentative}
	\bibinfo{author}{Bradley, V.~C.} \emph{et~al.}
	\newblock \bibinfo{title}{Unrepresentative big surveys significantly
		overestimated {US} vaccine uptake}.
	\newblock \emph{\bibinfo{journal}{Nature}} \textbf{\bibinfo{volume}{600}},
	\bibinfo{pages}{695--700} (\bibinfo{year}{2021}).
	
	\bibitem{allcott2017social}
	\bibinfo{author}{Allcott, H.} \& \bibinfo{author}{Gentzkow, M.}
	\newblock \bibinfo{title}{Social media and fake news in the 2016 election}.
	\newblock \emph{\bibinfo{journal}{Journal of Economic Perspectives}}
	\textbf{\bibinfo{volume}{31}}, \bibinfo{pages}{211--236}
	(\bibinfo{year}{2017}).
	
	\bibitem{grover2019polarization}
	\bibinfo{author}{Grover, P.}, \bibinfo{author}{Kar, A.~K.},
	\bibinfo{author}{Dwivedi, Y.~K.} \& \bibinfo{author}{Janssen, M.}
	\newblock \bibinfo{title}{Polarization and acculturation in {US Election} 2016
		outcomes--{Can} twitter analytics predict changes in voting preferences}.
	\newblock \emph{\bibinfo{journal}{Technological Forecasting and Social Change}}
	\textbf{\bibinfo{volume}{145}}, \bibinfo{pages}{438--460}
	(\bibinfo{year}{2019}).
	
	\bibitem{bail2020assessing}
	\bibinfo{author}{Bail, C.~A.} \emph{et~al.}
	\newblock \bibinfo{title}{Assessing the {Russian Internet Research Agency}'s
		impact on the political attitudes and behaviors of {American} {Twitter} users
		in late 2017}.
	\newblock \emph{\bibinfo{journal}{Proceedings of the National Academy of
			Sciences}} \textbf{\bibinfo{volume}{117}}, \bibinfo{pages}{243--250}
	(\bibinfo{year}{2020}).
	
	\bibitem{lazer2020computational}
	\bibinfo{author}{Lazer, D.~M.} \emph{et~al.}
	\newblock \bibinfo{title}{Computational social science: Obstacles and
		opportunities}.
	\newblock \emph{\bibinfo{journal}{Science}} \textbf{\bibinfo{volume}{369}},
	\bibinfo{pages}{1060--1062} (\bibinfo{year}{2020}).
	
	\bibitem{lorenzspreen2023systematic}
	\bibinfo{author}{Lorenz-Spreen, P.}, \bibinfo{author}{Oswald, L.},
	\bibinfo{author}{Lewandowsky, S.} \& \bibinfo{author}{Hertwig, R.}
	\newblock \bibinfo{title}{A systematic review of worldwide causal and
		correlational evidence on digital media and democracy}.
	\newblock \emph{\bibinfo{journal}{Nature Human Behaviour}}
	\textbf{\bibinfo{volume}{7}}, \bibinfo{pages}{74--101}
	(\bibinfo{year}{2023}).
	
	\bibitem{waller2021quantifying}
	\bibinfo{author}{Waller, I.} \& \bibinfo{author}{Anderson, A.}
	\newblock \bibinfo{title}{Quantifying social organization and political
		polarization in online platforms}.
	\newblock \emph{\bibinfo{journal}{Nature}} \textbf{\bibinfo{volume}{600}},
	\bibinfo{pages}{264--268} (\bibinfo{year}{2021}).
	
	\bibitem{eady2023exposure}
	\bibinfo{author}{Eady, G.} \emph{et~al.}
	\newblock \bibinfo{title}{Exposure to the {Russian Internet Research Agency}
		foreign influence campaign on {Twitter} in the 2016 {US} election and its
		relationship to attitudes and voting behavior}.
	\newblock \emph{\bibinfo{journal}{Nature Communications}}
	\textbf{\bibinfo{volume}{14}}, \bibinfo{pages}{62} (\bibinfo{year}{2023}).
	
	\bibitem{green2022online}
	\bibinfo{author}{Green, J.}, \bibinfo{author}{Hobbs, W.},
	\bibinfo{author}{McCabe, S.} \& \bibinfo{author}{Lazer, D.}
	\newblock \bibinfo{title}{Online engagement with 2020 election misinformation
		and turnout in the 2021 {Georgia} runoff election}.
	\newblock \emph{\bibinfo{journal}{Proceedings of the National Academy of
			Sciences}} \textbf{\bibinfo{volume}{119}}, \bibinfo{pages}{e2115900119}
	(\bibinfo{year}{2022}).
	
	\bibitem{echen2020dataset}
	\bibinfo{author}{Chen, E.}, \bibinfo{author}{Deb, A.} \&
	\bibinfo{author}{Ferrara, E.}
	\newblock \bibinfo{title}{{\#Election2020}: the first public {Twitter} dataset
		on the 2020 {US Presidential} election}.
	\newblock \emph{\bibinfo{journal}{Journal of Computational Social Science}}
	\textbf{\bibinfo{volume}{5}}, \bibinfo{pages}{1--18} (\bibinfo{year}{2022}).
	
	\bibitem{cinelli2021echo}
	\bibinfo{author}{Cinelli, M.}, \bibinfo{author}{De~Francisci~Morales, G.},
	\bibinfo{author}{Galeazzi, A.}, \bibinfo{author}{Quattrociocchi, W.} \&
	\bibinfo{author}{Starnini, M.}
	\newblock \bibinfo{title}{The echo chamber effect on social media}.
	\newblock \emph{\bibinfo{journal}{Proceedings of the National Academy of
			Sciences}} \textbf{\bibinfo{volume}{118}}, \bibinfo{pages}{e2023301118}
	(\bibinfo{year}{2021}).
	
	\bibitem{retweeting_code}
	\bibinfo{howpublished}{\url{https://github.com/wanqinjun0514/Retweeting-patterns-in-asymmetric-political-information-ecosystems-signal-electoral-outcomes}}.
	
	\bibitem{thurner2025more}
	\bibinfo{author}{Thurner, S.}, \bibinfo{author}{Hofer, M.} \&
	\bibinfo{author}{Korbel, J.}
	\newblock \bibinfo{title}{Why more social interactions lead to more
		polarization in societies}.
	\newblock \emph{\bibinfo{journal}{Proceedings of the National Academy of
			Sciences}} \textbf{\bibinfo{volume}{122}}, \bibinfo{pages}{e2517530122}
	(\bibinfo{year}{2025}).
	
	\bibitem{yang2023enhanced}
	\bibinfo{author}{Yang, Y.} \emph{et~al.}
	\newblock \bibinfo{title}{Enhanced brain structure-function tethering in
		transmodal cortex revealed by high-frequency eigenmodes}.
	\newblock \emph{\bibinfo{journal}{Nature Communications}}
	\textbf{\bibinfo{volume}{14}}, \bibinfo{pages}{6744} (\bibinfo{year}{2023}).
	
	\bibitem{guess2023social}
	\bibinfo{author}{Guess, A.~M.} \emph{et~al.}
	\newblock \bibinfo{title}{How do social media feed algorithms affect attitudes
		and behavior in an election campaign?}
	\newblock \emph{\bibinfo{journal}{Science}} \textbf{\bibinfo{volume}{381}},
	\bibinfo{pages}{398--404} (\bibinfo{year}{2023}).
	
	\bibitem{gonzalez2023asymmetric}
	\bibinfo{author}{Gonz{\'a}lez-Bail{\'o}n, S.} \emph{et~al.}
	\newblock \bibinfo{title}{Asymmetric ideological segregation in exposure to
		political news on {Facebook}}.
	\newblock \emph{\bibinfo{journal}{Science}} \textbf{\bibinfo{volume}{381}},
	\bibinfo{pages}{392--398} (\bibinfo{year}{2023}).
	
\end{thebibliography}

\end{document}


\title{Supplementary Material for ``Retweeting patterns reveal partisan asymmetries in information sharing and signal state-level electoral outcomes in the 2020 US presidential election''}
	
	\author[1]{\fnm{Dawei} \sur{Li}}
	\equalcont{These authors contributed equally to this work.}
	
	\author[1]{\fnm{Qinjun} \sur{Wan}}
	\equalcont{These authors contributed equally to this work.}
	
	\author*[2,4,5]{\fnm{Xin} \sur{Wang}}
	\email{wangxin\_1993@buaa.edu.cn}
	
	\author[2]{\fnm{Dahe} \sur{Jiang}}
	
	\author[2]{\fnm{Junyu} \sur{Lu}}
	
	\author[2]{\fnm{Shitong} \sur{Wang}}
	
	\author[1]{\fnm{Zhelun} \sur{Zhang}}
	
	\author[1]{\fnm{Zhenyu} \sur{Guan}}
	
	\author[2,3,6,7]{\fnm{Shaoting} \sur{Tang}}
	
	\author[8,9,10]{\fnm{Feng} \sur{Fu}}

	\affil[1]{\orgname{School of Cyber Science and Technology,
			Beihang University, Beijing, 100191, China}}
	
	\affil[2]{\orgname{School of Artificial Intelligence,
			Beihang University, Beijing, 100191, China}}
	
	\affil[3]{\orgname{Hangzhou International Innovation Institute,
			Beihang University, Hangzhou, 311115, China}}
	
	\affil[4]{\orgname{Key Laboratory of Mathematics, Informatics and Behavioral Semantics,
			Beihang University, Beijing, 100191, China}}
	
	\affil[5]{\orgname{Beijing Advanced Innovation Center for Future Blockchain and Privacy Computing,
			Beihang University, Beijing, 100191, China}}
	
	\affil[6]{\orgname{Institute of Trustworthy Artificial Intelligence,
			Zhejiang Normal University, Hangzhou, 310012, China}}
	
	\affil[7]{\orgname{Shandong Key Lab of Complex Medical Intelligence and Aging,
			Shandong Medical and Pharmaceutical University, Yantai, 264003, China}}
	
	\affil[8]{\orgname{Department of Mathematics,
			Dartmouth College, Hanover, NH, USA}}
	
	\affil[9]{\orgname{Department of Biomedical Data Science,
			Geisel School of Medicine at Dartmouth, Lebanon, NH 03756, USA}}
	
	\affil[10]{\orgname{Department of Applied Mathematics,
			School of Engineering \& Applied Science, Yale University,
			New Haven, CT 06520, USA}}

	\maketitle
	
	This Supplementary Information provides detailed methodological specifications, validation analyses, complementary results, and robustness analyses supporting the main text “Retweeting patterns in asymmetric political information ecosystems signal electoral outcomes in the 2020 US presidential election”. It includes detailed descriptions of the data sources and ideological entity dictionaries, user-level ideological scoring, temporal ideological transitions, and the construction and normalization of information-consumption matrices. The validity of incorporating quote tweets into the ideological analysis is examined through the stance classification analysis. We also provide the formulation of ideological permeability and the robust regression framework linking information consumption to subsequent ideological change, together with complementary analyses of scoring coverage and long-term-active users. Furthermore, sensitivity analyses under alternative parameter settings are presented to assess the robustness of the ideological-permeability results, while representative media and political entity dictionaries are reported.
	\clearpage
	\setcounter{tocdepth}{2}
	\tableofcontents
	\clearpage
	
	\begin{appendices}
		
		\renewcommand{\figurename}{Fig.}
		\renewcommand{\tablename}{Table}
		\setcounter{table}{0}
		\setcounter{figure}{0}
		\setcounter{equation}{0}
		\renewcommand{\thetable}{S\arabic{table}}
		\renewcommand{\thefigure}{S\arabic{figure}}
		\renewcommand{\theequation}{S\arabic{equation}}
		
		\phantomsection
		\section*{S1 Supplementary Methods}
		\addcontentsline{toc}{section}{S1 Supplementary Methods}
		\label{sec:supp_methods}
		
		\phantomsection
		\subsection*{S1.1 Data sources and notation}
		\addcontentsline{toc}{subsection}{S1.1 Data sources and notation}
		\label{subsec:s1_1_data_notation}
		
		The study uses 542,994,806 available non-reply retweet and quote-tweet records hydrated from the public Tweet ID collection for the 2020 U.S. presidential election. The records span December 2019 through February 2021. Records were divided according to URL metadata into External URLs (82,242,667 records; 15.15\%), Twitter URLs (95,775,521; 17.64\%), Without URLs (361,690,197; 66.61\%), and Multiple URLs (3,286,421; 0.61\%). The first three categories form the mutually exclusive information-source types used in downstream analysis. Multiple-URL records were retained in the corpus audit but excluded from ideological scoring because a unique information source could not be assigned without additional assumptions. Monthly source totals are reported in Supplementary Table~\ref{tab:supp_monthly_url_volume}.
		
		The notation used in the full methodological specification is collected in Supplementary Table~\ref{tab:supp_notation}.

		\begin{longtable}{
				@{}
				>{\raggedright\arraybackslash}p{0.25\textwidth}
				>{\raggedright\arraybackslash}p{0.69\textwidth}
				@{}
			}
			\caption{Key notation used in the supplementary methods.}
			\label{tab:supp_notation}\\
			
			\toprule
			\textbf{Symbol} & \textbf{Definition} \\
			\midrule
			\endfirsthead
			
			\multicolumn{2}{c}{
				\tablename\ \thetable{} -- continued from previous page
			}\\
			\toprule
			\textbf{Symbol} & \textbf{Definition} \\
			\midrule
			\endhead
			
			\midrule
			\multicolumn{2}{r}{Continued on next page}\\
			\endfoot
			
			\bottomrule
			\endlastfoot

			
			$u$
			&
			Index of a user.
			\\
			
			$r$
			&
			Index of a retweet or quote-tweet record. A successfully matched record is treated as one information-consumption event.
			\\
			
			$s$
			&
			Index of an information-source type.
			\\
			
			$m$
			&
			Index of a calendar month, with $m=1,\ldots,15$ for the 15-month study period.
			\\
			
			$q$
			&
			Index of one of the six consecutive analysis periods, with $q=1,\ldots,6$.
			\\

			
			$\mathcal{S}=\{E,T,W\}$
			&
			Set of the three information-source types, where $E$ denotes External URLs, $T$ denotes Twitter URLs, and $W$ denotes Without URLs.
			\\

			
			$\mathcal{R}^{\mathrm{M}}_{u,m,s}$
			&
			Set of retweet and quote-tweet records produced by user $u$ in month $m$ and information-source type $s$ that were successfully matched to a labeled media entity.
			\\
			
			$\mathcal{R}^{\mathrm{M}}_{u,q,s}$
			&
			Set of retweet and quote-tweet records produced by user $u$ in analysis period $q$ and information-source type $s$ that were successfully matched to a labeled media entity.
			\\

			
			$\mathcal{R}^{\mathrm{P}}_{u,m,s}$
			&
			Set of retweet and quote-tweet records produced by user $u$ in month $m$ and information-source type $s$ that were successfully matched to a labeled politician entity.
			\\
			
			$\mathcal{R}^{\mathrm{P}}_{u,q,s}$
			&
			Set of retweet and quote-tweet records produced by user $u$ in analysis period $q$ and information-source type $s$ that were successfully matched to a labeled politician entity.
			\\

			
			$\operatorname{MediaEntityScore}_{r}$
			&
			Numerical ideological score of the media entity matched to record $r$, taking a value in $\{-3,-2,-1,0,1,2,3\}$.
			\\
			
			$\operatorname{PoliticianEntityScore}_{r}$
			&
			Numerical ideological score of the politician entity matched to record $r$, taking a value in $\{-1,0,1\}$.
			\\

			
			$\operatorname{MediaScore}_{u,m}$
			&
			Overall media-based ideology score of user $u$ in month $m$, obtained by pooling successfully scored records across the three information-source types.
			\\
			
			$\operatorname{MediaScore}_{u,q}$
			&
			Overall media-based ideology score of user $u$ in analysis period $q$, obtained by pooling successfully scored records across the three information-source types.
			\\
			
			$\operatorname{MediaScore}_{u,q,s}$
			&
			Source-specific media-based ideology score of user $u$ in analysis period $q$ for information-source type $s$.
			\\

			
			$\operatorname{PoliticianScore}_{u,m}$
			&
			Overall politician-based ideology score of user $u$ in month $m$, obtained by pooling successfully scored records across the three information-source types.
			\\
			
			$\operatorname{PoliticianScore}_{u,q}$
			&
			Overall politician-based ideology score of user $u$ in analysis period $q$, obtained by pooling successfully scored records across the three information-source types.
			\\
			
		\end{longtable}
		
		\phantomsection

		\subsection*{S1.2 Quote-tweet sampling and stance classification}
		\addcontentsline{toc}{subsection}{S1.2 Quote-tweet sampling and stance classification}
		\label{subsec:s1_2_quote_validation}
		To examine how users responded to quoted political content, we extracted direct quote tweets from the hydrated dataset, excluding retweets and replies. This procedure yielded 37,059,671 valid quote records. After removing 956,270 duplicate records based on the unique quote-tweet id, the sampling population consisted of 36,103,401 unique quote tweets. We then drew a proportional stratified sample of 100,000 quote tweets by month, preserving the temporal distribution of the population from December 2019 to February 2021. A fixed deterministic sampling procedure was used to ensure reproducibility. For each sampled quote tweet, we classified the stance expressed in the quoting user's comment toward the claim or viewpoint conveyed in the quoted original tweet. Classification was performed through the DeepSeek API using the deepseek-v4-flash model, with three possible labels: support, oppose, and neutral. Support indicated that the quoting user agreed with or reinforced the original claim, whereas oppose required an explicit denial, contradiction, rejection, or counterargument. Cases that were unclear, unrelated, dependent on unavailable external context, or merely reacted to a person or event mentioned in the original tweet were classified as neutral.

		All 100,000 sampled quote tweets were successfully classified. Of these, 48,440 (48.44\%) were labelled support, 38,578 (38.58\%) were labelled neutral, and 12,982 (12.98\%) were labelled oppose. Thus, explicit opposition to the claim or viewpoint expressed in the quoted tweet accounted for a minority of quote-tweet interactions, whereas most quote tweets either reinforced the original content or did not express a clear opposing stance. Quote tweets also constituted only approximately 6.83\% of the 542,994,806 retweet and quote-tweet records in the full dataset. Based on the stratified sample, quote tweets expressing explicit opposition were therefore estimated to represent approximately 0.89\% of all records. The subsequent ideological analysis pooled retweets and quote tweets because it was designed to measure users' exposure to and engagement with ideologically labelled information sources, rather than to interpret every individual interaction as an endorsement of the referenced content. The stance analysis provided empirical support for this treatment by showing that quote tweets explicitly rejecting the referenced claim formed only a limited share of the full interaction dataset. Accordingly, the inclusion of quote tweets was unlikely to cause the ideology measures to be dominated by oppositional quoting.
		
		The complete classification prompt is shown as follows. For each sampled quote tweet, the quoting user's comment was denoted by \texttt{q}, and the text of the quoted original tweet was denoted by \texttt{o}. The model was instructed to classify the stance expressed by \texttt{q} toward the claim or viewpoint conveyed in \texttt{o}, rather than the sentiment expressed in either text independently. The complete instruction prompt was:
		
		\begin{verbatim}
			Classify q's stance toward the claim or viewpoint expressed by o.
			Labels:
			- support: q agrees with or reinforces o's claim or viewpoint.
			- oppose: q denies, contradicts, rejects, or argues against
			o's claim or viewpoint.
			- neutral: q only reacts to people or events mentioned in o, adds
			unrelated commentary, or lacks clear evidence of support or opposition.
			Rules:
			- Sentiment is not stance. Negative language alone is not oppose.
			- Criticizing a person mentioned in o is neutral unless q rejects o's claim.
			- If o reports that someone said or did something, reacting negatively to
			that person or event is neutral unless q disputes the report or an
			explicit claim quoted in o.
			- If q repeats or strengthens o's criticism, label support even when both
			texts are negative.
			- Insults, emojis, hashtags, missing image/link context, and unclear
			sarcasm alone are neutral.
			Return each id once as JSON:
			{"results":[{"id":"0","label":"support"}]}
		\end{verbatim}
		
		\phantomsection
		\subsection*{S1.3 Ideological labels for media and politicians}
		\addcontentsline{toc}{subsection}{S1.3 Ideological labels for media and politicians}
		\label{subsec:s1_3_dictionaries}

		To extend ideological labeling beyond external media links, we constructed an entity-level annotation framework covering both media organizations and political figures across different types of information sources. The framework consisted of four complementary dictionaries: media-domain labels, media-username labels, politician-related domain labels, and politician-username labels. Media entities were assigned ideological categories using Media Bias/Fact Check. Political entities were classified as Left, Center, or Right according to party affiliation, which was identified with GPT assistance. The following paragraphs describe the identification and ideological labeling procedures used to construct each dictionary.

		Media-domain labels were constructed from external URLs. The hostname of each URL was extracted and standardized by removing wrapping characters, adding a protocol header when absent, removing port numbers and the www. prefix, and converting the hostname to lowercase. The resulting domains were deduplicated and ranked by occurrence frequency. Domains corresponding to news media organizations were identified and labeled according to the Media Bias/Fact Check classification system, including Extreme Left, Left, Left-Leaning, Center, Right-Leaning, Right, and Extreme Bias Right. This procedure yielded 1,061 media-domain labels.
		
		Media-username labels were constructed from Twitter URLs. For each Twitter URL, the first path segment following "twitter.com/" was extracted as the username of the linked account. The extracted usernames were cleaned, deduplicated, and ranked according to their frequency of occurrence in Twitter URLs. The corresponding account IDs were also retained to support account matching. Candidate accounts were then manually verified to determine whether they represented official accounts of news media organizations. Once the corresponding media organization was identified, its Media Bias/Fact Check classification was assigned to the associated Twitter account. This procedure produced 663 media-username labels.
		
		Politician-username labels were constructed from records without URLs. For each retweet or quote, the user ID and username of the author of the original post were extracted. The number of times each original author appeared as the source of a retweet or quote was then counted across the dataset, and candidate accounts were ranked in descending order of occurrence frequency. Candidate accounts were reviewed to determine whether they belonged to identifiable politicians. For accounts associated with influencers, party affiliation was identified with GPT assistance and then mapped to political orientation: Democratic politicians were labeled Left, Republican politicians were labeled Right, and politicians affiliated with other parties or neither of the two major parties were labeled Center. Both the user ID and username were retained in the label dictionary. The user ID served as the primary identifier for account deduplication and identity verification, whereas the username was retained for matching with Twitter URLs. After account identification and deduplication,  4,023 politician-username labels were obtained.
		
		Politician-related domain labels were also constructed from External URLs. Following the same URL-standardization procedure used for media domains, the extracted hostnames were deduplicated, their frequencies of occurrence were counted, and domains were ranked in descending order of occurrence frequency. Domains corresponding to politicians' websites, campaign websites, and official websites were identified as politician domains. The party affiliation of the politician associated with each domain was identified with GPT assistance and mapped to a political label: Democratic politicians were labeled Left, Republican politicians were labeled Right, and politicians affiliated with other parties or neither of the two major parties were labeled Center. After domain identification, deduplication, and label consolidation, 35 politician-related domain labels were obtained.
		
		\phantomsection
		\subsection*{S1.4 User-level ideological scoring}
		\addcontentsline{toc}{subsection}{S1.4 User-level ideological scoring}
		\label{subsec:s1_4_user_scoring}
		
		After the four label dictionaries were constructed, ideological labels were assigned to retweet and quote records by matching the referenced website domain or Twitter username against the corresponding media or politician label dictionary. Records containing multiple URLs were excluded from ideological scoring because they accounted for only 0.61\% of the dataset and could not be assigned to a unique information source without introducing additional assumptions. After this exclusion, each remaining record was assigned to one of three mutually exclusive information-source types: External URLs, Twitter URLs, and Without URLs. The remaining records were evaluated separately in the media and politician dimensions.

		For records containing an external URL, the hostname was extracted and standardized using the same procedure applied during label construction. The standardized hostname was matched separately against the media-domain dictionary and the politician-related domain dictionary. For records containing a Twitter URL, the first path segment following "twitter.com/" was extracted as the username of the linked account and matched separately against the media-username dictionary and the politician-username dictionary. For records without URLs, the user ID of the author of the original retweeted or quoted post was extracted and matched against the account identifiers retained in the media-username and politician-username dictionaries. For each successful match, the political-orientation label associated with the matched entity was retrieved from the corresponding dictionary and converted into a numerical score, which was then assigned to that retweet or quote-tweet. Media labels were mapped to a seven-point scale. Extreme Left was assigned as -3, Left as -2, Left-Leaning as -1, Center as 0, Right-Leaning as 1, Right as 2, and Extreme Right as 3. Politician labels were mapped to a three-point scale, with Left assigned -1, Center 0, and Right 1. Media-based and politician-based matching was performed independently. A record was included in the media-based calculation whenever a media match was found and in the politician-based calculation whenever a politician match was found. Failure to match in one dimension did not affect eligibility in the other. Records that could not be matched to a labeled entity in a given dimension were excluded from both the score total and the valid count for that dimension.

		For each user, media-based and politician-based ideology scores were calculated separately for each monthly and quarterly period. Let $\mathcal{S}=\{E,T,W\}$ denote the three information-source types: External URLs, Twitter URLs, and Without URLs. Let $\mathcal{R}^{\mathrm{M}}_{u,m,s}$ and $\mathcal{R}^{\mathrm{M}}_{u,q,s}$ denote the sets of retweet and quote-tweet posts made by user $u$ during month $m$ and analysis period $q$, respectively, that belonged to source type $s$ and were successfully matched to a labeled media entity. Let $\operatorname{MediaEntityScore}_{r}$ denote the numerical media-orientation score assigned to record $r$ based on the media entity matched to that record. The media-based ideology scores of user $u$ during month $m$ and analysis period $q$ were calculated as
		
		\begin{equation}
			\begin{aligned}
				\operatorname{MediaScore}_{u,m}
				&=
				\frac{
					\displaystyle
					\sum_{s \in \mathcal{S}}
					\sum_{r \in \mathcal{R}^{\mathrm{M}}_{u,m,s}}
					\operatorname{MediaEntityScore}_{r}
				}{
					\displaystyle
					\sum_{s \in \mathcal{S}}
					\left|
					\mathcal{R}^{\mathrm{M}}_{u,m,s}
					\right|
				},
				\\[6pt]
				\operatorname{MediaScore}_{u,q}
				&=
				\frac{
					\displaystyle
					\sum_{s \in \mathcal{S}}
					\sum_{r \in \mathcal{R}^{\mathrm{M}}_{u,q,s}}
					\operatorname{MediaEntityScore}_{r}
				}{
					\displaystyle
					\sum_{s \in \mathcal{S}}
					\left|
					\mathcal{R}^{\mathrm{M}}_{u,q,s}
					\right|
				}.
			\end{aligned}
		\end{equation}

		For analyses requiring source-specific media-based ideology scores, we additionally calculated the user's average media-based ideology score separately within each information-source type. For user $u$, analysis period $q$, and source type $s$, the source-specific media-based ideology score was defined as
		
		\begin{equation}
			\operatorname{MediaScore}_{u,q,s}
			=
			\frac{
				\displaystyle
				\sum_{r \in \mathcal{R}^{\mathrm{M}}_{u,q,s}}
				\operatorname{MediaEntityScore}_{r}
			}{
				\left|
				\mathcal{R}^{\mathrm{M}}_{u,q,s}
				\right|
			},
		\end{equation}
		
		provided that $\left|\mathcal{R}^{\mathrm{M}}_{u,q,s}\right|>0$. If no record from source type $s$ was successfully matched to a labeled media entity for user $u$ in analysis period $q$, the corresponding source-specific score was treated as undefined. These source-specific scores were used where the analysis was performed separately by information-source type, whereas $\operatorname{MediaScore}_{u,q}$ defined above represents the overall media-based ideology score obtained by pooling successfully scored records across all three source types.

		Similarly, let $\mathcal{R}^{\mathrm{P}}_{u,m,s}$ and $\mathcal{R}^{\mathrm{P}}_{u,q,s}$ denote the sets of retweet and quote-tweet posts made by user $u$ during month $m$ and analysis period $q$, respectively, that belonged to source type $s$ and were successfully matched to a labeled politician entity. Let $\operatorname{PoliticianEntityScore}_{r}$ denote the numerical politician-orientation score assigned to record $r$ based on the politician entity matched to that record. The politician-based ideology scores of user $u$ during month $m$ and analysis period $q$ were calculated as
		
		\begin{equation}
			\begin{aligned}
				\operatorname{PoliticianScore}_{u,m}
				&=
				\frac{
					\displaystyle
					\sum_{s \in \mathcal{S}}
					\sum_{r \in \mathcal{R}^{\mathrm{P}}_{u,m,s}}
					\operatorname{PoliticianEntityScore}_{r}
				}{
					\displaystyle
					\sum_{s \in \mathcal{S}}
					\left|
					\mathcal{R}^{\mathrm{P}}_{u,m,s}
					\right|
				},
				\\[6pt]
				\operatorname{PoliticianScore}_{u,q}
				&=
				\frac{
					\displaystyle
					\sum_{s \in \mathcal{S}}
					\sum_{r \in \mathcal{R}^{\mathrm{P}}_{u,q,s}}
					\operatorname{PoliticianEntityScore}_{r}
				}{
					\displaystyle
					\sum_{s \in \mathcal{S}}
					\left|
					\mathcal{R}^{\mathrm{P}}_{u,q,s}
					\right|
				}.
			\end{aligned}
		\end{equation}
		
		For both dimensions, all successfully scored posts from the three information-source types were pooled before the user-level average ideology score was calculated. Specifically, the scores assigned to these posts were summed, and the resulting total was divided by the total number of successfully scored posts. Thus, each source type contributed to the user-level score in proportion to the number of successfully matched posts it contained, rather than the three source-specific mean scores being assigned equal weight. The denominator represented the number of successfully scored posts. If a user had no successfully scored post in a given dimension and period, the corresponding ideology score was treated as undefined rather than zero, and the user was not classified in that dimension for that period.

		\section*{S2 Complementary results and parameter settings}
		\addcontentsline{toc}{section}{S2 Complementary results and parameter settings}
		\label{sec:supp_complementary}
		
		\phantomsection
		\subsection*{S2.1 Complementary results for Fig. 1}
		\addcontentsline{toc}{subsection}{S2.1 Complementary results for Fig. 1}
		\label{subsec:s2_1_fig1}
		
		\paragraph*{(a) Monthly tweet volume in the three data sources.}
		The dataset covers December 2019 through February 2021. Replies were removed, while retweets and quote tweets were retained. The Multiple URLs column is retained as an additional category, whereas downstream analyses use the three mutually exclusive source categories: Without URL, Twitter URL, and External URL. Monthly tweet volumes varied over the study period, with activity increasing markedly during the later months of 2020 and remaining high into 2021. Across the full dataset, tweets without URLs accounted for the largest share, followed by tweets containing Twitter URLs and external URLs.

		\begin{table}[htbp]
			\centering
			\caption{Monthly tweet volume by URL type from December 2019 to February 2021.}
			\label{tab:supp_monthly_url_volume}
			\begin{tabular}{lrrrrr}
				\toprule
				Month & Without URL & External URL & Twitter URL & Multiple URLs & Total Tweets \\
				\midrule
				2019\_12 & 10,115,654 & 1,827,536  & 3,044,175  & 110,340  & 15,097,705 \\
				2020\_01 & 10,598,954 & 1,753,350  & 3,180,760  & 106,973  & 15,640,037 \\
				2020\_02 & 10,323,981 & 1,867,428  & 3,370,861  & 102,809  & 15,665,079 \\
				2020\_03 & 12,874,981 & 1,926,242  & 3,716,167  & 140,463  & 18,657,853 \\
				2020\_04 & 21,382,606 & 3,896,867  & 6,506,755  & 215,009  & 32,001,237 \\
				2020\_05 & 10,876,260 & 1,724,488  & 3,305,059  & 133,867  & 16,039,674 \\
				2020\_06 & 15,733,642 & 1,882,170  & 4,118,089  & 152,680  & 21,886,581 \\
				2020\_07 & 19,346,280 & 2,754,638  & 4,993,880  & 204,464  & 27,299,262 \\
				2020\_08 & 26,589,391 & 4,040,111  & 7,114,636  & 262,584  & 38,006,722 \\
				2020\_09 & 32,484,650 & 7,907,073  & 9,636,881  & 306,551  & 50,335,155 \\
				2020\_10 & 41,940,768 & 11,181,648 & 11,190,292 & 367,661  & 64,680,369 \\
				2020\_11 & 40,681,021 & 9,493,624  & 10,248,948 & 294,446  & 60,718,039 \\
				2020\_12 & 33,220,882 & 8,865,402  & 7,779,509  & 259,107  & 50,124,900 \\
				2021\_01 & 43,218,852 & 12,996,317 & 10,366,156 & 335,222  & 66,916,547 \\
				2021\_02 & 32,302,275 & 10,125,773 & 7,203,353  & 294,245  & 49,925,646 \\
				\midrule
				Total     & 361,690,197 & 82,242,667 & 95,775,521 & 3,286,421 & 542,994,806 \\
				\bottomrule
			\end{tabular}
		\end{table}

		\FloatBarrier
		\paragraph*{(b) Monthly scoring coverage under the media- and politician-based systems.}
		The two scoring systems showed complementary coverage patterns across the three source types. Media-based scoring had highest coverage for tweets containing external URLs, whereas politician-based scoring covered a much larger share of tweets without URLs and tweets containing Twitter URLs. The monthly figures show how the numbers of scored users and tweets varied over time within each source type.
		
		\begin{table}[htbp]
			\centering
			\caption{Summary of media-based scoring coverage by URL type.}
			\label{tab:supp_media_scoring_coverage_summary}
			\begin{tabular}{lrrr}
				\toprule
				URL type & Scored tweets & Total tweets & Tweet scoring coverage \\
				\midrule
				Without URL & 5,306,383 & 361,690,197 & 1.47\% \\
				Twitter URL & 13,008,910 & 95,775,521 & 13.58\% \\
				External URL & 50,313,771 & 82,242,667 & 61.18\% \\
				\bottomrule
			\end{tabular}
		\end{table}

		\begin{table}[htbp]
			\centering
			\setlength{\tabcolsep}{2pt} 
			\caption{Monthly users and tweets with media-based scores in the content-without-URL subset.}
			\begin{tabular}{lrrrrrr}
				\toprule
				Month & Scored Users & Total Users & User Score Coverage & Scored Tweets & Total Tweets & Tweet Score Coverage \\
				\midrule
				2019-12 & 90{,}845 & 1{,}297{,}839 & 0.0700 & 138{,}495 & 10{,}115{,}654 & 0.0137 \\
				2020-01 & 66{,}246 & 1{,}571{,}089 & 0.0422 & 118{,}037 & 10{,}598{,}954 & 0.0111 \\
				2020-02 & 63{,}172 & 1{,}429{,}587 & 0.0442 & 107{,}666 & 10{,}323{,}981 & 0.0104 \\
				2020-03 & 65{,}519 & 2{,}144{,}962 & 0.0305 & 89{,}124 & 12{,}874{,}981 & 0.0069 \\
				2020-04 & 41{,}046 & 1{,}863{,}831 & 0.0220 & 124{,}100 & 21{,}382{,}606 & 0.0058 \\
				2020-05 & 46{,}163 & 1{,}792{,}097 & 0.0258 & 67{,}415 & 10{,}876{,}260 & 0.0062 \\
				2020-06 & 69{,}281 & 2{,}316{,}720 & 0.0299 & 91{,}991 & 15{,}733{,}642 & 0.0058 \\
				2020-07 & 63{,}616 & 2{,}300{,}531 & 0.0277 & 91{,}819 & 19{,}346{,}280 & 0.0047 \\
				2020-08 & 220{,}994 & 2{,}888{,}647 & 0.0765 & 311{,}370 & 26{,}589{,}391 & 0.0117 \\
				2020-09 & 262{,}136 & 3{,}398{,}235 & 0.0771 & 519{,}520 & 32{,}484{,}650 & 0.0160 \\
				2020-10 & 321{,}954 & 4{,}655{,}193 & 0.0692 & 683{,}447 & 41{,}940{,}768 & 0.0163 \\
				2020-11 & 341{,}035 & 5{,}908{,}930 & 0.0577 & 676{,}136 & 40{,}681{,}021 & 0.0166 \\
				2020-12 & 220{,}290 & 3{,}360{,}885 & 0.0655 & 535{,}119 & 33{,}220{,}882 & 0.0161 \\
				2021-01 & 450{,}485 & 5{,}630{,}950 & 0.0800 & 995{,}142 & 43{,}218{,}852 & 0.0230 \\
				2021-02 & 292{,}594 & 3{,}724{,}358 & 0.0786 & 757{,}002 & 32{,}302{,}275 & 0.0234 \\
				\midrule
				Total & – & – & – & 5{,}306{,}383 & 361{,}690{,}197 & 0.0147 \\
				\bottomrule
			\end{tabular}
			\label{tab:supp_media_without_coverage}
		\end{table}
		
		\begin{table}[htbp]
			\centering
			\setlength{\tabcolsep}{2pt} 
			\caption{Monthly users and tweets with media-based scores in the Twitter-URL subset.}
			\begin{tabular}{lrrrrrr}
				\toprule
				Month & Scored Users & Total Users & User Score Coverage & Scored Tweets & Total Tweets & Tweet Score Coverage \\
				\midrule
				2019-12 & 177{,}737 & 603{,}947 & 0.2943 & 417{,}221 & 3{,}044{,}175 & 0.1371 \\
				2020-01 & 175{,}687 & 621{,}073 & 0.2829 & 368{,}382 & 3{,}180{,}760 & 0.1158 \\
				2020-02 & 160{,}215 & 630{,}328 & 0.2542 & 327{,}859 & 3{,}370{,}861 & 0.0973 \\
				2020-03 & 221{,}816 & 793{,}292 & 0.2796 & 446{,}908 & 3{,}716{,}167 & 0.1203 \\
				2020-04 & 173{,}942 & 723{,}089 & 0.2406 & 731{,}628 & 6{,}506{,}755 & 0.1124 \\
				2020-05 & 195{,}866 & 667{,}193 & 0.2936 & 395{,}086 & 3{,}305{,}059 & 0.1195 \\
				2020-06 & 199{,}135 & 959{,}021 & 0.2076 & 391{,}679 & 4{,}118{,}089 & 0.0951 \\
				2020-07 & 243{,}843 & 1{,}036{,}055 & 0.2354 & 574{,}765 & 4{,}993{,}880 & 0.1151 \\
				2020-08 & 407{,}237 & 1{,}304{,}215 & 0.3122 & 1{,}008{,}307 & 7{,}114{,}636 & 0.1417 \\
				2020-09 & 592{,}553 & 1{,}523{,}040 & 0.3891 & 1{,}553{,}053 & 9{,}636{,}881 & 0.1612 \\
				2020-10 & 644{,}201 & 1{,}971{,}444 & 0.3268 & 1{,}516{,}534 & 11{,}190{,}292 & 0.1355 \\
				2020-11 & 557{,}782 & 2{,}029{,}553 & 0.2748 & 1{,}263{,}372 & 10{,}248{,}948 & 0.1233 \\
				2020-12 & 401{,}956 & 1{,}168{,}863 & 0.3439 & 1{,}130{,}101 & 7{,}779{,}509 & 0.1453 \\
				2021-01 & 610{,}419 & 1{,}911{,}231 & 0.3194 & 1{,}587{,}897 & 10{,}366{,}156 & 0.1532 \\
				2021-02 & 455{,}207 & 1{,}341{,}402 & 0.3394 & 1{,}296{,}118 & 7{,}203{,}353 & 0.1799 \\
				\midrule
				Total & – & – & – & 13{,}008{,}910 & 95{,}775{,}521 & 0.1358 \\
				\bottomrule
			\end{tabular}
			\label{tab:supp_media_twitter_coverage}
		\end{table}
		
		\begin{table}[htbp]
			\centering
			\setlength{\tabcolsep}{2pt} 
			\caption{Monthly users and tweets with media-based scores in the external-URL subset.}
			\begin{tabular}{lrrrrrr}
				\toprule
				Month & Scored Users & Total Users & User Score Coverage & Scored Tweets & Total Tweets & Tweet Score Coverage \\
				\midrule
				2019-12 & 264{,}131 & 388{,}969 & 0.6791 & 1{,}086{,}419 & 1{,}827{,}536 & 0.5945 \\
				2020-01 & 250{,}915 & 389{,}641 & 0.6440 & 948{,}178 & 1{,}753{,}350 & 0.5408 \\
				2020-02 & 281{,}779 & 471{,}879 & 0.5971 & 970{,}514 & 1{,}867{,}428 & 0.5197 \\
				2020-03 & 322{,}865 & 503{,}708 & 0.6410 & 1{,}049{,}035 & 1{,}926{,}242 & 0.5446 \\
				2020-04 & 323{,}373 & 474{,}111 & 0.6821 & 2{,}359{,}063 & 3{,}896{,}867 & 0.6054 \\
				2020-05 & 277{,}707 & 404{,}134 & 0.6872 & 1{,}052{,}771 & 1{,}724{,}488 & 0.6105 \\
				2020-06 & 337{,}066 & 503{,}301 & 0.6697 & 1{,}114{,}830 & 1{,}882{,}170 & 0.5923 \\
				2020-07 & 395{,}837 & 581{,}745 & 0.6804 & 1{,}546{,}993 & 2{,}754{,}638 & 0.5616 \\
				2020-08 & 508{,}793 & 809{,}445 & 0.6286 & 2{,}257{,}209 & 4{,}040{,}111 & 0.5587 \\
				2020-09 & 814{,}323 & 1{,}208{,}513 & 0.6738 & 4{,}906{,}714 & 7{,}907{,}073 & 0.6205 \\
				2020-10 & 1{,}013{,}715 & 1{,}631{,}051 & 0.6215 & 6{,}682{,}474 & 11{,}181{,}648 & 0.5976 \\
				2020-11 & 1{,}033{,}900 & 1{,}735{,}996 & 0.5956 & 5{,}553{,}313 & 9{,}493{,}624 & 0.5850 \\
				2020-12 & 854{,}029 & 1{,}256{,}601 & 0.6796 & 5{,}733{,}070 & 8{,}865{,}402 & 0.6467 \\
				2021-01 & 1{,}314{,}845 & 1{,}999{,}367 & 0.6576 & 8{,}382{,}425 & 12{,}996{,}317 & 0.6450 \\
				2021-02 & 938{,}323 & 1{,}463{,}523 & 0.6411 & 6{,}670{,}763 & 10{,}125{,}773 & 0.6588 \\
				\midrule
				Total & – & – & – & 50{,}313{,}771 & 82{,}242{,}667 & 0.6118 \\
				\bottomrule
			\end{tabular}
			\label{tab:supp_media_external_coverage}
		\end{table}
		
		\begin{table}[htbp]
			\centering
			\caption{Summary of politician-based scoring coverage by URL type.}
			\label{tab:supp_politician_scoring_coverage_summary}
			\begin{tabular}{lrrr}
				\toprule
				URL type & Scored tweets & Total tweets & Tweet scoring coverage \\
				\midrule
				Without URL & 214,495,937 & 361,690,197 & 59.30\% \\
				Twitter URL & 56,595,626 & 95,775,521 & 59.09\% \\
				External URL & 3,745,948 & 82,242,667 & 4.55\% \\
				\bottomrule
			\end{tabular}
		\end{table}
		
		\begin{table}[htbp]
			\centering
			\setlength{\tabcolsep}{2pt} 
			\caption{Monthly users and tweets with politician-based scores in the content-without-URL subset.}
			\begin{tabular}{lrrrrrr}
				\toprule
				Month & Scored Users & Total Users & User Score Coverage & Scored Tweets & Total Tweets & Tweet Score Coverage \\
				\midrule
				2019-12 & 885{,}720 & 1{,}297{,}839 & 0.6825 & 6{,}299{,}544 & 10{,}115{,}654 & 0.6228 \\
				2020-01 & 1{,}060{,}430 & 1{,}571{,}089 & 0.6750 & 6{,}563{,}320 & 10{,}598{,}954 & 0.6192 \\
				2020-02 & 949{,}893 & 1{,}429{,}587 & 0.6645 & 6{,}068{,}808 & 10{,}323{,}981 & 0.5878 \\
				2020-03 & 1{,}535{,}455 & 2{,}144{,}962 & 0.7158 & 8{,}128{,}458 & 12{,}874{,}981 & 0.6313 \\
				2020-04 & 1{,}263{,}217 & 1{,}863{,}831 & 0.6778 & 12{,}932{,}668 & 21{,}382{,}606 & 0.6048 \\
				2020-05 & 1{,}265{,}274 & 1{,}792{,}097 & 0.7060 & 6{,}927{,}636 & 10{,}876{,}260 & 0.6370 \\
				2020-06 & 1{,}605{,}117 & 2{,}316{,}720 & 0.6928 & 10{,}188{,}124 & 15{,}733{,}642 & 0.6475 \\
				2020-07 & 1{,}628{,}417 & 2{,}300{,}531 & 0.7078 & 12{,}711{,}055 & 19{,}346{,}280 & 0.6570 \\
				2020-08 & 1{,}891{,}926 & 2{,}888{,}647 & 0.6550 & 17{,}769{,}962 & 26{,}589{,}391 & 0.6683 \\
				2020-09 & 1{,}924{,}268 & 3{,}398{,}235 & 0.5663 & 20{,}114{,}164 & 32{,}484{,}650 & 0.6192 \\
				2020-10 & 2{,}641{,}250 & 4{,}655{,}193 & 0.5674 & 25{,}972{,}539 & 41{,}940{,}768 & 0.6193 \\
				2020-11 & 3{,}033{,}942 & 5{,}908{,}930 & 0.5135 & 23{,}466{,}500 & 40{,}681{,}021 & 0.5768 \\
				2020-12 & 1{,}710{,}828 & 3{,}360{,}885 & 0.5090 & 19{,}938{,}806 & 33{,}220{,}882 & 0.6002 \\
				2021-01 & 2{,}694{,}380 & 5{,}630{,}950 & 0.4785 & 21{,}983{,}105 & 43{,}218{,}852 & 0.5086 \\
				2021-02 & 1{,}389{,}085 & 3{,}724{,}358 & 0.3730 & 15{,}431{,}248 & 32{,}302{,}275 & 0.4777 \\
				\midrule
				Total & – & – & – & 214{,}495{,}937 & 361{,}690{,}197 & 0.5930 \\
				\bottomrule
			\end{tabular}
			\label{tab:supp_politician_without_coverage}
		\end{table}
		
		\begin{table}[htbp]
			\centering
			\setlength{\tabcolsep}{2pt} 
			\caption{Monthly users and tweets with politician-based scores in the Twitter-URL subset.}
			\begin{tabular}{lrrrrrr}
				\toprule
				Month & Scored Users & Total Users & User Score Coverage & Scored Tweets & Total Tweets & Tweet Score Coverage \\
				\midrule
				2019-12 & 389{,}249 & 603{,}947 & 0.6445 & 1{,}710{,}974 & 3{,}044{,}175 & 0.5620 \\
				2020-01 & 440{,}739 & 621{,}073 & 0.7096 & 1{,}899{,}235 & 3{,}180{,}760 & 0.5971 \\
				2020-02 & 465{,}321 & 630{,}328 & 0.7382 & 2{,}065{,}263 & 3{,}370{,}861 & 0.6127 \\
				2020-03 & 557{,}135 & 793{,}292 & 0.7023 & 2{,}278{,}960 & 3{,}716{,}167 & 0.6133 \\
				2020-04 & 529{,}643 & 723{,}089 & 0.7325 & 4{,}310{,}120 & 6{,}506{,}755 & 0.6624 \\
				2020-05 & 488{,}074 & 667{,}193 & 0.7315 & 2{,}181{,}209 & 3{,}305{,}059 & 0.6600 \\
				2020-06 & 727{,}631 & 959{,}021 & 0.7587 & 2{,}827{,}676 & 4{,}118{,}089 & 0.6866 \\
				2020-07 & 816{,}537 & 1{,}036{,}055 & 0.7881 & 3{,}324{,}901 & 4{,}993{,}880 & 0.6658 \\
				2020-08 & 916{,}861 & 1{,}304{,}215 & 0.7030 & 4{,}414{,}321 & 7{,}114{,}636 & 0.6205 \\
				2020-09 & 1{,}013{,}828 & 1{,}523{,}040 & 0.6657 & 5{,}755{,}655 & 9{,}636{,}881 & 0.5973 \\
				2020-10 & 1{,}328{,}778 & 1{,}971{,}444 & 0.6740 & 6{,}580{,}718 & 11{,}190{,}292 & 0.5881 \\
				2020-11 & 1{,}217{,}082 & 2{,}029{,}553 & 0.5997 & 5{,}766{,}164 & 10{,}248{,}948 & 0.5626 \\
				2020-12 & 724{,}718 & 1{,}168{,}863 & 0.6200 & 4{,}451{,}525 & 7{,}779{,}509 & 0.5722 \\
				2021-01 & 1{,}149{,}467 & 1{,}911{,}231 & 0.6014 & 5{,}563{,}296 & 10{,}366{,}156 & 0.5367 \\
				2021-02 & 688{,}165 & 1{,}341{,}402 & 0.5130 & 3{,}465{,}609 & 7{,}203{,}353 & 0.4811 \\
				\midrule
				Total & – & – & – & 56{,}595{,}626 & 95{,}775{,}521 & 0.5909 \\
				\bottomrule
			\end{tabular}
			\label{tab:supp_politician_twitter_coverage}
		\end{table}
		
		\begin{table}[htbp]
			\centering
			\setlength{\tabcolsep}{2pt} 
			\caption{Monthly users and tweets with politician-based scores in the external-URL subset.}
			\begin{tabular}{lrrrrrr}
				\toprule
				Month & Scored Users & Total Users & User Score Coverage & Scored Tweets & Total Tweets & Tweet Score Coverage \\
				\midrule
				2019-12 & 39{,}226 & 388{,}969 & 0.1008 & 73{,}531 & 1{,}827{,}536 & 0.0402 \\
				2020-01 & 47{,}366 & 389{,}641 & 0.1216 & 94{,}721 & 1{,}753{,}350 & 0.0540 \\
				2020-02 & 55{,}024 & 471{,}879 & 0.1166 & 105{,}995 & 1{,}867{,}428 & 0.0568 \\
				2020-03 & 82{,}212 & 503{,}708 & 0.1632 & 158{,}423 & 1{,}926{,}242 & 0.0822 \\
				2020-04 & 68{,}891 & 474{,}111 & 0.1453 & 268{,}428 & 3{,}896{,}867 & 0.0689 \\
				2020-05 & 60{,}240 & 404{,}134 & 0.1491 & 127{,}473 & 1{,}724{,}488 & 0.0739 \\
				2020-06 & 79{,}710 & 503{,}301 & 0.1584 & 156{,}343 & 1{,}882{,}170 & 0.0831 \\
				2020-07 & 104{,}444 & 581{,}745 & 0.1795 & 280{,}954 & 2{,}754{,}638 & 0.1020 \\
				2020-08 & 128{,}487 & 809{,}445 & 0.1587 & 298{,}573 & 4{,}040{,}111 & 0.0739 \\
				2020-09 & 167{,}108 & 1{,}208{,}513 & 0.1383 & 467{,}814 & 7{,}907{,}073 & 0.0592 \\
				2020-10 & 222{,}298 & 1{,}631{,}051 & 0.1363 & 634{,}880 & 11{,}181{,}648 & 0.0568 \\
				2020-11 & 225{,}859 & 1{,}735{,}996 & 0.1301 & 428{,}422 & 9{,}493{,}624 & 0.0451 \\
				2020-12 & 80{,}758 & 1{,}256{,}601 & 0.0643 & 185{,}458 & 8{,}865{,}402 & 0.0209 \\
				2021-01 & 90{,}375 & 1{,}999{,}367 & 0.0452 & 224{,}812 & 12{,}996{,}317 & 0.0173 \\
				2021-02 & 90{,}312 & 1{,}463{,}523 & 0.0617 & 240{,}121 & 10{,}125{,}773 & 0.0237 \\
				\midrule
				Total & – & – & – & 3{,}745{,}948 & 82{,}242{,}667 & 0.0455 \\
				\bottomrule
			\end{tabular}
			\label{tab:supp_politician_external_coverage}
		\end{table}

\begin{figure}[htbp]
			\centering
			\includegraphics[width=0.95\linewidth,height=0.78\textheight,keepaspectratio]{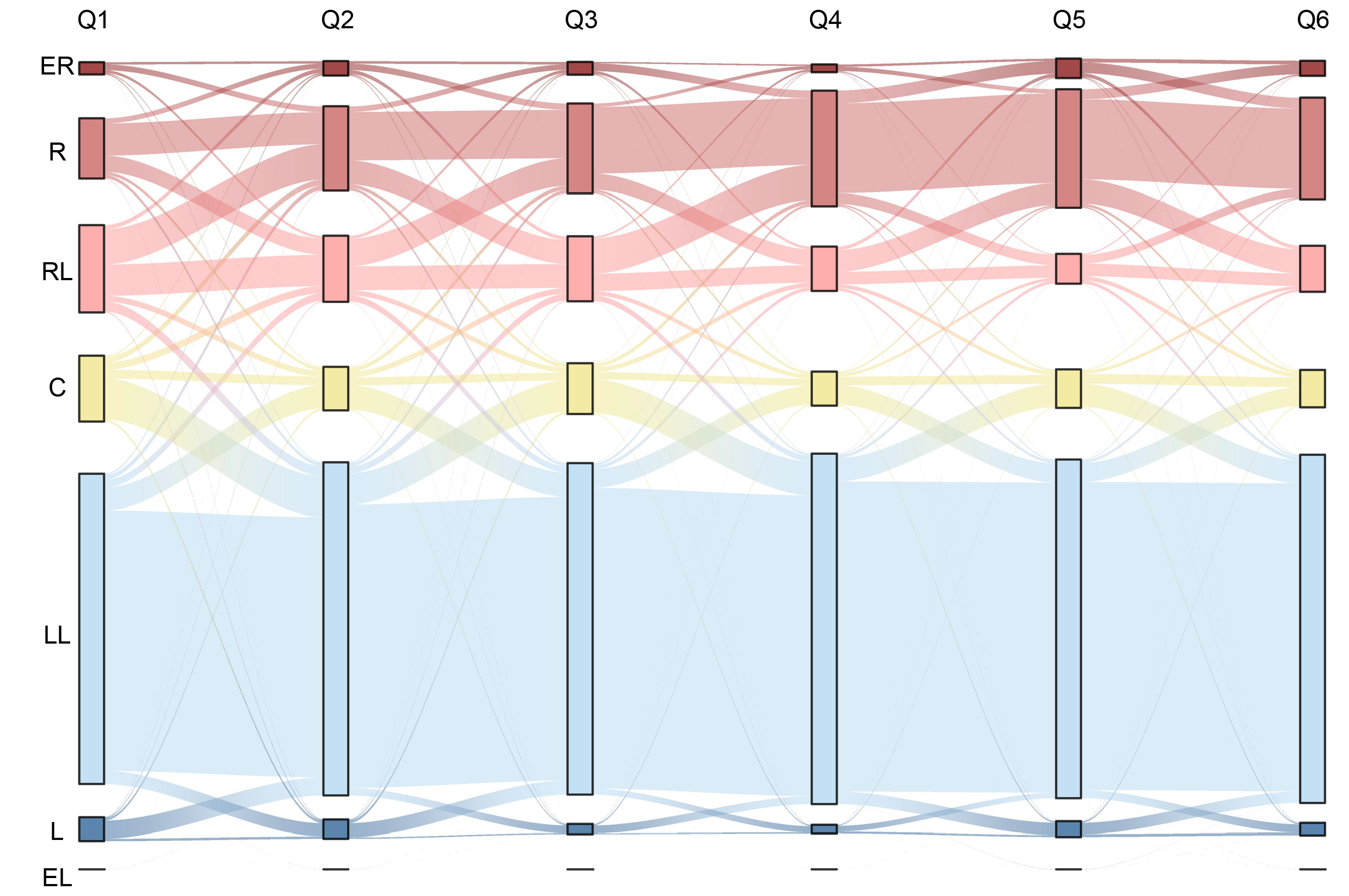}
			\caption{Six-quarter ideological transitions among long-term-active users with valid media-based ideological scores in all six quarters.}
			\label{fig:supp_long_term_active_sankey}
		\end{figure}

		\FloatBarrier
		\paragraph*{(c) Long-term-active user transitions.}
		Let $\mathcal{U}_q$ denote the users with a valid overall media-based ideological score in analysis period $q$. Long-term-active users were defined as
		\[
		\mathcal{U}_{\mathrm{LTA}}=\bigcap_{q=1}^{6}\mathcal{U}_q.
		\]
		A user was retained only if a valid score was observed in all six analysis periods, yielding 194,340 long-term-active users. Supplementary Fig.~\ref{fig:supp_long_term_active_sankey} shows ideological transitions across the six periods within this fixed group, and Supplementary Table~\ref{tab:supp_long_term_active_composition} reports the corresponding ideological composition in each period.

		Across the six quarters, the ideological composition of long-term-active users remained relatively stable, with left-leaning users consistently representing the largest group. The proportion of left-leaning users increased from 55.56\% in Q1 to 61.83\% in Q6, indicating a gradual concentration of the active user population toward the left-leaning category. Meanwhile, the share of center users declined from 12.20\% to 7.61\%, suggesting a reduction in the relative size of moderate users over time. On the right side, right-leaning users showed a decreasing trend overall (from 14.91\% in Q1 to 9.54\% in Q6), whereas right users increased from 11.24\% to 16.15\%. Extreme ideological categories remained rare throughout the observation period, each accounting for less than 2.5\% of users. Overall, the stable composition across quarters suggests that the long-term-active users exhibit persistent ideological characteristics, while moderate changes mainly occur within adjacent ideological categories.
		
		\begin{table}[htbp]
			\centering
			\caption{Quarterly ideological composition among long-term-active users.
				Values represent the proportion of users in each ideological category
				within each quarter. EL, extreme left; L, left; LL, left-leaning;
				C, center; RL, right-leaning; R, right; ER, extreme right.}
			\label{tab:supp_long_term_active_composition}
			\begin{tabular}{lrrrrrrr}
				\toprule
				Quarter & EL & L & LL & C & RL & R & ER \\
				\midrule
				Q1 & 0.0016 & 0.0415 & 0.5556 & 0.1220 & 0.1491 & 0.1124 & 0.0178 \\
				Q2 & 0.0021 & 0.0433 & 0.5895 & 0.0844 & 0.1145 & 0.1441 & 0.0220 \\
				Q3 & 0.0008 & 0.0232 & 0.6008 & 0.0950 & 0.1324 & 0.1319 & 0.0159 \\
				Q4 & 0.0006 & 0.0199 & 0.6257 & 0.0688 & 0.1018 & 0.1740 & 0.0091 \\
				Q5 & 0.0013 & 0.0353 & 0.5968 & 0.0808 & 0.0684 & 0.1979 & 0.0195 \\
				Q6 & 0.0007 & 0.0284 & 0.6183 & 0.0761 & 0.0954 & 0.1615 & 0.0196 \\
				\bottomrule
			\end{tabular}
		\end{table}
		
		\phantomsection

		\FloatBarrier
		\subsection*{S2.2 Complementary results for Fig. 2}
		\addcontentsline{toc}{subsection}{S2.2 Complementary results for Fig. 2}
		\label{subsec:s2_2_fig2}

		\paragraph*{(a) Aggregate information-consumption matrix.}
		
		\begin{figure}[!htbp]
			\centering
			\includegraphics[width=0.95\textwidth]{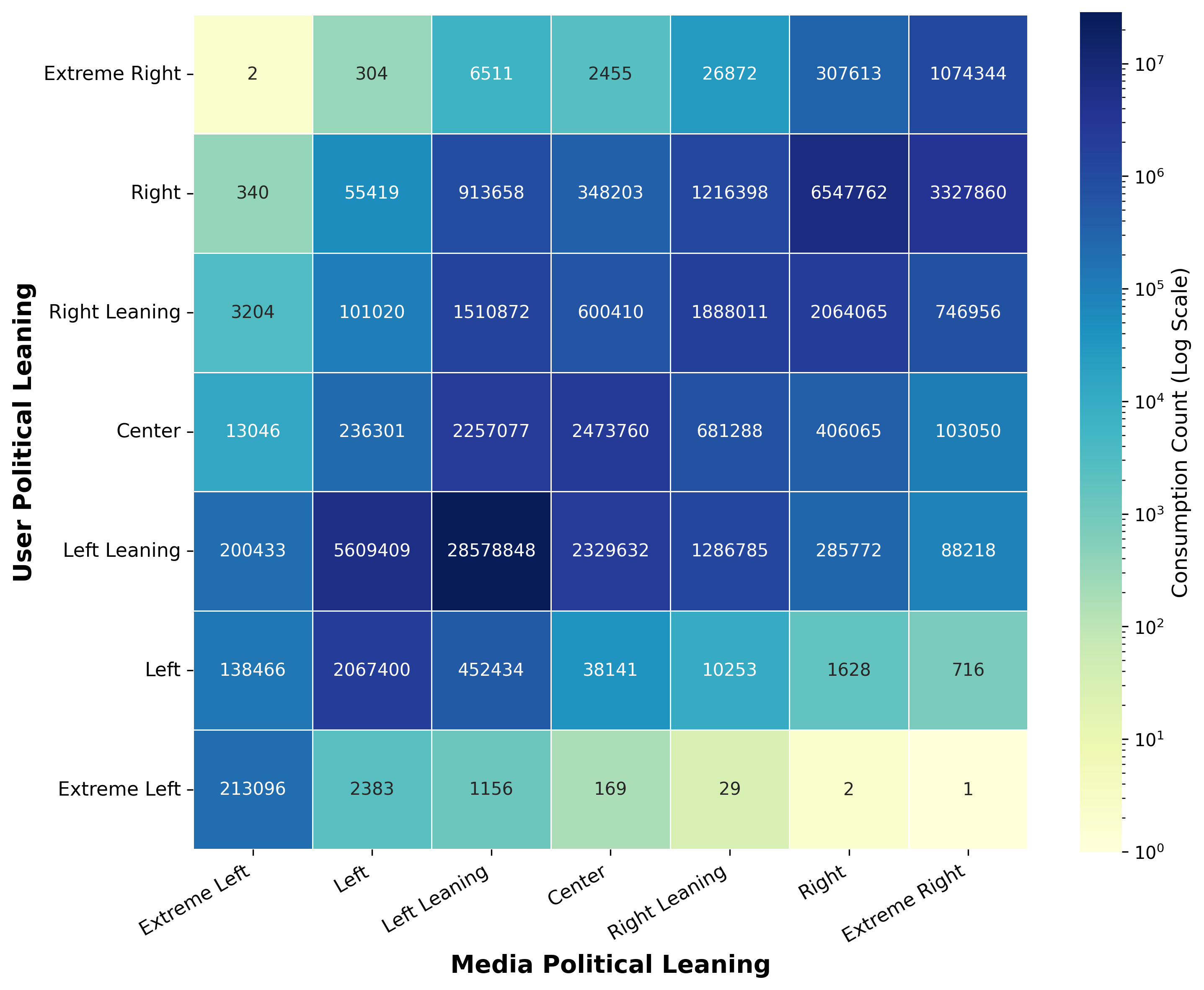}
			\caption{\textbf{Aggregate information-consumption matrix with raw counts.}
				Heatmap of the total numbers of successfully matched retweet and quote-tweet records between user ideological categories (rows) and media ideological categories (columns), aggregated across all study months and across the three information-source types: External URLs, Twitter URLs, and Without URLs. This figure displays the underlying raw counts for each user-media ideological pairing.}
			\label{fig:supp_heatmap_raw_counts}
		\end{figure}

		Supplementary Fig.~\ref{fig:supp_heatmap_raw_counts} presents the aggregate information-consumption matrix underlying Fig.~2a using raw counts. The matrix was constructed by summing the corresponding cells of the monthly information-consumption count matrices across all months and all three information-source types. Rows represent the ideological categories of users, and columns represent the ideological categories of the media entities with which they were matched. Cell values indicate the total numbers of successfully matched retweet and quote-tweet records in each user-media ideological pairing over the full period. The three source-type counts reported in Table~\ref{tab:supp_media_scoring_coverage_summary} sum to 68,629,064 records. The media-scoring statistics reported in Table~\ref{tab:supp_media_scoring_coverage_summary} and the information-consumption matrix were generated through two separate processing pipelines. Both the media-scoring and information-consumption matrix pipelines involved URL-parsing and matching steps from the raw retweet records, with the latter used to identify information-source types. Because URLs occur in diverse formats, differences in URL parsing and matching across the two pipelines resulted in different record counts. The resulting matrix contains 68,217,837 records, compared with 68,629,064 scored tweets reported in Table~\ref{tab:supp_media_scoring_coverage_summary}, a difference of 411,227 records.

		\FloatBarrier
		\paragraph*{(b) Sensitivity to the ideological-distance weighting multiplier.}
		Ideological distance was given additional weight when users consumed media information from the opposite side of the ideological spectrum. The ideological distance for consumption events was multiplied by 2. To examine whether the results depended on this weighting choice, we repeated the analysis using alternative multipliers of 1.0, 1.1, 1.2, 1.3, 1.4, and 1.5. For each multiplier, ideological permeability was recalculated separately for External URLs, Twitter URLs, and Without URLs for every user and analysis period. The weighted ideological-distance scores and the corresponding numbers of information-consumption events were then combined across the three information-source types. The overall ideological permeability score was calculated by dividing the summed weighted ideological distance by the total number of information-consumption events across the three source types. Using the ideological-permeability scores obtained under each alternative multiplier, we repeated the robust regressions of subsequent ideological change. Regressions were conducted separately for each initial ideological category and each pair of adjacent analysis periods. All calculation and regression procedures were kept unchanged except for the weighting multiplier. The resulting regression estimates were compared across the alternative multiplier settings to evaluate whether the observed relationship between ideological permeability and subsequent ideological change was sensitive to this parameter choice.

\begin{figure}[htbp]
			\centering
			\includegraphics[width=\textwidth]{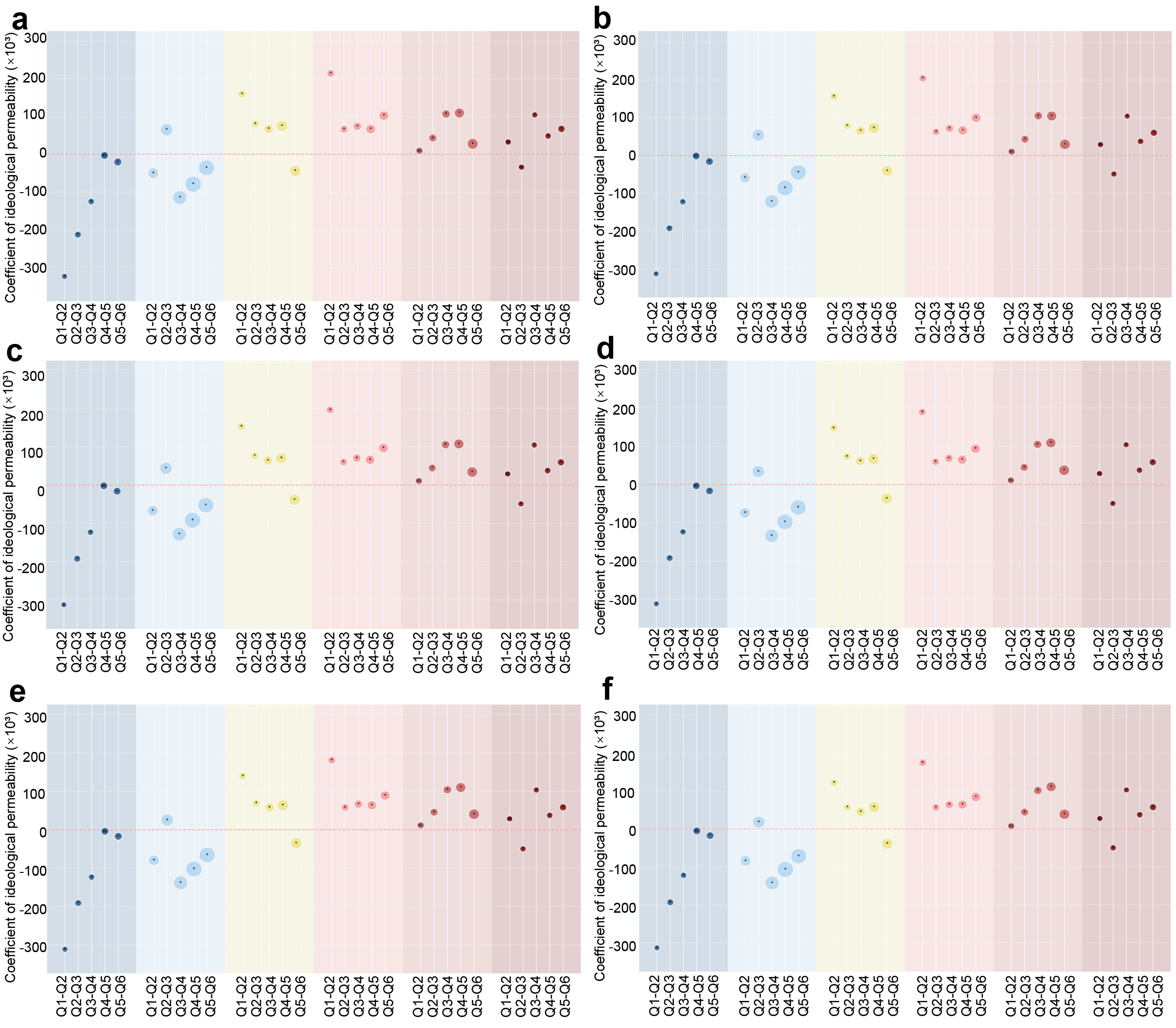}
			\caption{\textbf{Sensitivity of the ideological-permeability regressions to alternative weighting multipliers.}
				Robust regression coefficients relating ideological permeability in the initial analysis period to subsequent bias change are shown for weighting multipliers of
				\textbf{a,} 1.0;
				\textbf{b,} 1.1;
				\textbf{c,} 1.2;
				\textbf{d,} 1.3;
				\textbf{e,} 1.4; and
				\textbf{f,} 1.5.
				Positive coefficients indicate relatively more rightward subsequent change and negative coefficients indicate relatively more leftward subsequent change. The overall ideological pattern remains stable across alternative weighting settings.}
			\label{fig:supp_multiplier_robustness}
		\end{figure}

		The relationship between ideological permeability and subsequent ideological change was highly consistent across alternative weighting multipliers ranging from 1.0 to 1.5 (Supplementary Fig.~\ref{fig:supp_multiplier_robustness}). Although the magnitude of individual regression coefficients varied as the weighting multiplier changed, the overall ideological pattern remained stable. Left and Left-Leaning users predominantly showed negative coefficients, indicating that greater ideological permeability was associated with relatively more leftward subsequent change. In contrast, Right-Leaning, Right, and Extreme Right users predominantly showed positive coefficients, indicating relatively more rightward subsequent change. The coefficients for Center users were generally smaller and showed greater variation across analysis periods. Importantly, changing the weighting assigned to ideologically distant information did not reverse the broader left-right asymmetry observed in the main text. These results indicate that the association reported in the main text is not driven by the specific weighting multiplier used in the construction of ideological permeability.
		
\begin{figure}[htbp]
			\centering
			\includegraphics[width=0.85\textwidth]{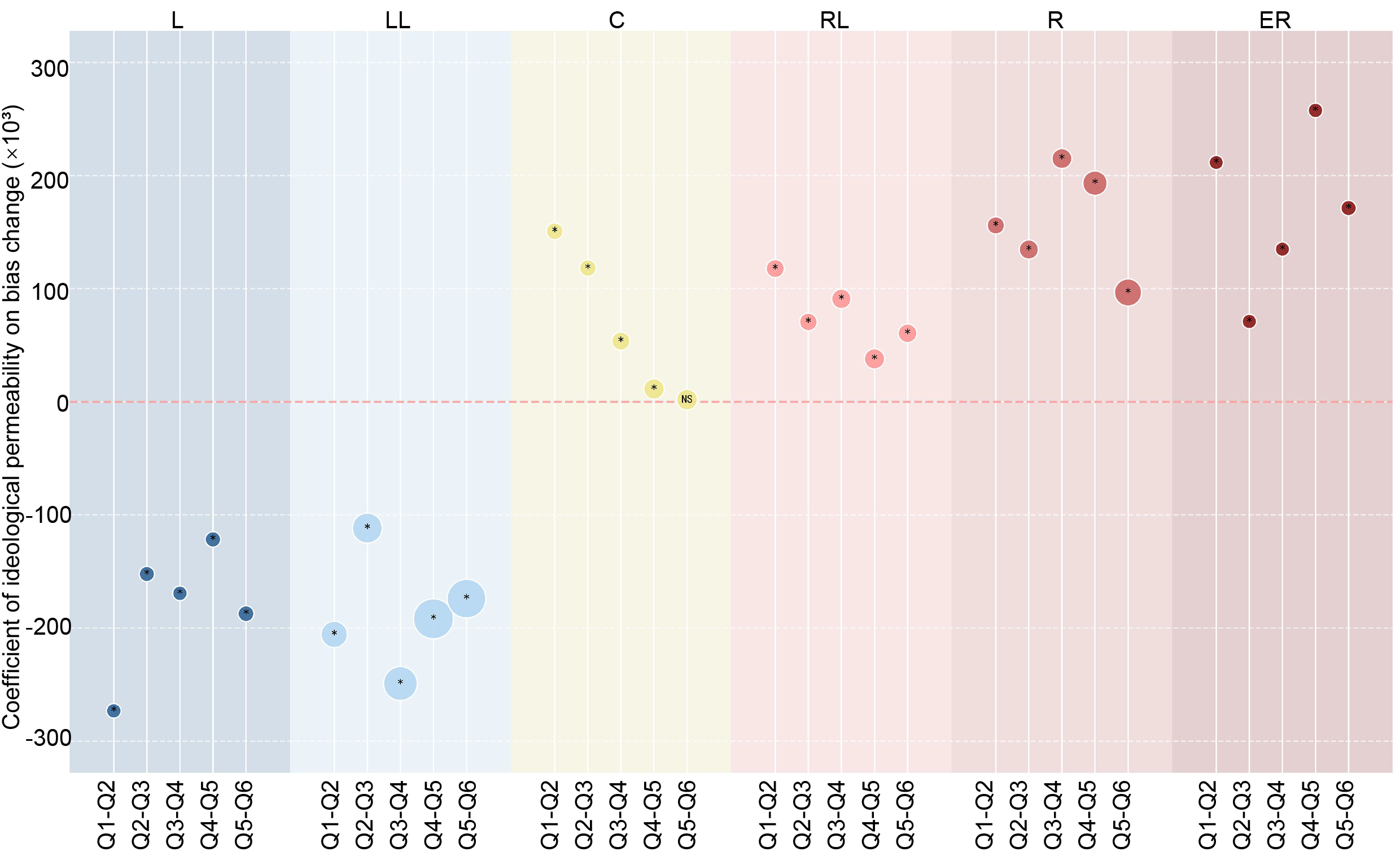}
			\caption{\textbf{Robustness of the ideological-permeability regressions using a minimum of three information-consumption events.}
				The regressions were repeated after restricting the sample to user-period observations with at least three valid information-consumption events, compared with the threshold of at least two events used in the main text. Positive coefficients indicate relatively more rightward subsequent change and negative coefficients indicate relatively more leftward subsequent change. The main ideological pattern remains consistent under the stricter activity threshold.}
			\label{fig:supp_appear3_robustness}
		\end{figure}

		\FloatBarrier
		\paragraph*{(c) Sensitivity to the minimum number of information-consumption events.}
		We further examined whether the regression results were affected by users with very few observed information-consumption events. The main text retained user-period observations with at least two valid information-consumption events. As a robustness check, we increased this requirement to at least three events and repeated the regressions (Supplementary Fig.~\ref{fig:supp_appear3_robustness}). The resulting pattern closely reproduced that of the main text despite the more restrictive sample. Negative associations remained concentrated among Left and Left-Leaning users, whereas positive associations remained predominant among Right-Leaning, Right, and Extreme Right users. Center users again showed comparatively weaker and less consistent coefficients across periods. Thus, excluding observations based on only two information-consumption events did not materially alter the substantive pattern of the results.
		
		Together, the two robustness analyses show that the observed association between ideological permeability and subsequent bias change is stable to both the weighting used for ideologically distant information and a stricter requirement for the number of observed information-consumption events. 
		
		\FloatBarrier
		\paragraph*{(d) Top- versus bottom-quartile ideological permeability and subsequent bias change.}
		To examine whether the ideological transitions observed in Fig. 1c varied with the degree of ideological permeability in Fig. 2b,c, we compared the distributions of subsequent bias change between users in the top and bottom quartiles of ideological permeability in Supplementary Fig.~\ref{fig:supp_top25_bottom25_boxplot}. Within each initial ideological group and adjacent-period transition, users were ranked by their ideological permeability to select the highest and lowest 25\%, pooled across transitions for each group. Focusing on three intermediate ideological categories, we examined Left-Leaning (LL), Center (C), and Right-Leaning (RL) users. Bias change is a signed measure. A positive mean indicates an average rightward transition in the subsequent analysis period, whereas a negative mean indicates an average leftward transition. For Left-Leaning users, the top-permeability group showed a lower distribution of bias change compared with the bottom-permeability group, indicating relatively more leftward subsequent changes among users with higher ideological permeability. In contrast, Right-Leaning users exhibited the opposite pattern, with the top-permeability group showing a higher distribution of bias change, suggesting relatively more rightward subsequent changes associated with higher ideological permeability. Center users displayed an intermediate pattern: although both permeability groups showed predominantly negative bias changes, indicating overall leftward shifts, the top-permeability group exhibited a less negative distribution, suggesting a weaker leftward shift compared with the bottom-permeability group. Overall, higher ideological permeability is associated with directional changes that differ by users' initial ideological positions: relatively more leftward changes among Left-Leaning users, but relatively more rightward changes among Center and Right-Leaning users. These patterns are consistent with the regression coefficients observed in Fig. 2c, where the association between ideological permeability and subsequent bias change varies across ideological categories.

		\begin{figure}[htbp]
			\centering
			\includegraphics[width=0.95\textwidth]{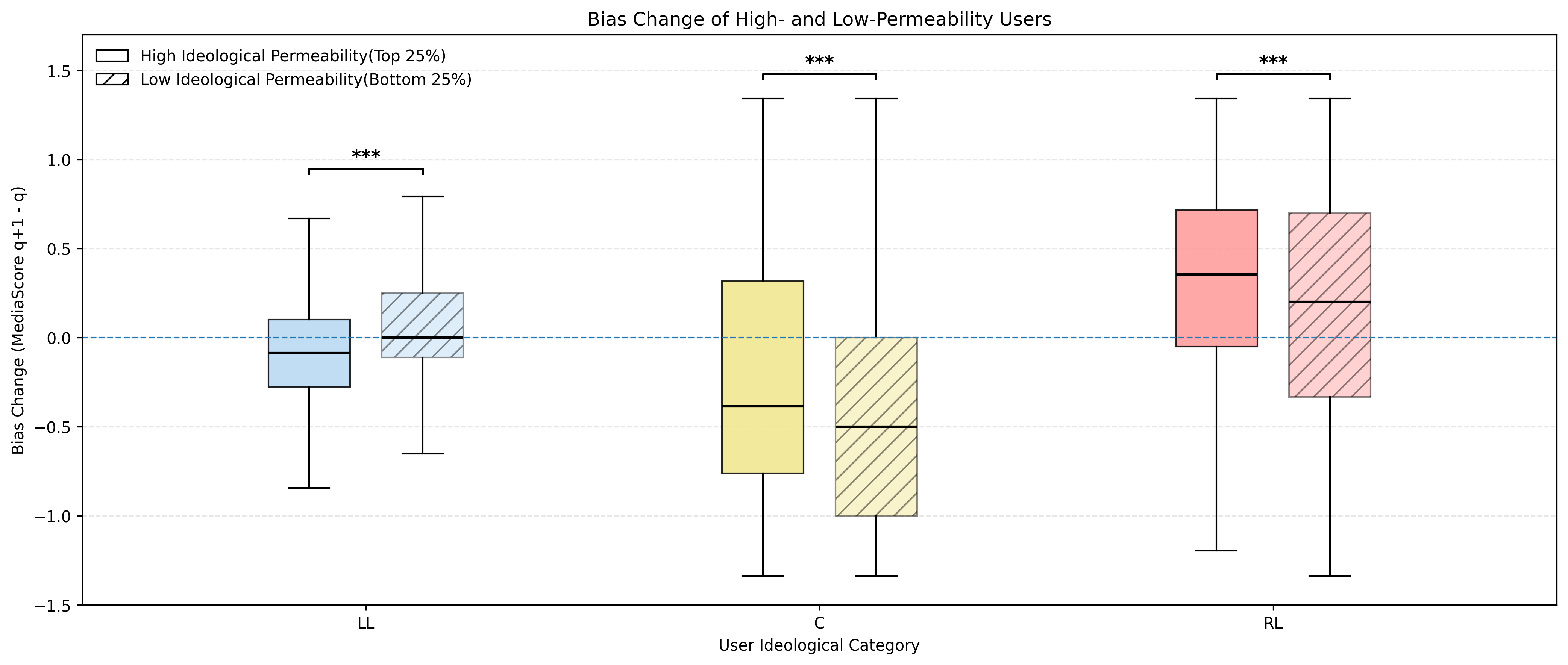}
			
			\caption{\textbf{Subsequent bias change among users in the top and bottom quartiles of ideological permeability.}
				Boxplots compare the distributions of subsequent bias change between the top 25\% and bottom 25\% ideological-permeability groups within Left-Leaning (LL), Center (C), and Right-Leaning (RL)  ideological categories. Bias change is defined as the difference between the user’s media-based ideological score in the next quarter and that in the current quarter. Higher values indicate more rightward subsequent change, and lower values indicate more leftward subsequent change. Significant differences between the top and bottom quartiles for all three categories are indicated by *** ($p < 0.001$).}
			
			\label{fig:supp_top25_bottom25_boxplot}
		\end{figure}

		To further assess the statistical significance of the differences in subsequent ideological change between high- and low-permeability users, we conducted two-sided Mann-Whitney $U$ tests using the same pooled top- and bottom-quartile samples (Supplementary Table~\ref{tab:supp_top_bottom_tests}). The Mann-Whitney tests showed differences in the distributions of bias change between the top and bottom quartiles of ideological permeability across the three examined ideological categories (LL, C, and RL; all $p<0.001$, as denoted by *** in Supplementary Fig.~\ref{fig:supp_top25_bottom25_boxplot}). The direction of the mean differences was consistent with the descriptive analysis: the mean difference between the top and bottom quartiles was -0.132 for Left-Leaning users, indicating relatively more leftward subsequent changes among high-permeability users, whereas the differences were 0.150 and 0.130 for Center and Right-Leaning users, respectively, indicating relatively more rightward subsequent changes among high-permeability users. Together, these results support the observed directional relationship between ideological permeability and subsequent ideological change.

		\begin{table}[htbp]
			\centering
			\caption{\textbf{Comparison of subsequent bias change between the top and bottom quartiles of ideological permeability.}
				Top 25\% and Bottom 25\% users were selected separately within each adjacent-period transition and initial ideological group and were subsequently pooled across transitions within each ideological group. Mean bias change is reported separately for the two groups. Difference is calculated as Top 25\% minus Bottom 25\%. Negative values therefore indicate relatively more leftward subsequent change among high-permeability users, whereas positive values indicate relatively more rightward subsequent change. Two-sided Mann-Whitney $U$ tests compare the distributions of bias change between the two groups.}
			\label{tab:supp_top_bottom_tests}
			
			\resizebox{\textwidth}{!}{
				\begin{tabular}{lrrrrrr}
					\toprule
					Ideological group &
					Top 25\% mean &
					Bottom 25\% mean &
					Difference &
					Mann-Whitney $U$ &
					Mann-Whitney $p$ \\
					\midrule
					
					Left-Leaning
					& -0.071
					& 0.061
					& -0.132
					& $3.568\times10^{10}$
					& $<0.001$\\
					
					Center
					& -0.216
					& -0.366
					& 0.150
					& $2.030\times10^{9}$
					& $<0.001$\\
					
					Right-Leaning
					& 0.292
					& 0.162
					& 0.130
					& $1.295\times10^{9}$
					& $<0.001$\\
					
					\bottomrule
				\end{tabular}
			}
			
		\end{table}
		
		\FloatBarrier
		\phantomsection

		\FloatBarrier
		\phantomsection
		\subsection*{S2.3 Complementary results for Fig. 3}
		\addcontentsline{toc}{subsection}{S2.3 Complementary results for Fig. 3}
		\label{subsec:s2_3_fig3}
		
		\begin{figure}[htbp]
			\centering
			\includegraphics[width=\textwidth]{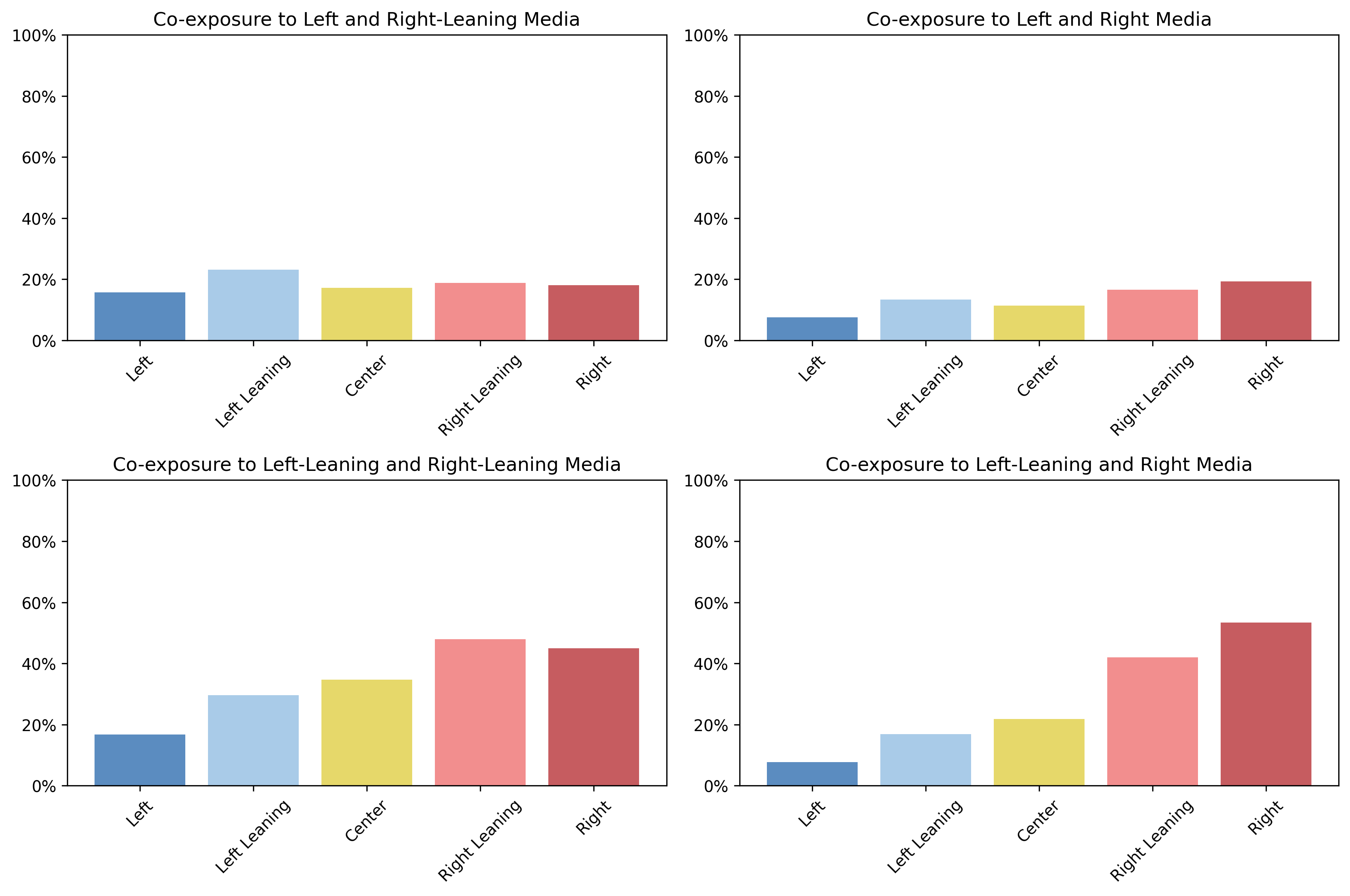}
			\caption{\textbf{Normalized user ideological composition of shared audiences across four cross-camp media-category combinations.}
				Each panel represents one of four combinations formed by pairing a Left or Left-Leaning media category with a Right-Leaning or Right media category. The analysis was restricted to media sources included in the final co-exposure network shown in Fig.~3. For each combination, we identified users who retweeted at least one media account from each of the two categories during the full observation period. The shared-audience set was constructed by taking the union of the pairwise overlaps between accounts from the two categories, with each user counted only once within each combination. Bars show the pooled proportion of users in each ideological category who belonged to the corresponding full-period shared-audience set, calculated across the six analysis periods. For each ideological category, the numerator was the sum of the period-specific counts of users belonging to the shared-audience set, and the denominator was the sum of the total number of users in that category across the same six periods. This normalization accounts for differences in the sizes of ideological groups. Users appearing in multiple periods were counted once in each period in which they appeared. Five user ideological categories are shown: Left, Left-Leaning, Center, Right-Leaning, and Right. The resulting proportions represent the prevalence of membership in the full-period shared-audience set.}
			\label{fig:s2_3_crosscamp}
		\end{figure}
		
		To account for differences in the sizes of user ideological groups, we normalized shared-audience counts by the corresponding group populations. Using the media accounts retained in the co-exposure network, we examined four cross-camp media-category combinations, pairing each of the two left-side categories (Left and Left-Leaning) with each of the two right-side categories (Right-Leaning and Right). For each combination, we identified the users shared by each pair of media accounts, with one account from each category, and took the union of these pairwise shared-audience sets. Duplicate user IDs were removed so that each user was counted only once within each media-category combination. These shared-audience sets were constructed over the full observation period. We then matched each set to users’ quarter-specific ideological classifications across the six analysis periods. For each ideological group, we summed the quarter-specific counts of users belonging to the full-period shared-audience set across the six quarters and divided this total by the sum of the total number of users in that ideological group across six quarters. This pooled rate represents the proportion of quarter-specific user observations in each ideological group that belonged to the full-period shared-audience set.

		Figure~3 shows a clear asymmetry in the structure of the left- and right-side media ecosystems in the co-exposure network. Influence on the left is more dispersed, whereas influence on the right is more concentrated. This media-level concentration appears to contrast with the relatively high ideological permeability observed among right-side users in Fig.~2b. To further examine this, we analyzed four cross-camp media combinations formed by pairing two left-side media categories (Left and Left-Leaning) with two right-side categories (Right-Leaning and Right), and identified unique users appearing in at least one pairwise overlap between media accounts from the two categories over the full observation period. To account for differences in the sizes of ideological groups, the number of users belonging to each shared-audience set was normalized by the corresponding group population, pooled across the six analysis periods, as shown in Supplementary Fig.~\ref{fig:s2_3_crosscamp}. The results reveal a notable pattern among Right-Leaning and Right users. Their normalized shared-audience rates were particularly high in combinations involving Left-Leaning media. For Right-Leaning users, the rates reached 42.03\% and 47.98\% in the LL+R and LL+RL combinations, respectively, compared with 16.52\% and 18.78\% in the corresponding L+R and L+RL combinations. Right users showed a similar pattern, with rates of 53.38\% for LL+R and 44.93\% for LL+RL, compared with 19.27\% and 18.06\% for L+R and L+RL, respectively. Center users also showed higher rates in combinations involving Left-Leaning media, reaching 21.81\% for LL+R and 34.64\% for LL+RL. Although the rates among Left and Left-Leaning users were generally lower, they also increased in combinations involving Left-Leaning rather than Left media. For Left users, the rates were 7.73\% in LL+R and 16.76\% in LL+RL, compared with 7.46\% in L+R and 15.67\% in L+RL. For Left-Leaning users, the corresponding rates were 16.86\% and 29.65\%, compared with 13.36\% and 23.07\%, respectively. This contrast indicates that the relatively high normalized shared-audience rates among Left-Leaning users were concentrated in combinations involving Left-Leaning media. At the same time, Right and Right-Leaning users also showed particularly high normalized shared-audience rates in combinations involving Left-Leaning media. Figure~3 in the main text captures the top-down organization of media influence through the co-exposure network, whereas this analysis reveals that Right and Right-Leaning users more frequently occupied the overlapping audiences of left- and right-side media. These findings provide complementary evidence for the higher ideological permeability observed among Right-Leaning users in Fig.~2b.

		\FloatBarrier
		\phantomsection
		\subsection*{S2.4 Complementary results for Fig. 5}
		\addcontentsline{toc}{subsection}{S2.4 Complementary results for Fig. 5}
		\label{subsec:s2_4_complementary_results}
		\begin{figure}[htbp]
			\centering
			\includegraphics[width=\textwidth]{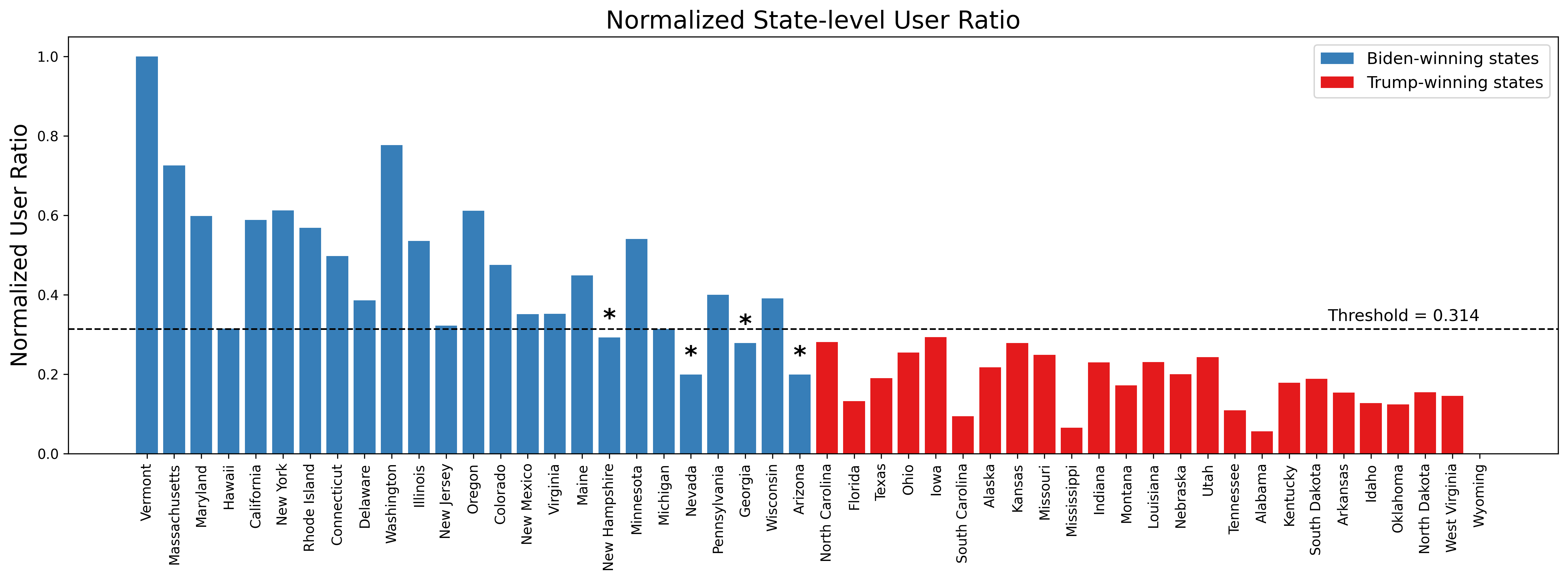}
			\caption{
				\textbf{Normalized state-level user ratio.}
				The ratio is calculated as the number of left-wing users divided by the number of right-wing users within each state, followed by normalization of the resulting ratio. States are ordered according to the Democratic vote share in the 2020 U.S. presidential election, from highest to lowest, with blue representing Biden and red representing Trump. The dashed horizontal lines indicate the empirical thresholds used to categorize states into left-wing, neutral, and right-wing groups. States marked with an asterisk ($*$) indicate the incorrect states where the user ratio deviates from the actual voting outcome in this comparison. This metric characterizes the relative ideological composition of Twitter users within each state and is used to examine state-level differences in online political information environments.
			}
			\label{fig:s2_4_state_ratio}
		\end{figure}
		
		As a complement to Fig. 5a in the main text, the normalized state-level user ratio illustrates the relationship between the ideological composition of Twitter users and state-level electoral patterns as shown in Supplementary Fig.~\ref{fig:s2_4_state_ratio}. States are ordered according to Democratic vote share in the 2020 U.S. presidential election, from highest to lowest. The ratio generally decreases along this electoral gradient, with Democratic-leaning states exhibiting higher proportions of left-wing users and Republican-leaning states exhibiting lower proportions. For example, strongly Democratic states such as Vermont, Massachusetts, California, and Washington show higher left-wing user ratios, whereas strongly Republican states such as Alabama, South Dakota, Idaho, and Wyoming show much lower ratios. Swing states located between these two groups exhibit intermediate values, reflecting a gradual transition in the online ideological composition of Twitter users across the electoral spectrum. The four states marked with asterisks, New Hampshire, Nevada, Georgia, and Arizona, were won by Biden yet have user ratios below the threshold; these are the states in which the online composition, taken alone, would misclassify the electoral outcome. However, the state-level user ratio does not indicate that Twitter users constitute a representative sample of the electorate. It characterizes differences in the ideological composition of the online political environment across states.

		\begin{figure}[htbp]
			\centering
			\includegraphics[width=\textwidth]{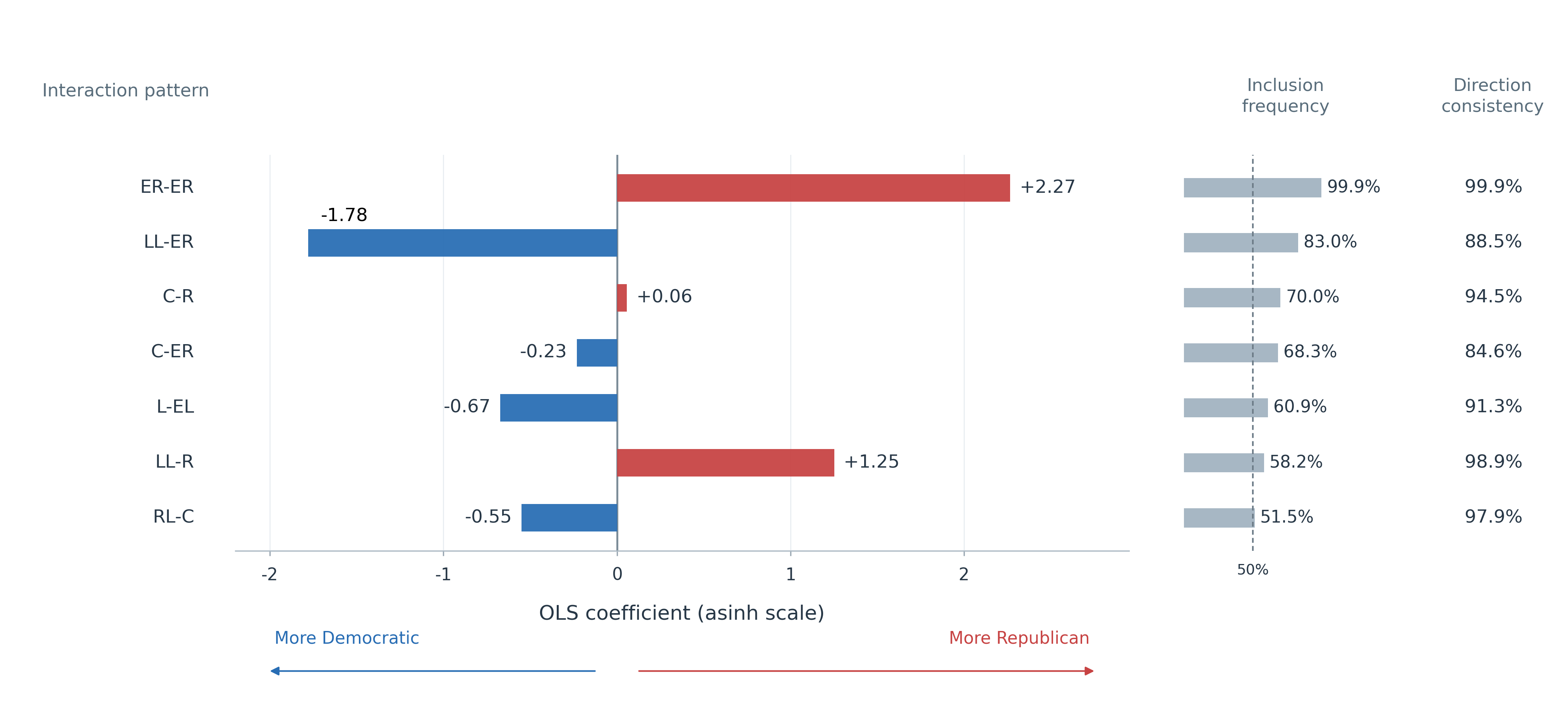}
			\caption{
				\textbf{Post-selection model coefficients and Bootstrap-Lasso selection.}
				The left panel shows the OLS coefficients of the seven predictors retained by the Bootstrap-Lasso procedure with an inclusion frequency exceeding 50\%, estimated from the post-selection OLS model. Positive coefficients indicate interaction patterns associated with larger Republican margins, while negative coefficients indicate patterns associated with larger Democratic margins. The right panel displays the inclusion frequency, represented by gray bars with a dashed vertical line marking the 50\% threshold. Direction consistency is calculated as the percentage of resamples in which the predictor's coefficient has the same sign as in the final model, considering only resamples in which the predictor was selected.
			}
			\label{fig:s2_4_model_fit_coef}
		\end{figure}
		
		As a complement to Fig. 5c in the main text, while the user ratio summarizes the overall ideological composition of each state's online population, we next ask whether the structure of user-media interactions carries additional electoral signals. For each state, we constructed 49 interaction features describing the engagement between each of the seven user groups and each of the seven media groups. Interaction features accounting for less than 0.1\% of all user–media interactions were excluded, leaving 30 features for analysis. Because raw interaction frequencies vary in scale across states and across patterns, each feature was CLR-transformed, so that it reflects the relative enrichment or depletion of an interaction within the state's overall interaction structure rather than its absolute volume. A pattern with a small absolute frequency can therefore still carry predictive information if its relative variation across states consistently tracks electoral differences. We applied a Bootstrap-Lasso procedure with 10,000 iterations, repeatedly resampling the states and fitting a Lasso regression of the Republican-minus-Democratic vote-share margin on these 30 features in each resample. The Bootstrap-Lasso procedure identifies such features based on their stability across repeated resampling and their conditional association with electoral margins. Seven features exceeded the 50\% inclusion threshold: ER-ER (99.9\%), LL-ER (83.0\%), C-R (70.0\%), C-ER (68.3\%), L-EL (60.9\%), LL-R (58.2\%), and RL-C (51.5\%). Direction consistency was calculated among resamples in which each feature was selected, as the proportion of those resamples in which its coefficient sign agreed with that of the corresponding coefficient in the final model. This conditional direction consistency exceeded 84\% for all seven features, indicating that their estimated associations were directionally stable across bootstrap resamples. 
		
		Supplementary Fig.~\ref{fig:s2_4_model_fit_coef} evaluates the coefficient structure of the state-level regression model. Because Lasso coefficients are shrunk toward zero and therefore biased, the seven retained features were refit in an ordinary least squares (OLS) estimated on the 38-state sample, with feature transformation and standardization re-estimated on the same sample. ER-ER exhibits the largest positive coefficient ($+2.27$): states in which engagement between extreme-right users and extreme-right media is relatively more pronounced tend to show larger Republican margins. LL-ER exhibits the largest negative coefficient ($-1.78$), indicating that relatively stronger engagement of left-leaning users with extreme-right media is associated with Democratic advantages. The second-largest positive coefficient, LL-R ($+1.25$), contrasts with LL-ER: engagement of the same user group with right-wing rather than extreme-right media is associated with Republican advantages, suggesting that the electoral signal depends on the specific configuration of cross-ideological exposure rather than on cross-ideological exposure per se. L-EL ($-0.67$) and RL-C ($-0.55$) are negatively associated with the margin, whereas C-ER ($-0.23$) and C-R ($+0.06$) show weak associations. Together with their high direction consistency, these coefficients indicate that the selected retweeting patterns represent stable state-level signals associated with electoral differences.

		\begin{figure}[htbp]
			\centering
			\includegraphics[height=0.85\textheight, keepaspectratio]{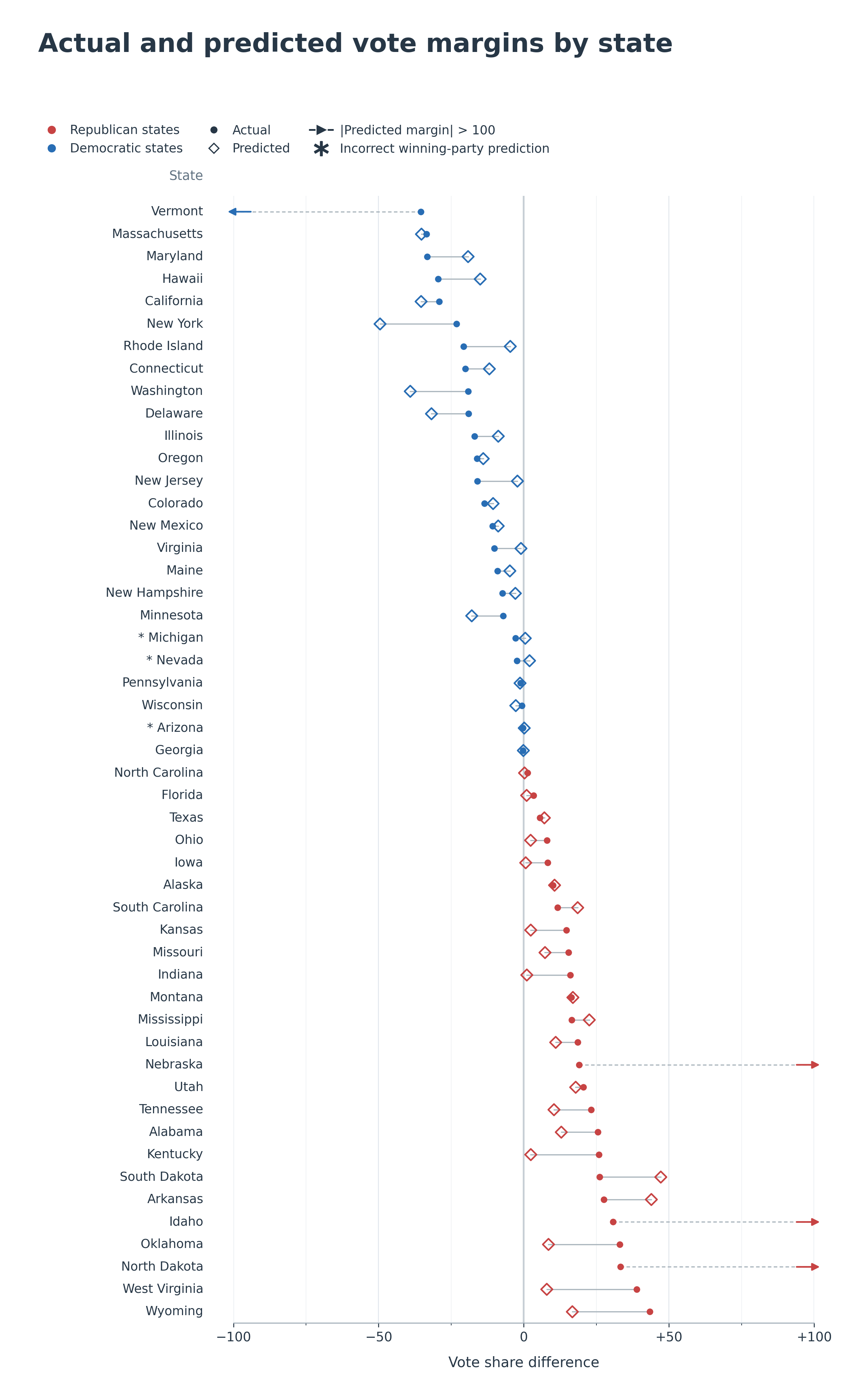}
			\caption{
				\textbf{Observed and predicted Republican-Democratic vote-share margins across states.}
				For each state, the observed electoral margin marked by a filled circle and the model-predicted margin marked by an open diamond are shown. States are ordered by the observed margin from most Democratic at the top to most Republican at the bottom and are colored by the winning party. Gray segments connect the observed and predicted values and visualize the magnitude and direction of the prediction deviation. Asterisks mark states for which the model predicts the wrong winning party, and dashed arrows indicate predictions exceeding the plotted range with an absolute predicted margin greater than 100.
			}
			\label{fig:s2_4_prediction_states}
		\end{figure}
		
		Finally, we compared the observed and predicted Republican-Democratic vote-share margins for each state (Supplementary Fig.~\ref{fig:s2_4_prediction_states}). The predictions were generated from the post-selection OLS model using the seven retained predictors, fitted on the 38-state sample. The predicted margins align closely with the ordering of states along the Democratic-Republican electoral spectrum (Spearman $\rho = 0.92$, $P = 1.3 \times 10^{-21}$). States won by Democrats are predicted to have negative margins, whereas states won by Republicans are predicted to have positive margins, and 47 of the 50 states are correctly classified in terms of the winning party. The three misclassified states, namely Michigan, Nevada, and Arizona, marked with asterisks, were all closely contested states with observed margins smaller than 3 percentage points; Nevada and Arizona were also misclassified by the user-ratio threshold in Supplementary Fig.~\ref{fig:s2_4_state_ratio}. For these states, the observed and predicted margins differ by only a few percentage points. Deviations remain moderate elsewhere, as illustrated by Alabama, which has an observed margin of $+25.46$ percentage points and a predicted margin of $+12.93$ percentage points, yielding an absolute deviation of $12.53$ percentage points, with both values indicating a Republican advantage. Overall, the agreement between the observed and predicted signs and relative magnitudes across 47 of 50 states supports the conclusion that aggregated user-media interaction patterns contain signals associated with state-level electoral outcomes.

		\clearpage
		\phantomsection
		\section*{S3 Ideological scoring labels}
		\addcontentsline{toc}{section}{S3 Ideological scoring labels}
		\label{sec:supp_entity_dictionaries}

		\phantomsection
		\subsection*{S3.1 Political labels in the co-retweet network}
		\addcontentsline{toc}{subsection}{S3.1 Political labels in the co-retweet network}
		\label{subsec:s3_1_political_entities}
		The political co-retweet network shown in the main text contains 207 political-entity nodes. 
		Each node was matched to the final political-entity dictionary and assigned an ideological 
		label of Left, Center, or Right. Table~\ref{tab:supp_political_network_entities} reports 
		the public display name, screen name, and ideological label of all 207 nodes included in 
		the network. For privacy and data-minimization considerations, persistent numerical Twitter 
		user IDs used internally for record linkage are not reported; entities are identified only 
		by their public display names and screen names. Retweet counts are not reported because 
		this table is intended to document the identities and ideological labels of the network 
		nodes rather than their activity volumes.
		
		\begin{longtable}{
				@{}
				>{\raggedright\arraybackslash}p{0.48\textwidth}
				>{\raggedright\arraybackslash}p{0.30\textwidth}
				>{\raggedright\arraybackslash}p{0.14\textwidth}
				@{}
			}
			\caption{Political entities included in the co-retweet network and their ideological labels.}
			\label{tab:supp_political_network_entities}\\
			
			\toprule
			\textbf{Political entity} & \textbf{Username} & \textbf{Ideological label} \\
			\midrule
			\endfirsthead
			
			\multicolumn{3}{c}{
				\tablename\ \thetable{} -- continued from previous page
			}\\
			\toprule
			\textbf{Political entity} & \textbf{Username} & \textbf{Ideological label} \\
			\midrule
			\endhead
			
			\midrule
			\multicolumn{3}{r}{Continued on next page}\\
			\endfoot
			
			\bottomrule
			\endlastfoot
			
			Donald J. Trump & realDonaldTrump & Right \\
			Joe Biden & JoeBiden & Left \\
			Bernie Sanders & BernieSanders & Left \\
			Kamala Harris & KamalaHarris & Left \\
			Andrew Yang & AndrewYang & Left \\
			The Lincoln Project & ProjectLincoln & Left \\
			Pete Buttigieg & PeteButtigieg & Left \\
			Elizabeth Warren & ewarren & Left \\
			Vice President Kamala Harris & VP & Left \\
			Charlie Kirk & charliekirk11 & Right \\
			BrooklynDad\_Defiant! & mmpadellan & Left \\
			Don Winslow & donwinslow & Left \\
			Team Trump (Text TRUMP to 88022) & TeamTrump & Right \\
			The White House 45 Archived & WhiteHouse45 & Right \\
			MeidasTouch & MeidasTouch & Left \\
			Scott Dworkin & funder & Left \\
			Donald Trump Jr. & DonaldJTrumpJr & Right \\
			Kyle Griffin & kylegriffin1 & Left \\
			Terrence K. Williams & w\_terrence & Right \\
			Barack Obama & BarackObama & Left \\
			Joe Walsh & WalshFreedom & Right \\
			Robert Reich & RBReich & Left \\
			Jeff Tiedrich & itsJeffTiedrich & Left \\
			Aaron Rupar & atrupar & Left \\
			Hillary Clinton & HillaryClinton & Left \\
			Eugene Gu, MD & eugenegu & Left \\
			Jack Poso & JackPosobiec & Right \\
			GOP & GOP & Right \\
			Stacey Abrams & staceyabrams & Left \\
			Kayleigh McEnany & kayleighmcenany & Right \\
			President Biden & POTUS & Left \\
			Amy Klobuchar & amyklobuchar & Left \\
			Trump War Room & TrumpWarRoom & Right \\
			NoelCaslerComedy & caslernoel & Left \\
			Brian Tyler Cohen & briantylercohen & Left \\
			Andrea Junker & Strandjunker & Left \\
			Ryan Fournier & RyanAFournier & Right \\
			Jon Cooper & joncoopertweets & Left \\
			Rudy W. Giuliani & RudyGiuliani & Right \\
			George Conway & gtconway3d & Right \\
			Ari Berman & AriBerman & Left \\
			Candace Owens & RealCandaceO & Right \\
			Dan Rather & DanRather & Left \\
			Tammy Duckworth & SenDuckworth & Left \\
			Rex Chapman & RexChapman & Left \\
			Tulsi Gabbard & TulsiGabbard & Left \\
			Elizabeth Warren & SenWarren & Left \\
			Catturd & catturd2 & Right \\
			Jenna Ellis & JennaEllisEsq & Right \\
			Rep. Jim Jordan & Jim\_Jordan & Right \\
			George Takei & GeorgeTakei & Left \\
			Ronna McDaniel & RonnaMcDaniel & Right \\
			Eric Trump & EricTrump & Right \\
			Taylor Swift & taylorswift13 & Left \\
			\#ThePersistence & ScottPresler & Right \\
			Rep. Matt Gaetz & RepMattGaetz & Right \\
			Jim Acosta & Acosta & Left \\
			Dinesh D'Souza & DineshDSouza & Right \\
			Joe Lockhart & joelockhart & Left \\
			LeoTerrell & TheLeoTerrell & Right \\
			Kayleigh McEnany 45 Archived & PressSec45 & Right \\
			Mike Pence & Mike\_Pence & Right \\
			David J Harris Jr & DavidJHarrisJr & Right \\
			Anthony Scaramucci & Scaramucci & Center \\
			Lin Wood & LLinWood & Right \\
			Jake Tapper & jaketapper & Center \\
			Kaitlan Collins & kaitlancollins & Center \\
			Anonymous & YourAnonCentral & Center \\
			Amy Siskind & Amy\_Siskind & Left \\
			Dan Scavino Jr. & DanScavino & Right \\
			Daniel Dale & ddale8 & Center \\
			Stephen King & StephenKing & Left \\
			Alex Cole & acnewsitics & Left \\
			Rep. Eric Swalwell & RepSwalwell & Left \\
			Rob Reiner & robreiner & Left \\
			Benny Johnson & bennyjohnson & Right \\
			Marianne Williamson & marwilliamson & Left \\
			Lindsey Graham & LindseyGrahamSC & Right \\
			Sarah Cooper & sarahcpr & Left \\
			Former Congresswoman Val Demings & RepValDemings & Left \\
			Kevin Sorbo & ksorbs & Right \\
			Really American & ReallyAmerican1 & Left \\
			Ken Olin & kenolin1 & Left \\
			Adam Parkhomenko & AdamParkhomenko & Left \\
			Brigitte Gabriel & ACTBrigitte & Right \\
			Ana Navarro-Cárdenas & ananavarro & Right \\
			Lauren Boebert & laurenboebert & Right \\
			James Woods & RealJamesWoods & Right \\
			Cory Booker & CoryBooker & Left \\
			Tom Fitton & TomFitton & Right \\
			Steve Scalise & SteveScalise & Right \\
			Chip Franklin.com & chipfranklin & Left \\
			Steve Schmidt & SteveSchmidtSES & Left \\
			Yamiche Alcindor & Yamiche & Left \\
			Matt Couch & RealMattCouch & Right \\
			Marjorie Taylor Greene & mtgreenee & Right \\
			James O'Keefe & JamesOKeefeIII & Right \\
			Robby Starbuck & robbystarbuck & Right \\
			Brian J. Karem & BrianKarem & Left \\
			Fred Guttenberg & fred\_guttenberg & Left \\
			Michael Beschloss & BeschlossDC & Left \\
			Bill Kristol & BillKristol & Right \\
			Senator Ted Cruz & SenTedCruz & Right \\
			Sara A. Carter & SaraCarterDC & Right \\
			Adam Schiff & RepAdamSchiff & Left \\
			David Axelrod & davidaxelrod & Left \\
			Alexandria Ocasio-Cortez & AOC & Left \\
			Alyssa Milano & Alyssa\_Milano & Left \\
			Secretary Pompeo & SecPompeo & Right \\
			Lou Dobbs & LouDobbs & Right \\
			Andrew Pollack & AndrewPollackFL & Right \\
			Mike Bloomberg & MikeBloomberg & Left \\
			Kambree & KamVTV & Right \\
			John Cardillo & johncardillo & Right \\
			Steven Beschloss & StevenBeschloss & Left \\
			DC\_Draino & DC\_Draino & Right \\
			Brandon Straka & BrandonStraka & Right \\
			Juanita Broaddrick & atensnut & Right \\
			Julián Castro & JulianCastro & Left \\
			Paul Sperry & paulsperry\_ & Right \\
			Rick Wilson & TheRickWilson & Right \\
			Chris Murphy & ChrisMurphyCT & Left \\
			Chuck Schumer & SenSchumer & Left \\
			Chris Hayes & chrislhayes & Left \\
			Eric Swalwell & ericswalwell & Left \\
			Ivanka Trump & IvankaTrump & Right \\
			Cernovich & Cernovich & Right \\
			MURRAY & murray\_nyc & Center \\
			The White House & WhiteHouse & Center \\
			Cari Kelemen & KelemenCari & Right \\
			Ricky Davila & TheRickyDavila & Left \\
			Claire McCaskill & clairecmc & Left \\
			Jim Sciutto & jimsciutto & Left \\
			Chuck Callesto & ChuckCallesto & Right \\
			Geoff Bennett & GeoffRBennett & Left \\
			Tea Pain & TeaPainUSA & Left \\
			Chris D. Jackson & ChrisDJackson & Left \\
			Beto O'Rourke & BetoORourke & Left \\
			Mkay & JoyAnnReid & Left \\
			Dena Grayson, MD, PhD & DrDenaGrayson & Left \\
			A Worried Citizen & ThePubliusUSA & Center \\
			Laurence Tribe & tribelaw & Left \\
			Geraldo Rivera & GeraldoRivera & Center \\
			Mark R. Levin & marklevinshow & Right \\
			Andy Ostroy & AndyOstroy & Left \\
			David Corn & DavidCornDC & Left \\
			Glenn Kirschner & glennkirschner2 & Left \\
			Angela Belcamino & AngelaBelcamino & Left \\
			Power to the People & ProudSocialist & Left \\
			The Hoarse Whisperer & TheRealHoarse & Left \\
			Manu Raju & mkraju & Center \\
			John O. Brennan & JohnBrennan & Left \\
			Tim Young & TimRunsHisMouth & Right \\
			Steve Guest & SteveGuest & Right \\
			David Leavitt Pax East & David\_Leavitt & Left \\
			Bernard B. Kerik & BernardKerik & Right \\
			Nick Adams (Alpha Male) & NickAdamsinUSA & Right \\
			Mark Meadows & MarkMeadows & Right \\
			Melania Trump 45 Archived & FLOTUS45 & Right \\
			George Papadopoulos & GeorgePapa19 & Right \\
			Bishop Talbert Swan & TalbertSwan & Left \\
			Nancy Pelosi & SpeakerPelosi & Left \\
			John Pavlovitz & johnpavlovitz & Left \\
			David Rothkopf & djrothkopf & Left \\
			Hugo Lowell & hugolowell & Left \\
			Lance Gooden & Lancegooden & Right \\
			Scott Adams & ScottAdamsSays & Right \\
			Andrew Feinberg & AndrewFeinberg & Left \\
			Herschel Walker & HerschelWalker & Right \\
			Evan Kilgore & EvanAKilgore & Right \\
			Mary L Trump & MaryLTrump & Left \\
			Students For Trump & TrumpStudents & Right \\
			Hoodlum & NotHoodlum & Left \\
			Elie Mystal & ElieNYC & Left \\
			Billy Baldwin & BillyBaldwin & Left \\
			Paul the other one... & paulcshipley & Left \\
			Daniel Goldman & danielsgoldman & Left \\
			Laura Ingraham & IngrahamAngle & Right \\
			Narendra Modi & narendramodi & Center \\
			Republican Voters Against Trump & AccountableGOP & Left \\
			Charlotte Clymer & cmclymer & Left \\
			Kurt Eichenwald & kurteichenwald & Left \\
			Jon Favreau & jonfavs & Left \\
			Nick Knudsen & NickKnudsenUS & Left \\
			Jennifer Truth Over Phony Balance Rubin & JRubinBlogger & Left \\
			Lindy Li & lindyli & Left \\
			Dr. Rob Davidson MD, MPH & DrRobDavidson & Left \\
			Emerald Robinson & EmeraldRobinson & Right \\
			Sidney Powell Attorney \& Author & SidneyPowell1 & Right \\
			Brad Parscale & parscale & Right \\
			Ben Shapiro & benshapiro & Right \\
			Rand Paul & RandPaul & Right \\
			Lisa Boothe & LisaMarieBoothe & Right \\
			The Spectator Index & spectatorindex & Center \\
			Ana Cabrera & AnaCabrera & Left \\
			Republicans against Trump & RpsAgainstTrump & Left \\
			Cenk Uygur & cenkuygur & Left \\
			Hamish Mitchell & H\_MitchellPhoto & Center \\
			D Villella & dvillella & Left \\
			(((DeanObeidallah))) & DeanObeidallah & Left \\
			Andrew Weinstein & Weinsteinlaw & Center \\
			Graham Allen & GrahamAllen\_1 & Right \\
			Ron & CodeMonkeyZ & Right \\
			Dave Wasserman & Redistrict & Center \\
			Tomi Lahren & TomiLahren & Right \\
			Jo & JoJoFromJerz & Left \\
			Michael McFaul & McFaul & Left \\
			
		\end{longtable}

		\phantomsection
		\subsection*{S3.2 Media labels in the co-retweet network}
		\addcontentsline{toc}{subsection}{S3.2 Media labels in the co-retweet network}
		\label{subsec:s3_2_media_entities}
		
		The media co-retweet network shown in the main text contains 152 media nodes. Table~\ref{tab:supp_media_network_entities} reports all media accounts included in the network and their ideological labels. The entities are ordered by their retweet counts from highest to lowest. 
		
		\begin{longtable}{
				@{}
				>{\raggedright\arraybackslash}p{0.70\textwidth}
				>{\raggedright\arraybackslash}p{0.24\textwidth}
				@{}
			}
			\caption{Media accounts included in the co-retweet network and their ideological labels.}
			\label{tab:supp_media_network_entities}\\
			
			\toprule
			\textbf{Media accounts} & \textbf{Ideological label} \\
			\midrule
			\endfirsthead
			
			\multicolumn{2}{c}{
				\tablename\ \thetable{} -- continued from previous page
			}\\
			\toprule
			\textbf{Media accounts} & \textbf{Ideological label} \\
			\midrule
			\endhead
			
			\midrule
			\multicolumn{2}{r}{Continued on next page}\\
			\endfoot
			
			\bottomrule
			\endlastfoot
			
			Cable News Network & Left-Leaning \\
			The New York Times & Left-Leaning \\
			The Washington Post & Left-Leaning \\
			The Hill & Center \\
			National Broadcasting Company News & Left-Leaning \\
			Columbia Broadcasting System News & Left-Leaning \\
			American Broadcasting Company & Left-Leaning \\
			Politico & Left-Leaning \\
			Fox News Channel & Right \\
			The Associated Press & Left-Leaning \\
			Axios & Left-Leaning \\
			Thomson Reuters & Center \\
			The New York Post & Right-Leaning \\
			Microsoft and National Broadcasting Company & Left \\
			Consumer News and Business Channel & Left-Leaning \\
			The Washington Examiner & Right \\
			Breitbart News Network & Extreme Right \\
			The Wall Street Journal & Right-Leaning \\
			Bloomberg L.P. & Left-Leaning \\
			The Recount & Center \\
			The Daily Caller & Right \\
			The Gateway Pundit & Extreme Right \\
			USA TODAY & Left-Leaning \\
			National Public Radio & Left-Leaning \\
			The Guardian & Left-Leaning \\
			Newsweek & Right-Leaning \\
			Yahoo News & Left-Leaning \\
			The Huffington Post & Left \\
			Business Insider & Left-Leaning \\
			NowThis News & Extreme Left \\
			Forbes & Right-Leaning \\
			The Daily Beast & Left \\
			TIME Magazine & Left-Leaning \\
			The Atlantic & Left-Leaning \\
			Los Angeles Times & Left-Leaning \\
			Daily Mail & Extreme Right \\
			British Broadcasting Corporation, BBC & Left-Leaning \\
			Raw Story & Left \\
			Vox Media & Left \\
			Newsmax Media, Inc & Extreme Right \\
			The Independent & Left-Leaning \\
			The Epoch Times & Right \\
			The Daily Show & Left-Leaning \\
			The Daily Wire & Right \\
			The Washington Times & Right-Leaning \\
			Cable-Satellite Public Affairs Network & Center \\
			Mediaite & Left \\
			Palmer Report & Left \\
			Mother Jones & Left-Leaning \\
			BuzzFeed News & Left-Leaning \\
			Breaking911 & Extreme Right \\
			Vanity Fair & Left \\
			Al Jazeera Media Network & Left-Leaning \\
			Entertainment and Sports Programming Network & Left-Leaning \\
			One America News Network & Right \\
			National Review Online & Right \\
			RightSide Broadcasting Network & Right \\
			PoliticusUSA & Left \\
			TheBlaze & Right \\
			Popular Information & Left \\
			Election Wizard & Right-Leaning \\
			Townhall & Right \\
			NPR Politics & Left \\
			Judicial Watch, Inc. & Extreme Right \\
			The Washington Free Beacon & Extreme Right \\
			Disclose.tv & Right \\
			Agence France-Presse (AFP) & Center \\
			The Daily Telegraph & Right \\
			Rasmussen Reports & Right-Leaning \\
			The First TV & Right \\
			The Jerusalem Post & Right-Leaning \\
			Thirty Mile Zone & Right-Leaning \\
			RedState & Right \\
			Reagan Battalion & Right-Leaning \\
			The Associated Press\_Politics & Center \\
			The American Independent & Left \\
			The Daily Edge & Right-Leaning \\
			OutKick & Right-Leaning \\
			SCOTUSblog & Center \\
			New Tang Dynasty Television & Right \\
			Democratic National Convention & Left \\
			Mediavenir & Center \\
			Baller Alert & Center \\
			LifeNews.com & Extreme Right \\
			The Tennessee Holler & Left \\
			Asian News International & Right-Leaning \\
			Media Research Center TV & Right-Leaning \\
			The Conservative Treehouse & Extreme Right \\
			Zero Hedge & Extreme Right \\
			WRIC-TV & Center \\
			Iowa Starting Line & Left \\
			Fxhedgers & Center \\
			Democracy Now & Left \\
			Agencia EFE Noticias & Center \\
			WSPA-TV & Right-Leaning \\
			Independent Television News & Right-Leaning \\
			XXL Magazine & Center \\
			Advertising Age & Center \\
			KTVN 2 News & Center \\
			BNO News & Center \\
			WBNS-10TV & Center \\
			RT (Russia Today) & Right-Leaning \\
			Business Frequency Modulation TV & Center \\
			The Young, Black, and Fabulous & Center \\
			Media Matters for America & Left \\
			Just Security & Left \\
			Real America's Voice & Right \\
			The Insider Paper & Right-Leaning \\
			The Federalist & Right \\
			The Economist & Center \\
			Texas Signal & Left \\
			Global Times & Left \\
			PanAm Post & Right \\
			Upward News & Right-Leaning \\
			O Globo & Right-Leaning \\
			Rolling Stone & Left \\
			The American Prospect & Left \\
			Breaking News & Extreme Right \\
			PopWrapped & Center \\
			Financial Times & Center \\
			Stars and Stripes & Center \\
			Latinus & Left \\
			FOX 13 News & Center \\
			KTLA & Center \\
			Status Coup & Left \\
			Democracy Docket & Left \\
			The National Pulse & Extreme Right \\
			WEAR-TV & Right-Leaning \\
			Crain's Chicago Business & Right-Leaning \\
			Yonhap News Agency & Right-Leaning \\
			El Mundo & Right-Leaning \\
			NewsBusters & Extreme Right \\
			Total News World & Right \\
			Jezebel & Extreme Left \\
			The Post Millennial & Right \\
			Not the Bee & Right \\
			Right Side Broadcasting Network & Right \\
			Just the News & Right \\
			Ami Magazine & Right \\
			Sputnik News & Right \\
			NowThis News\textsuperscript{*} & Extreme Left \\
			Valeurs Actuelles & Extreme Right \\
			Occupy Democrats & Extreme Left \\
			Blunt Force Truth & Extreme Right \\
			Bossip & Extreme Left \\
			The Western Journal & Extreme Right \\
			Political Flare & Extreme Left \\
			NOQ Report & Extreme Right \\
			True Pundit & Extreme Right \\
			WorldNetDaily & Extreme Right \\
			The RightScoop & Extreme Right \\
			The DC Patriot & Extreme Right \\
			
		\end{longtable}

		\phantomsection 
		\subsection*{S3.3 Political URL labels} 
		\addcontentsline{toc}{subsection}{S3.3 Political URL labels} 
		\label{subsec:s4_3_political_dictionary} 
		
		A total of 35 political domains were identified and labeled in the political URL dictionary. All 35 entries are reported below together with their corresponding political entities and ideological labels. Table~\ref{tab:supp_political_domains} presents the original entries in the political-domain dictionary, preserving their recorded capitalization and any `www.` prefixes. These records are shown in their original form for documentation. Hostname standardization and deduplication were applied during URL matching.
		
		\begin{table}[htbp] 
			\centering 
			\small 
			\caption{Political domain labels used by the politician-based scoring system.} 
			\label{tab:supp_political_domains} 
			\begin{tabular}{llcl} 
				\toprule 
				\textbf{Domain} & \textbf{Full Name} & \textbf{Bias} \\ 
				\midrule 
				www.whitehouse.gov & The White House & Left \\ 
				joe.link & Joe Biden & Left \\ 
				hann.it & Sean Hannity & Right \\ 
				joebiden.com & Joe Biden & Left \\ 
				saraacarter.com & Sara A. Carter & Right \\ 
				Vote.DonaldJTrump.com & Donald J. Trump & Right \\ 
				45.wh.gov & The White House & Right \\ 
				JoeBiden.com & Joe Biden & Left \\ 
				nestorgomezstoryteller.com & Nestor Gomez & Center \\ 
				davidharrisjr.com & David James Harris Jr. & Right \\ 
				www.donaldjtrump.com & Donald J. Trump & Right \\ 
				djhjmedia.com & David James Harris Jr. & Right \\ 
				hannity.com & Sean Hannity & Right \\ 
				berniesanders.com & Bernie Sanders & Left \\ 
				gjarr.it & Giuseppe Conte & Center \\ 
				GAsenate.com & Georgia State Senate & Center \\ 
				vote.donaldjtrump.com & Donald J. Trump & Right \\ 
				buildbackbetter.gov & The White House & Left \\ 
				secure.joebiden.com & Joe Biden & Left \\ 
				www.justice.gov & U.S. Department of Justice & Center \\ 
				act.betoorourke.com & Beto O'Rourke & Left \\ 
				GASenate.com & Georgia State Senate & Center \\ 
				ewar.ren & Elizabeth Warren & Left \\ 
				www.speaker.gov & Nancy Pelosi & Left \\ 
				live.berniesanders.com & Bernie Sanders & Left \\ 
				lincolnproject.us & The Lincoln Project & Left \\ 
				events.donaldjtrump.com & Donald J. Trump & Right \\ 
				go.joebiden.com & Joe Biden & Left \\ 
				djt45.co & Donald J. Trump & Right \\ 
				mikebloom.bg & Mike Bloomberg & Left \\ 
				DJTVote.co & Donald J. Trump & Right \\ 
				www.hsgac.senate.gov & U.S. Senate Committee & Center \\ 
				gop.com & Republican National Committee & Right \\ 
				buildbackbetter.com & Build Back Better & Left \\ 
				ArmyForTrump.com & Donald J. Trump & Right \\ 
				\bottomrule 
			\end{tabular} 
		\end{table}

		\phantomsection
		\subsection*{S3.4 Media username labels}
		\addcontentsline{toc}{subsection}{S3.4 Media username labels}
		\label{subsec:s3_4_media_username_dictionary}
		
		The media username dictionary contained 663 username entries. Some media accounts appeared more than once in the dictionary because the same media organization could be associated with multiple accounts. Representative accounts were selected separately within each ideological category, prioritizing accounts with larger retweet counts. The complete dictionary contained only four unique Extreme Left accounts, all of which are therefore shown. Five accounts from each of the remaining ideological categories are included, giving 34 representative accounts in total.
		
		\begin{longtable}{
				@{}
				>{\raggedright\arraybackslash}p{0.32\textwidth}
				>{\raggedright\arraybackslash}p{0.20\textwidth}
				>{\raggedright\arraybackslash}p{0.28\textwidth}
				>{\raggedright\arraybackslash}p{0.16\textwidth}
				@{}
			}
			\caption{Representative media username labels used in the media-based ideological scoring system.}
			\label{tab:supp_media_username_examples}\\
			
			\toprule
			\textbf{Media organization} &
			\textbf{Username} &
			\textbf{Display name} &
			\textbf{Ideological label} \\
			\midrule
			\endfirsthead
			
			\multicolumn{4}{c}{
				\tablename\ \thetable{} -- continued from previous page
			}\\
			\toprule
			\textbf{Media organization} &
			\textbf{Username} &
			\textbf{Display name} &
			\textbf{Ideological label} \\
			\midrule
			\endhead
			
			\midrule
			\multicolumn{4}{r}{Continued on next page}\\
			\endfoot
			
			\bottomrule
			\endlastfoot
			
			Jezebel & Jezebel & Jezebel & Extreme Left \\
			NowThis News & nowthisimpact & NowThis Impact & Extreme Left \\
			Occupy Democrats & OccupyDemocrats & Occupy Democrats & Extreme Left \\
			Political Flare & nicolejames & Nicole Hickman James & Extreme Left \\
			
			NPR Politics & nprpolitics & NPR Politics & Left \\
			Vox Media & voxdotcom & Vox & Left \\
			Microsoft and National Broadcasting Company & DeadlineWH & Deadline White House & Left \\
			Iowa Starting Line & iastartingline & Iowa Starting Line & Left \\
			The Huffington Post & blackvoices & HuffPost BlackVoices & Left \\
			
			Cable News Network & cnnpolitics & CNN Politics & Left-Leaning \\
			The Associated Press & AP & The Associated Press & Left-Leaning \\
			Consumer News and Business Channel & CNBCnow & CNBCnow & Left-Leaning\\	
			Channel 4 News & Channel4News & Channel 4 News & Left-Leaning \\
			Public Broadcasting Service & NewsHour & PBS NewsHour & Left-Leaning \\

			Cable-Satellite Public Affairs Network & cspan & CSPAN & Center \\
			The Associated Press\_Politics & AP\_Politics & AP Politics & Center \\
			SCOTUSblog & scotusblog & SCOTUSblog & Center \\
			Baller Alert & balleralert & Baller Alert & Center \\
			Agence France-Presse (AFP) & AFP & AFP News Agency & Center \\
			
			The New York Post & nypost & New York Post & Right-Leaning \\
			The Washington Times & washtimes & The Washington Times & Right-Leaning \\
			Rasmussen Reports & Rasmussen\_Poll & Rasmussen Reports & Right-Leaning \\
			Newsweek & Newsweek & Newsweek & Right-Leaning \\
			The Wall Street Journal & WSJ & The Wall Street Journal & Right-Leaning \\

			The Washington Examiner & dcexaminer & Washington Examiner & Right \\
			Fox News Channel & wattersworld & Watters' World & Right \\
			Disclose.tv & disclosetv & Disclose.tv & Right \\
			The Daily Wire & realdailywire & Daily Wire & Right \\
			The Epoch Times & epochtimes & The Epoch Times & Right \\
			
			Breaking911 & Breaking911 & Breaking911 & Extreme Right \\
			Breitbart News Network & BreitbartNews & Breitbart News & Extreme Right \\
			The Gateway Pundit & gatewaypundit & The Gateway Pundit & Extreme Right \\
			Breaking News & BreakingNews & Breaking News & Extreme Right \\
			LifeNews.com & lifenewshq & LifeNews.com & Extreme Right \\
			
		\end{longtable}

	\end{appendices}